\documentclass[
  aps,
  prd,
  reprint,
  amsmath,amssymb,
  superscriptaddress,
  nofootinbib,
  floatfix,
]{revtex4-2}

\usepackage{graphicx}
\usepackage{mathrsfs}
\usepackage{slashed}
\usepackage{bm}
\usepackage{hyperref}
\hypersetup{colorlinks=true,linkcolor=blue,citecolor=blue,urlcolor=blue}
\usepackage{xcolor}
\newcommand{\blue}{\color{blue}}

\newcommand{\tr}{\mathrm{tr}}
\newcommand{\Tr}{\mathrm{Tr}}
\newcommand{\Id}{\mathbf{1}}
\newcommand{\Unit}{\mathbb{I}}   
\newcommand{\eps}{\varepsilon}    

\newcommand{\msbar}{\overline{\text{MS}}}
\newcommand{\vev}[1]{\langle #1 \rangle}

\begin{document}

\title{Renormalization of meson susceptibilities and
RG-invariant symmetry ratios in QCD}

\author{Ting-Wai Chiu}

\affiliation{Department of Physics, National Taiwan Normal University, Taipei~11677, Taiwan}
\affiliation{Institute of Physics, Academia Sinica, Taipei~11529, Taiwan}
\affiliation{Department of Physics, National Taiwan University, Taipei~10617, Taiwan}
\affiliation{Physics Division, National Center for Theoretical Sciences, Taipei~10617, Taiwan}
\affiliation{Nuclear Science Division, Lawrence Berkeley National Laboratory, Berkeley, CA 94720, USA}

\date{\today}

\begin{abstract}
We analyze the ultraviolet divergence structure of meson
susceptibilities in finite-temperature QCD, for lattice formulations
with exact chiral symmetry.
The bare susceptibility separates into additive divergences and a
multiplicative renormalization $Z_\Gamma^{2}$.
The additive divergences are temperature-independent, and are removed by
the temperature subtraction.
They consist of the leading power divergence $\alpha_\Gamma/(2a^2)$ from
the identity operator, together with a mass-dependent logarithmic term
$\propto m^2\ln(1/(am))$.
Exact chiral symmetry forbids all mass-dependent \emph{power} divergences
of the susceptibility.
The multiplicative factor $Z_\Gamma^{2}$ has a logarithmic dependence on
the lattice spacing, controlled by the operator anomalous dimension.
We show that the symmetry ratio
$\kappa_{AB} = |\chi_A^{\rm reg} - \chi_B^{\rm reg}|/
(|\chi_A^{\rm reg}| + |\chi_B^{\rm reg}|)$,
built from temperature-subtracted susceptibilities of symmetry partners,
is exactly renormalization-group invariant and scheme-independent.
The additive divergence is removed by the subtraction, and the
multiplicative factor cancels through the equality $Z_A = Z_B$.
This equality holds for any number of flavors and any quark masses in a
mass-independent scheme, unaffected by spontaneous symmetry breaking or
the $U(1)_A$ anomaly.
We derive the complete $Z$-factor chains for all meson channels and
contrast the divergence structure with that of Wilson fermions, for which
the explicit chiral-symmetry breaking induces a chiral-odd power-divergent
mixing and spoils the equality $Z_A = Z_B$ on which the construction relies.
\end{abstract}

\maketitle

\section{Introduction}
\label{sec:intro}

The restoration of chiral symmetry and $U(1)_A$ axial symmetry
at high temperature is a fundamental question in QCD.
Meson susceptibilities, integrated Euclidean correlators of
quark bilinear operators, provide natural probes of symmetry
breaking and restoration.
The degeneracy of susceptibilities in symmetry-related channels signals
effective symmetry restoration.

However, bare meson susceptibilities contain ultraviolet divergences
that complicate their interpretation.
There are two kinds.
One is the additive divergences arising from short distances.
The other is a multiplicative renormalization factor $Z_A^2$, which is
logarithmically divergent through the anomalous dimensions of the
composite operators.
The proper treatment of these divergences is essential for any
quantitative study of symmetry restoration.

We consider the symmetry ratio
\begin{equation}
\kappa_{AB} \equiv \frac{|\chi_A^{\rm reg} - \chi_B^{\rm reg}|}
{|\chi_A^{\rm reg}| + |\chi_B^{\rm reg}|},
\label{eq:kappa_def}
\end{equation}
where $\chi_A^{\rm reg}(T;T_r) \equiv \chi_A(T) - \chi_A(T_r)$
is the regularized susceptibility obtained by subtracting a
reference temperature $T_r$ at which the symmetry is effectively restored,
in the sense that the partner susceptibilities are degenerate within the
precision of the data (at nonzero quark mass the degeneracy is never exact,
see Sec.~\ref{subsec:T_sub}).
By the triangle inequality $0\le\kappa_{AB}\le1$, with $\kappa_{AB}=0$
only if $\chi_A^{\rm reg}=\chi_B^{\rm reg}$. When $\chi_A^{\rm reg}$ and
$\chi_B^{\rm reg}$ share a sign, as symmetry partners do away from the
crossover, $\kappa_{AB}=|(\chi_A^{\rm reg}-\chi_B^{\rm reg})/
(\chi_A^{\rm reg}+\chi_B^{\rm reg})|$. If they have opposite signs,
$\kappa_{AB}=1$, and the splitting is then as large as the
susceptibilities themselves. The operator phases are those for which the
symmetry implies $C_A=+C_B$, as confirmed by the degeneracy of the partners
at the reference temperature, and $\kappa_{AB}$ is defined relative to this
convention. Which partner is the larger
is recorded by the sign of the splitting
$\chi_A^{\rm reg}-\chi_B^{\rm reg}$.
This ratio probes the restoration of $SU(2)_L \times SU(2)_R$ and
$U(1)_A$ symmetries through the degeneracy of symmetry partners $A$ and $B$,
and is studied numerically in Ref.~\cite{Chiu:2026sxy}.
Although the restoration probed in Ref.~\cite{Chiu:2026sxy} is that of
the two-flavor chiral group of the light sector, the effective chiral symmetry is expected to
enlarge hierarchically as the temperature rises, from
$SU(2)_L\times SU(2)_R$ for $(u,d)$ to $SU(3)_L\times SU(3)_R$ for
$(u,d,s)$, $SU(4)_L\times SU(4)_R$ for $(u,d,s,c)$, and
$SU(5)_L\times SU(5)_R$ for $(u,d,s,c,b)$ as successive quark thresholds
are crossed,
together with the axial $U(1)_A$~\cite{Chiu:2023hnm,Chiu:2024jyz,Chiu:2024bqx}.
Since the temperature for effective chiral restoration depends on the
number of flavors, we write $T_c^{ud}$, $T_c^{uds}$, $T_c^{udsc}$, and
$T_c^{udscb}$ for the chiral crossover of the sector with the
corresponding flavor content, and similarly $T_1^{ud}$, $T_1^{uds}$,
$T_1^{udsc}$, and $T_1^{udscb}$ for the effective restoration of
$U(1)_A$~\cite{Chiu:2026tot}.
For the $(u,d)$ sector we simply write $T_c$ and $T_1$.
This hierarchy is one reason the renormalization results below are
established for general $N_f$ rather than only for $N_f=2$: the same
construction of $\kappa_{AB}$ applies unchanged to the partner channels
that become degenerate at each stage.
The present paper establishes the renormalization properties that
underpin the construction.
In particular, we show that $\kappa_{AB}$ is exactly
renormalization-group invariant and scheme independent in any mass-independent scheme.

For this construction to be well-defined, three technical questions
must be addressed.
(i)~Does the temperature subtraction remove all additive divergences?
(ii)~Does the equality $Z_A = Z_B$ hold in general, including in the
presence of spontaneous symmetry breaking, the $U(1)_A$ anomaly, and
nondegenerate quark masses?
(iii)~Does the use of local bilinear operators (as opposed to the
nonlocal conserved Noether currents of the lattice chiral symmetry)
introduce finite renormalizations that could affect the construction?

The purpose of this paper is to provide a self-contained
treatment of the UV structure of meson susceptibilities,
resolving these questions. We show that:
\begin{itemize}
\item All additive divergences are temperature-independent, and
cancel exactly in the temperature subtraction.
These are the power divergence and the mass-dependent $m^2\ln(1/(am))$ term.
The remaining logarithmic UV dependence is multiplicative, and resides
entirely in $Z_A^{2}$ (and $Z_B^{2}$).
\item The multiplicative renormalization factor $Z_A^2$ cancels in the
ratio $\kappa_{AB}$ when $Z_A = Z_B$.
\item The equality $Z_A = Z_B$ holds for all symmetry partners, for any
$N_f$ and any quark masses, in any mass-independent scheme with chirally
symmetric regularization.
\item
Whether the currents are those
built from the domain-wall boundary modes or the nonlocal Noether currents
of the lattice chiral symmetry, they give the same
correlator formula after Wick contraction.
Together with the equality $Z_A = Z_B$
proved in Sec.~\ref{subsec:ZAZB_lat} for both DWF and overlap, this settles question~(iii).
\item These results hold for any regularization that preserves
chiral symmetry (Ginsparg-Wilson fermions~\cite{Ginsparg:1981bj} on the lattice,
realized by overlap~\cite{Neuberger:1997fp,Neuberger:1998wv} and
DWF~\cite{Kaplan:1992bt,Kaplan:1992sg,Shamir:1993zy,Borici:1999zw,Chiu:2002ir,Brower:2012vk}),
and extend to continuum QCD.
\end{itemize}

In outline, the argument is a short chain. The bare susceptibility carries
additive divergences, a power divergence and a mass logarithm, both
temperature-independent, so the temperature subtraction removes them. What
survives is the multiplicative factor $Z_\Gamma^2$, which cancels in
$\kappa_{AB}$ because $Z_A = Z_B$ for symmetry partners. The equality
$Z_A = Z_B$ follows from the symmetry that relates the partners, holding for
any $N_f$ and any quark masses in any mass-independent scheme. Hence
$\kappa_{AB}$ is exactly RG-invariant.

The paper is organized as follows.
Section~\ref{sec:correlator} establishes that the meson correlator
$C(t)$ at fixed $t \neq 0$ is free of additive UV divergences.
Section~\ref{sec:susceptibility} presents the complete UV analysis of
the integrated susceptibility.
The additive divergences are identified from the short-distance behavior
of the correlator and organized by the operator product expansion.
The temperature subtraction is shown to remove the
temperature-independent additive divergences, both the power divergence
and the mass-dependent logarithm.
The remaining UV dependence is multiplicative. The renormalization factor $Z_\Gamma$
is derived from the renormalization group equation,
and shown to resum the large logs controlled by the anomalous dimension $\gamma_\Gamma$.
Section~\ref{sec:chains} establishes the equality $Z_A = Z_B$ for
symmetry partners and works out the complete $Z$-factor chains
for all meson channels.
Section~\ref{sec:kappa} proves the RG invariance of $\kappa_{AB}$.
Section~\ref{sec:sym_rel} collects the symmetry relations between the
partner correlators and identifies which $\kappa_{AB}$ probes $U(1)_A$
and which probes the chiral group.
Section~\ref{sec:wilson} contrasts the UV structure with that of
Wilson fermions.
Section~\ref{sec:continuum} discusses the extension to continuum QCD.
Section~\ref{sec:summary} summarizes the results.
Appendix~\ref{app:thermal} collects the free thermal correlator and the
tree-level divergence and thermal coefficients, including the massive and
nondegenerate case.

\section{UV structure of the meson correlator}
\label{sec:correlator}

Consider the meson correlator defined with VEV subtraction:
\begin{equation}
C_\Gamma(t) = \int d^3x \left[
\vev{O_\Gamma^a(\vec{x},t)\,O_\Gamma^{a\dagger}(0)}
- \big|\vev{O_\Gamma^a}\big|^2 \right],
\label{eq:Ct}
\end{equation}
where $O_\Gamma^a(x) = \bar{q}(x)\,\Gamma\,t^a\,q(x)$ is a quark bilinear
with Dirac structure $\Gamma$ and flavor structure $t^a$ (no sum over $a$),
and $O_\Gamma^{a\dagger}=\bar q\,\bar\Gamma\,t^a q$ with
\begin{equation}
\bar\Gamma\equiv\gamma_4\Gamma^\dagger\gamma_4=\pm\Gamma .
\label{eq:Gammabar}
\end{equation}
The sign in eq.~(\ref{eq:Gammabar}) is what makes $C_\Gamma(t)$ positive in
every channel; it is the ``manifestly positive normalization'' referred to
below. The flavor index $a$ is suppressed where no confusion arises.
Here $t^a$ ($a = 1,\ldots,N_f^2-1$) are the generators of $SU(N_f)$ for the
nonsinglets ($t^a=\tau^a/2$ for $N_f=2$) and
$t^0 = \Id/\sqrt{2N_f}$ for the singlet, with the normalization
$\tr_F(t^a t^b) = \delta^{ab}/2$, where $\tr_F$ denotes the trace over
flavor indices.

In lattice QCD with exact chiral symmetry,
the quark fields $\bar{q}(x)$ and $q(x)$ (local in $x$)
can be constructed with the boundary modes in domain-wall fermions, see
Ref.~\cite{Furman:1994ky} for Shamir/M{\"o}bius DWF
and Refs.~\cite{Chiu:2003ir,Chen:2012jya} for optimal DWF,
yielding local currents with exact chiral symmetry (in the limit $N_s \to \infty$).
(Note that the fifth dimension with subscript $s$ in DWF is an internal degree of freedom,
like color and Dirac indices, not related to the locality in physical space-time.)
On the other hand, one can construct nonlocal currents according to
the exact chiral symmetry on the 4-dimensional lattice \cite{Luscher:1998pqa},
$O_\Gamma^a(x) = \bar{q}(x)\,\Gamma\,t^a (1-raD) q(x)$,
where $D$ is the massless overlap operator
(in the normalization of Ref.~\cite{Luscher:1998pqa} this factor reads $1-aD/2$).
Nevertheless, in both cases,
after Wick contraction, the correlator gives the same expression in terms of
the valence quark propagator $(D_c + m)^{-1}_{x,y}$~\cite{Chiu:1998eu},
where $D_c=D(1- a r D)^{-1}$~\cite{Chiu:1998gp} is the chirally symmetric Dirac operator
satisfying $D_c \gamma_5 + \gamma_5 D_c = 0$, and
$(D_c + m)^{-1}$ goes to $[\gamma_\mu ( \partial_\mu + i T^a A_\mu^a) + m]^{-1} $
in the continuum limit. The local currents of DWF and the nonlocal currents
of the overlap thus give the same meson correlator after Wick contraction,
and hence the same susceptibilities. In practice, the numerical values at finite lattice spacing
depend on which DWF is used (Shamir, M\"obius, or optimal) and on its parameter
choices, and likewise for the overlap with its kernel and parameters: at finite
$N_s$ each DWF variant realizes a different approximation to the chirally
symmetric $D_c$, and different kernels give different $D_c$ at finite $a$. All
of them become consistent, up to numerical precision, only in the limit
$N_s \to \infty$ and in the continuum limit. Moreover, the local quark fields
defined in terms of the boundary modes of DWF obey the usual chiral projection
rule in the continuum, independent of the gauge fields, so any observable
constructed with the quark fields manifests the symmetries exactly as those of
its counterpart in the continuum. In the following, it is understood that the
quark fields in the local current (operator) are either in the continuum or
those built from the boundary modes of DWF.

One practical point deserves emphasis. At finite $N_s$ the DWF, or the
overlap with a rational approximation of the sign function, reproduces
the index and the exact zero modes of the gauge field only to the extent
that the sign function is exact on the low-lying modes of the kernel
$H_w$. A truncation whose window excludes the small eigenvalues of $H_w$
that carry the index leaves the would-be zero modes of the valence
operator at eigenvalues of order the approximation error rather than of
order $m_{\rm res}$, and this is invisible in $m_{\rm res}$ itself, which
averages over all modes. For observables governed by the $1/m^2$ poles of
these modes, the pseudoscalar and scalar susceptibilities near $T_c$ in
particular (Sec.~\ref{subsec:disconnected}), the low modes of $H_w$ must be
projected out and treated exactly. The vector- and tensor-type channels
have no such pole, but their partner splittings are controlled by the
density of near-zero modes, which the same truncation distorts. All statements of this paper refer to the exact
chiral limit $N_s\to\infty$, or equivalently to a valence Dirac operator with
exact low-mode treatment.

The underlying relationship between the two formulations can be traced
through the generating functional. The DWF action consists of the fermion
fields and the Pauli--Villars (pseudofermion) fields, which carry color and
Dirac indices but obey Bose statistics. Integrating out both,
the generating functional $ W $ for connected
$n$-point Green's functions of the quark fields takes the form~\cite{Chiu:2003ir}
\begin{align}
\label{eq:ZW}
e^{W[J,\bar J]} =
\frac{ \int [dU] e^{-{\cal A}_g} \det D(m) e^{ \bar J ( D_c + m )^{-1} J } }
     { \int [dU] e^{-{\cal A}_g} \det D(m) },
\end{align}
where
$D(m) = (D_c + m)(1 + r a D_c)^{-1}$
is the effective 4-dimensional Dirac operator of DWF,
${\cal A}_g$ is the gauge action,
and $\bar J$ and $J$ are the Grassmann sources of $q$ and $\bar q$ respectively.
The Pauli--Villars fields supply the factor $(1+raD_c)^{-1}$ in the effective
action of the sea quarks, whose Dirac operator is thus exactly the overlap
operator, while the valence propagator, read off from the source term, is
$(D_c+m)^{-1}$. The two formulations therefore share
the same sea-quark determinant and the same valence propagator, and differ
only in which fields the operators are built from. The overlap works with
the sea-quark fields: the exact symmetries are the Ginsparg--Wilson
(L\"uscher) transformations, the conserved currents carry the nonlocal
factor $(1-raD)=(1+raD_c)^{-1}$, and the dressed bilinears reproduce the
valence propagator after Wick contraction. The DWF works with the
boundary-mode fields: they retain the chiral projection properties of the
continuum, so the operators are local and transform as in the continuum,
at the expense that the local currents are not conserved at finite $a$.

The identical form of the correlators for DWF and overlap is, however,
only part of the answer to
question~(iii) of Sec.~\ref{sec:intro}. The construction of $\kappa_{AB}$
requires in addition the equality $Z_A = Z_B$ of the renormalization constants
of the two operators entering the ratio. This equality is proved
nonperturbatively in Sec.~\ref{subsec:ZAZB_lat}, for the local operators of DWF
and for the nonlocal operators of the overlap with their factor $(1-raD)$
alike: in each case the chiral rotation, ordinary for DWF and in the
asymmetric L\"uscher form for the overlap, maps the operator onto its symmetry
partner with no residual factor. Only with that result in hand is question~(iii) answered
completely in the negative: the use of local bilinear operators introduces no
finite renormalization that could affect the construction, and the entire
construction of $\kappa_{AB}$ is common to domain-wall and overlap fermions.

We now show that $C_\Gamma(t)$ at fixed $t \neq 0$ is free of ultraviolet
divergences, and that the divergences of the susceptibility come entirely
from the coincident point $t=0$.
At fixed Euclidean time $t \neq 0$, the two operators in $C_\Gamma(t)$
are separated by a nonzero distance.
The short-distance behavior of the integrand is governed by the operator
product expansion (OPE) of $O_\Gamma(\vec x,t)\,O_\Gamma^\dagger(0)$.
The leading singularity comes from the identity operator $\Unit$ in the OPE
(the operator $\Unit$ of vanishing dimension, written with a distinct
symbol to avoid confusion with the Dirac structure $\Gamma=\Id$ of the
scalar channel), and its coefficient is
determined by the free-fermion propagator
$S(x) \sim \slashed{x}/x^4$:
\begin{align}
\vev{O_\Gamma(\vec{x},t)\,O_\Gamma^\dagger(0)}
&\sim \mathrm{tr}\!\left[\Gamma\,S(x)\,\bar\Gamma\,S(-x)\right] + \cdots \notag\\
&\sim \frac{A_\Gamma}{(|\vec{x}|^2 + t^2)^3} + \cdots,
\label{eq:OPE}
\end{align}
where the coefficient $A_\Gamma$ depends on the Dirac structure
(through $\mathrm{tr}[\Gamma\gamma^\mu\bar\Gamma\gamma^\nu]$) and the
number of colors, but the \emph{power} of the leading singularity
is universal across all bilinears
($\Gamma = \Id,\,\gamma_5,\,\gamma_\mu,\,\gamma_5\gamma_\mu,\,\sigma_{\mu\nu}$,
$\gamma_5\sigma_{\mu\nu}$).
The spatial integral is convergent and evaluates to
\begin{equation}
\int d^3x\,(|\vec{x}|^2 + t^2)^{-3} = \frac{\pi^2}{4\,|t|^3},
\end{equation}
finite for any $t > 0$.

Therefore, for \emph{any} Dirac structure $\Gamma$, $C_\Gamma(t)$ at
fixed $t \neq 0$ is \textbf{free of all UV divergences}.

The situation is different at the coincident point $t = 0$, where the
integrand is singular. In the continuum, the spatial integral of the
leading OPE term~\eqref{eq:OPE} gives the short-distance form
\begin{equation}
C_\Gamma(t) \sim \frac{\alpha_\Gamma}{2\,|t|^3},
\label{eq:Ct_short}
\end{equation}
even about $t=0$, with $\alpha_\Gamma$ evaluated explicitly in
Section~\ref{subsec:additive}. Each $C_\Gamma(t)$ at $t\neq 0$ is finite,
but the $1/|t|^3$ singularity is not integrable at $t=0$, and its
accumulation upon integration over $t$ generates the additive divergence
of the susceptibility. On the lattice the coincident point is the $t=0$
time slice itself, whose contribution $a\,C_\Gamma(0)\sim 1/a^2$ is a
contact term present already in the free theory, and the same
singularity shows up as the growth $C_\Gamma(t)\sim 1/|t|^3$ on the
neighbouring slices $t = a, 2a, \ldots$. Both aspects are analyzed in the
next section.

\section{UV structure of the meson susceptibility}
\label{sec:susceptibility}

The bare susceptibility is the integral of the correlator over the thermal
circle. On the lattice it is the sum over all time slices,
\begin{equation}
\chi_\Gamma(T) = a\sum_{t=0}^{1/T-a} C_\Gamma(t,T),
\label{eq:chi_bare_lat}
\end{equation}
and in the continuum the corresponding integral over
one period,
\begin{equation}
\chi_\Gamma(T) = \int_{0}^{1/T} dt\, C_\Gamma(t,T),
\label{eq:chi_bare}
\end{equation}
regulated by excluding a short-distance region $|t|<\eps$ around the
coincident point, i.e.\ by an ultraviolet cutoff $\Lambda = 1/\eps$, with
$\eps\to 0$ taken only after the temperature subtraction and the
multiplicative renormalization. Throughout, $a$ denotes the lattice
spacing and $\eps$ the short-distance cutoff of the continuum integrals.
The two regularizations have the same divergence structure: on the lattice
the role of $\eps$ is played by $a$, the coefficient of a power divergence
changes by a regulator-dependent factor of order one, and a logarithmic
divergence changes by a finite constant. None of this matters below, since
all additive divergences are removed by the temperature subtraction
whatever their coefficients. Where a result is quoted for the lattice
susceptibility $\chi_\Gamma^{\rm bare}(T,a)$, the continuum cutoff has been
identified with the lattice spacing in this sense.
The same definition applies to the quark-connected and quark-disconnected
contributions alike (Sec.~\ref{subsec:disconnected}).

An equally admissible prescription omits the $t=0$ slice, the contact term of
the two coincident operators,
\begin{equation}
\chi_\Gamma(T) = a\sum_{t=a}^{1/T-a} C_\Gamma(t,T).
\label{eq:chi_bare_lat_ne0}
\end{equation}
The two definitions differ by
$a\,C_\Gamma(0,T)$ of the omitted slice, and they agree in the
continuum limit. As we show in Sec.~\ref{subsec:prescription}, the entire
renormalization analysis of this paper holds unchanged for either, because the
contact term is an additive contribution that does not affect the
multiplicative renormalization. We take eq.~(\ref{eq:chi_bare_lat}) as the
default and return to the equivalence once the renormalization has been
established. The correlator may also be summed in modulus, $\sum_t |C_\Gamma(t,T)|$,
as is convenient when a channel changes sign along the thermal circle.
Since $Z_\Gamma$ is real and the divergent short-distance region has a
definite sign, this changes neither the additive-divergence structure
nor the multiplicative renormalization. Every statement below applies to
that definition as well.

The endpoints $t=0$ and $t=1/T$ are the same point on the thermal circle, the
coincident point of the two operators, and $C_\Gamma$ is singular there. Every
ultraviolet divergence of $\chi_\Gamma$ originates in this short-distance
region: in the continuum, from the accumulation of the coincident-point
singularity as $t\to 0$ and $t\to 1/T$, and on the lattice additionally from
the $t=0$ slice itself, whose weight is $a\,C_\Gamma(0,T)$. We show below that
all of them are generated by the zero-temperature propagator at short
distances, and are therefore temperature-independent, so that all are removed
by the subtraction~(\ref{eq:chi_reg}). The coefficients quoted below are those
of the continuum divergences. The lattice contact term
adds a further temperature-independent constant of the same order, which the
subtraction removes along with them.

For the analysis it is convenient to place the coincident point at the centre
of the integration region. Shifting the integration window back by half a
period turns the regulated form of eq.~(\ref{eq:chi_bare}) into two
half-intervals meeting at the coincident point,
\begin{equation}
\chi_\Gamma(T) = \left(\int_{-1/(2T)}^{-\eps} + \int_{\eps}^{\,1/(2T)}\right)
dt\, C_\Gamma(t,T),
\label{eq:chi_bare_two}
\end{equation}
with outer limits at the regular midpoints $\pm 1/(2T)$ between neighbouring
coincident points. Each interval carries the short-distance singularity of a
single coincident point, so the short-distance expansion applies directly on
each, and at $T\to 0$ the limits open to the full axis,
$(-\infty,-\eps]\cup[\eps,\infty)$.

\subsection{Additive divergences}
\label{subsec:additive}

We derive the additive UV divergences of $\chi_\Gamma(T)$ at tree
level, both the leading massless term and the mass-dependent terms. The
temperature subtraction that removes them, the residual multiplicative
structure, and its renormalization-group treatment are developed in
Sec.~\ref{subsec:subtraction}.

At zero temperature, the meson correlator at tree level is given by (after Wick contraction)
\begin{equation}
\vev{O_\Gamma^a(x)\,O_\Gamma^{a\dagger}(0)} = -N_c\,\tr_F[t^a t^a]\,
\tr\!\left[\Gamma\,S(x)\,\bar\Gamma\,S(-x)\right],
\label{eq:tree_corr}
\end{equation}
where $\bar\Gamma=\pm\Gamma$ is defined in eq.~(\ref{eq:Gammabar}), $S(x)$ is
the free fermion propagator and the trace $\tr$ is over Dirac
indices only: the free propagator is trivial in color, so the color trace has
been carried out, giving the overall factor $N_c$. Throughout the divergence
analysis $\tr$ denotes the Dirac trace, with the color factor $N_c$ explicit.

Equation~(\ref{eq:tree_corr}) is the quark-connected contraction. A flavor
singlet receives in addition a quark-disconnected contraction, but it vanishes
identically at tree level. In the free theory the propagator is a fixed
c-number matrix, so the single-propagator loop $\Tr[\Gamma\,S]$ carries no
fluctuation, and the disconnected susceptibility, which is the variance of
that loop (Sec.~\ref{subsec:disconnected}), is zero. The disconnected
contribution is generated entirely by gauge-field fluctuations and is invisible
in free-field perturbation theory. The analysis of this section therefore gives
the complete tree-level divergence structure for every channel, singlet and
nonsinglet alike.

In the continuum limit the massless
free fermion propagator is
\begin{equation}
S(x) = \frac{1}{2\pi^2}\frac{\slashed{x}}{(x^2)^2},
\label{eq:free_prop}
\end{equation}
where $x^2 = |\vec x|^2 + t^2$. Substituting (\ref{eq:free_prop}) into
(\ref{eq:tree_corr}) and using $S(-x) = -S(x)$:
\begin{align}
\tr[\Gamma\,S(x)\,\bar\Gamma\,S(-x)]
&= -\frac{1}{(2\pi^2)^2}\,\frac{\tr[\Gamma\slashed{x}\bar\Gamma\slashed{x}]}{(x^2)^4}  \notag \\
&= -\frac{4\,\eta_\Gamma^2\,x^2}{(2\pi^2)^2(x^2)^4} = -\frac{1}{\pi^4 (x^2)^3},
\label{eq:trace_eval}
\end{align}
where, for the scalar and pseudoscalar channels, the Dirac trace is
isotropic,
\begin{equation}
\tr[\Gamma\,\slashed{x}\,\Gamma\,\slashed{x}] = 4\,\eta_\Gamma\,x^2,
\qquad \eta_{\Id} = +1,\quad \eta_{\gamma_5} = -1,
\end{equation}
and $\bar\Gamma=\eta_\Gamma\Gamma$ for these two channels, so that
$\tr[\Gamma\slashed{x}\bar\Gamma\slashed{x}]=4\eta_\Gamma^2x^2=4x^2$.
The vector, axial-vector, tensor vector, and axial-tensor vector channels have
anisotropic traces, which are not proportional to $x^2$. After the spatial
integration below, each nonetheless yields the same $1/t^3$ behavior with a
channel-dependent constant. The explicit traces, the spatial projections,
and the resulting constants are given in Appendix~\ref{app:thermal}.
Substituting the scalar or pseudoscalar case back
into (\ref{eq:tree_corr}):
\begin{equation}
\vev{O_\Gamma(x)\,O_\Gamma^\dagger(0)}
= \frac{N_c}{\pi^4(x^2)^3}\,\tr_F[t^a t^a]
\equiv \frac{B_\Gamma}{(x^2)^3},
\end{equation}
where $B_\Gamma \equiv N_c\,\tr_F[t^a t^a]/\pi^4$ collects
all channel-dependent constants.
With the $O_\Gamma^\dagger$ of eq.~(\ref{eq:Ct}), the sign of $\bar\Gamma$
relative to $\Gamma$ combines with the minus sign of the Wick contraction to
make the correlator, and hence the susceptibility $\chi_\Gamma$, positive in
every channel. We refer to this as the manifestly positive normalization.
The power-divergence coefficient $\alpha_\Gamma$ is positive in every
channel, and the relative constants quoted in Appendix~\ref{app:thermal}
are the corresponding magnitudes.

The temporal correlator is obtained by spatial integration:
\begin{equation}
C_\Gamma(t) = \int d^3x\,\frac{B_\Gamma}{(|\vec x|^2 + t^2)^3}.
\end{equation}
The spatial integral is elementary. With $r = |\vec x|$ and
$\int d^3x = 4\pi\int_0^\infty r^2\,dr$:
\begin{align}
\int d^3x\,\frac{1}{(r^2 + t^2)^3}
&= 4\pi\int_0^\infty\frac{r^2\,dr}{(r^2 + t^2)^3} \notag\\
&= 4\pi\cdot\frac{\pi}{16\,|t|^3} = \frac{\pi^2}{4\,|t|^3},
\end{align}
where the last integral is evaluated by setting $r = t\tan\theta$ and
using $\int_0^{\pi/2}\sin^2\theta\,d\theta = \pi/4$.

Therefore, at zero temperature,
\begin{equation}
C_\Gamma(t)\big|_{T=0} = \frac{\alpha_\Gamma}{2\,|t|^3}, \qquad
\alpha_\Gamma \equiv \frac{\pi^2 B_\Gamma}{2}
= \frac{N_c\,|\eta_\Gamma|\,\tr_F[t^a t^a]}{2\pi^2}.
\label{eq:Ct_short_explicit}
\end{equation}
The coefficient $\alpha_\Gamma$ is \emph{temperature-independent} (the
UV is insensitive to $T$) and channel-dependent. The expression above is
the scalar and pseudoscalar value. The constants for the vector,
axial-vector, tensor vector, and axial-tensor vector channels are given in
Appendix~\ref{app:thermal}, and all are temperature-independent.

At finite temperature the free correlator is built from the thermal image
sums of the two propagators (Appendix~\ref{app:thermal}). The image pairs
with $n=n'$ reproduce the zero-temperature
form~(\ref{eq:Ct_short_explicit}) at shifted times, and the pairs with
$n\neq n'$ are regular for all $t$, so
\begin{equation}
C_\Gamma(t,T) = \frac{\alpha_\Gamma}{2}
\sum_{n=-\infty}^{\infty}\frac{1}{|t+n/T|^3} \;+\; \Delta C_\Gamma(t,T),
\label{eq:Ct_thermal}
\end{equation}
even in $t$ and symmetric about $t=1/2T$. The image sum carries the entire
short-distance singularity, with the channel dependence of the singular part
carried by $\alpha_\Gamma$. The cross-image remainder $\Delta C_\Gamma$ is
finite for all $t$, contributes only to the finite part, and is given in
Appendix~\ref{app:thermal}. Only the $n=0$ image
is singular at the coincident point, where it reduces to the zero-temperature
form $\alpha_\Gamma/(2|t|^3)$ of eq.~(\ref{eq:Ct_short_explicit}). The images
with $n\neq 0$ are regular there and, together with $\Delta C_\Gamma$,
contribute a temperature-dependent but ultraviolet-finite remainder.

The power divergence follows by inserting this correlator into the two
intervals of eq.~(\ref{eq:chi_bare_two}). The coincident-point singularity
$\alpha_\Gamma/(2|t|^3)$, being even, sits at the near end of each interval
and contributes equally from either side,
\begin{equation}
\int_{\eps}^{1/(2T)}\frac{\alpha_\Gamma}{2\,|t|^3}\,dt
= \int_{-1/(2T)}^{-\eps}\frac{\alpha_\Gamma}{2\,|t|^3}\,dt
= \frac{\alpha_\Gamma}{4\eps^2} + (\text{finite}),
\label{eq:half_interval}
\end{equation}
while the $n\neq 0$ images and the cross-image remainder $\Delta C_\Gamma$
are regular and integrate to finite,
cutoff-independent contributions. Summing the two intervals,
\begin{align}
\chi_\Gamma^{\rm bare}(T,\eps) &=
\left(\int_{-1/(2T)}^{-\eps} + \int_{\eps}^{\,1/(2T)}\right) C_\Gamma(t,T)\,dt
\notag \\
&= 2\times\frac{\alpha_\Gamma}{4\eps^2} + (\text{finite})
= \frac{\alpha_\Gamma}{2\eps^2} + (\text{finite}),
\label{eq:tree_div}
\end{align}
the divergence being carried entirely by the coincident-point singularity and
hence temperature-independent. Evaluated directly in the
definition~(\ref{eq:chi_bare}), the same result follows from the $n=0$ image
supplying the divergence at $t\to \eps$ and the $n=-1$ image at $t\to 1/T-\eps$,
the two being the same coincident-point singularity since $t=0$ and $t=1/T$
are the same point on the thermal circle. The entire
temperature dependence therefore resides in the finite part. On the lattice,
with $\eps\to a$, this is the power divergence $\alpha_\Gamma/(2a^2)$ of
the bare susceptibility $\chi_\Gamma^{\rm bare}(T,a)$.

\medskip
\noindent\textit{Mass-dependent terms.}
The derivation so far used the massless propagator~(\ref{eq:free_prop}).
For nonzero quark mass $m$ the free propagator is known
in closed form.
In four Euclidean dimensions,
\begin{equation}
S^{(m)}(x) = \frac{m^2}{4\pi^2}
\left[\,K_2(m r)\,\frac{\slashed{x}}{r^2} + K_1(m r)\,\frac{1}{r}\,\right],
\quad r = \sqrt{x^2},
\label{eq:massive_prop}
\end{equation}
with $K_\nu$ the modified Bessel functions. As $m\to 0$, $K_2(z)\to 2/z^2$
and the vector term reduces to the massless
propagator~(\ref{eq:free_prop}), while $K_1(z)\to 1/z$ leaves the scalar term
$m/(4\pi^2 r^2)$.

Two features of eq.~(\ref{eq:massive_prop}) are essential.
First, the vector term is odd in $x$ and the scalar term is even,
so in the Dirac trace $\tr[\Gamma\,S^{(m)}(x)\,\bar\Gamma\,S^{(m)}(-x)]$ the cross
terms carry an odd number of $\gamma$-matrices and vanish identically. Only
even powers of $m$ survive. The divergence linear in $m$, which would give an
$m/a^3$ term in $\chi_\Gamma$, is therefore forbidden by the Dirac trace.
This is chiral symmetry
protection at work: the same $\gamma_5$ symmetry that forbids the $1/a^3$
divergence of $\vev{\bar qq}$ forbids any divergence of $\chi_\Gamma$ odd in
$m$. Second, $K_\nu(mr)$ decays as $e^{-mr}$, so the correlator is cut off
exponentially at $r\sim 1/m$. At zero-temperature the mass therefore supplies
its own infrared cutoff for the logarithm of eq.~(\ref{eq:mass_div}).

Carrying out the trace with $\tr[\Gamma\slashed{x}\bar\Gamma\slashed{x}]
= 4x^2$ and $\tr[\Gamma\bar\Gamma] = 4\eta_\Gamma$ for $\Gamma=\Id,\gamma_5$,
the radial kernel is
\begin{equation}
\frac{m^4}{4\pi^4 r^2}
\left[\,K_1(m r)^2 - \eta_\Gamma\,K_2(m r)^2\,\right],
\label{eq:massive_kernel}
\end{equation}
whose short-distance expansion, using
$K_1(z) = 1/z + O(z\ln z)$ and $K_2(z) = 2/z^2 - 1/2 + O(z^2\ln z)$, is
\begin{equation}
-\,\frac{\eta_\Gamma}{\pi^4 r^6}
\;+\; \frac{1 + 2\eta_\Gamma}{4\pi^4 r^4}\;m^2
\;+\; O(m^4\ln mr).
\label{eq:massive_expansion}
\end{equation}
The $m^0$ term reproduces eq.~(\ref{eq:trace_eval}) and hence
$\alpha_\Gamma$. The $m^1$ term is absent, as anticipated. The leading
mass-dependent singularity is the $m^2$ term, $\sim m^2/(x^2)^2$.

The spatial integral of the $m^2$ term gives the short-distance form
$\int d^3x\, m^2/(x^2)^2 = \pi^2 m^2/|t|$, even in $t$. Over the two intervals of
eq.~(\ref{eq:chi_bare_two}) the integral of $1/|t|$ is
\emph{logarithmically} divergent, the two intervals contributing equally,
running from the cutoff $t\sim \eps$ up to $t\sim 1/m$, where the exponential
decay of eq.~(\ref{eq:massive_prop}) closes it off.

The finite temperature does not change this. The antiperiodic massive
propagator is the sum over thermal images of
eq.~(\ref{eq:massive_prop}). At the coincident point $x\to 0$ only the $n=0$
image is singular, so the $m^2$ divergence is the zero-temperature one. Its
short-distance expansion is $1/|t| + O(T^2 t)$
(Appendix~\ref{app:thermal_mass}). The leading
$1/|t|$, picked up at both coincident-point ends, gives
the logarithm, while the temperature-dependent corrections
integrate to a finite remainder.
The competition between the scales $1/m$ and $1/T$ at the
infrared end of the logarithm resides entirely in this finite part, as the
$m^2\ln(m/T)$ piece. It never enters the divergent coefficient.
We therefore have
\begin{equation}
\chi_\Gamma^{\rm bare}(T,a) \supset
\frac{\alpha_\Gamma}{2a^2}
+ c_m^\Gamma\, m^2\,\ln\frac{1}{a m},
\label{eq:mass_div}
\end{equation}
with a channel-dependent coefficient $c_m^\Gamma$ (the superscript
distinguishes it from the perturbative Wilson coefficients $c_n$ of the
massless analysis in Sec.~\ref{subsec:subtraction} below). The explicit coefficients for all
channels are computed at tree level in Appendix~\ref{app:cm_coeff}. For the
scalar and pseudoscalar $c_m^S = -3N_c\,\tr_F[t^a t^a]/(2\pi^2)$ and
$c_m^P = -N_c\,\tr_F[t^a t^a]/(2\pi^2)$, and the vector coefficient vanishes,
$c_m^V = 0$. For every partner pair the coefficients differ,
$c_m^A\neq c_m^B$, so the mass logarithm does not cancel in the partner
difference. Here and below, $\alpha_\Gamma$ and $c_m^\Gamma$ denote the full
coefficients. The values quoted above are their tree-level parts. Higher
orders, including the quark-disconnected contractions of
Sec.~\ref{subsec:disconnected}, modify the coefficients but introduce no new
divergence structure.
The symbol $\supset$ indicates that only the ultraviolet-divergent part of
$\chi_\Gamma^{\rm bare}$ is displayed. The finite, temperature-dependent
remainder carries the physics, and is treated in
Sec.~\ref{subsec:UV_decomp}.

Two points are essential. First, the $m^2$ term is \emph{not} forbidden
by chiral symmetry.
The quark mass is a spurion of chiral-symmetry
breaking and $m^2$ is chirally even, so this logarithmic divergence is
present even for Ginsparg--Wilson fermions (and indeed already in the
continuum theory).
By contrast, exact chiral
symmetry forbids \emph{all} mass-dependent \emph{power} divergences of the
susceptibility.
The only dimension-two candidate, $\propto m/a$, has identically vanishing
coefficient by the Dirac-trace argument above. For formulations without
exact chiral symmetry (e.g.\ Wilson fermions) the chirally even part of
the propagator generated by the symmetry-breaking terms removes this
protection.
We return to this distinction in Sec.~\ref{sec:wilson}.
Second, like the leading term $\alpha_\Gamma/(2a^2)$,
the coefficient in (\ref{eq:mass_div}) is a short-distance quantity,
built from the fixed bare inputs $a$ and $m$,
and is therefore temperature-independent (Sec.~\ref{subsec:subtraction}).
Consequently both additive divergences, $\alpha_\Gamma/(2a^2)$ and
$c_m^\Gamma\, m^2\ln(1/(am))$, cancel exactly in the temperature
subtraction~(\ref{eq:chi_reg}),
and neither survives in $\chi_\Gamma^{\rm reg}$.

\subsection{Temperature subtraction and multiplicative structure}
\label{subsec:subtraction}

The additive divergences found at tree level, the power divergence
$\alpha_\Gamma/(2a^2)$ and the mass-dependent logarithm
$c_m^\Gamma\, m^2\ln(1/(am))$ of eq.~(\ref{eq:mass_div}), are
\emph{temperature-independent}. This holds not only at tree level but to
all orders in the QCD coupling, for every channel $\Gamma$, and is a
well-established property of thermal field
theory~\cite{Kislinger:1975ab,Landsman:1986uw,Dolan:1973qd,Weinberg:1974hy,Bernard:1974bq}:
the ultraviolet divergences at finite temperature are identical to those
at zero temperature. The reason is that a UV divergence is a
short-distance ($|x|\to 0$) property of the correlator, whereas the
temperature enters only through the dimensionless combination $T|x|$,
which vanishes as $|x|\to 0$. The short-distance singularity that
generates the divergence is therefore exactly its $T=0$ form, and the
temperature can enter only the ultraviolet-finite part. It is this
property that licenses an arbitrary \emph{thermal} reference temperature
in the subtraction below. We may
therefore remove all additive divergences at once, and to all orders, by
subtracting the susceptibility at a single reference temperature $T_r$,
\begin{equation}
\chi_\Gamma^{\rm reg}(T;T_r,a) \equiv
\chi_\Gamma^{\rm bare}(T,a) - \chi_\Gamma^{\rm bare}(T_r,a),
\label{eq:chi_reg}
\end{equation}
where the choice of reference temperature $T_r$ is discussed in
Sec.~\ref{subsec:T_sub}. Obviously there is no need to evaluate
$\alpha_\Gamma$ and $c_m^\Gamma$, at tree level or any order, since the
subtraction removes them whatever their values.
For the scalar (chiral) susceptibility, the temperature subtraction of the
additive divergences was proposed in Ref.~\cite{Aoki:2006br},
which subtracts the zero-temperature susceptibility, $\chi(T)-\chi(T=0)$.
Here $T_r$ plays that role, and the construction is applied to all
meson channels.

The temperature subtraction is possible because the additive divergences are
properties of the \emph{integrated product}
$\int d^4x\,\vev{O_\Gamma(x)O_\Gamma^\dagger(0)}$ at short distance, not of the
operator $O_\Gamma$ itself.
They arise from the coefficient of the identity operator $\Unit$ in the
operator product expansion of $O_\Gamma(x)O_\Gamma^\dagger(0)$, and this
coefficient is present for every channel.
The product carries flavor structure $(t^a)^2$, whose trace
$\tr_F[(t^a)^2]=\tfrac12$ is nonzero for all $a$, singlet ($t^0$) and
nonsinglet alike.

Although free of additive divergences, $\chi_\Gamma^{\rm reg}$ is not yet
finite.
The composite operator $O_\Gamma$ still carries a multiplicative
(logarithmic) renormalization associated with its anomalous dimension
$\gamma_\Gamma$.
To exhibit it we use the general renormalization of the bilinear.
For any Dirac structure $\Gamma$ and flavor structure $t^a$, the
renormalized operator is
\begin{equation}
O_\Gamma^{R}(\mu) = Z_\Gamma^{-1}(\mu,a)\big[\,O_\Gamma^{\rm bare}(a)
- b_\Gamma(a)\,\Unit\,\big],
\label{eq:O_renorm_def}
\end{equation}
a triangular mixing with the identity operator~\cite{Bochicchio:1985xa}.
Here $Z_\Gamma$ is the multiplicative (diagonal) renormalization.
It depends only on the Dirac structure.
The additive (off-diagonal) coefficient $b_\Gamma$ is a $c$-number, and it
is nonzero only for the scalar singlet
$O_S^0 = \bar q\, t^0 q$ (the superscript is the flavor index, with
$t^0 = \Id/\sqrt{2N_f}$), the sole bilinear with vacuum quantum numbers.
There it is the power-divergent condensate,
$b\sim c_3/a^3 + c_1\, m/a^2 + \cdots$, with the chirally odd $c_3$ present
for Wilson and forbidden for Ginsparg--Wilson.
For every other channel $b_\Gamma = 0$: there is no mixing, and the
renormalization is purely multiplicative (diagonal).
For the scalar singlet, the mixing does occur, and
it is triangular:
$O_S^0$ mixes into the identity operator, but the identity operator
renormalizes trivially, $\Unit^R = \Unit$, and does not mix into $O_S^0$.
In matrix form,
\begin{equation}
\begin{pmatrix} O_S^{0,R} \\ \Unit \end{pmatrix}
=
\begin{pmatrix} Z_S^{-1} & -Z_S^{-1}\, b_S \\ 0 & \Unit \end{pmatrix}
\begin{pmatrix} O_S^{0,\rm bare} \\ \Unit \end{pmatrix},
\label{eq:mixing_matrix}
\end{equation}
where the nonzero upper-right entry is the mixing of $O_S^0$ into the
identity operator, and the vanishing lower-left entry is the triangularity,
$\Unit^R = \Unit$.

The susceptibility is a VEV-subtracted quantity [eq.~(\ref{eq:Ct})],
and the additive $c$-number $b_\Gamma\,\Unit$ drops out of it.
A variance is unchanged by a constant shift of the operator, so with
$O_\Gamma^{\rm bare} = Z_\Gamma\,O_\Gamma^{R} + b_\Gamma\,\Unit$,
\begin{align}
&\vev{O_\Gamma^{\rm bare}(x)\,O_\Gamma^{{\rm bare}\dagger}(0)} - \big|\vev{O_\Gamma^{\rm bare}}\big|^2 \notag \\
=& Z_\Gamma^{2}\big[\vev{O_\Gamma^{R}(x)\,O_\Gamma^{R\dagger}(0)} - \big|\vev{O_\Gamma^{R}}\big|^2\big],
\label{eq:vev_sub_mult}
\end{align}
and the $b_\Gamma$-dependent terms cancel identically.
The multiplicative factor $Z_\Gamma$ renormalizes the operator, not the
thermal state, so it is $T$-independent and common to both temperatures.
Integrating over $x$ and forming the temperature difference,
\begin{equation}
\chi_\Gamma^{\rm reg}(T;T_r,a) = Z_\Gamma^{2}(\mu,a)\,
\big[\chi_\Gamma^{R}(\mu,T) - \chi_\Gamma^{R}(\mu,T_r)\big],
\label{eq:chi_reg_renorm}
\end{equation}
where $\chi_\Gamma^{R}(\mu,T)$ is the finite renormalized susceptibility
at the renormalization scale $\mu$.
This holds for \emph{every} channel, singlet and nonsinglet alike. The
scalar singlet's operator mixing, which prevents $\bar qq$ itself from
being multiplicatively renormalizable, is invisible in its susceptibility.

The most notable feature of eq.~(\ref{eq:chi_reg_renorm}) is that the
right-hand side decomposes into two parts, one at $T$ and one at $T_r$,
with no cross dependence between the two temperatures: the same finite
function $\chi_\Gamma^{R}(\mu,\cdot)$ is evaluated at the measured
temperature and at the reference temperature. It is the $T$-independence of
the additive divergences that makes this single function well defined for
all $T$ at once, up to one common $T$-independent constant, which cancels
in the difference. The entire dependence on $T_r$ is therefore an additive shift:
changing the reference temperature changes $\chi_\Gamma^{\rm reg}$ by a
constant in $T$, leaving temperature differences and derivatives unchanged.
Ratios such as $\kappa_{AB}$, in contrast, retain a genuine dependence on
$T_r$, addressed in Sec.~\ref{subsec:T_sub}. The decomposition is also
turned into reference-free determinations of characteristic temperatures in
Sec.~\ref{subsec:T_char}.

\subsubsection*{Renormalization group equations}

Using the standard renormalization-group treatment of composite
operators~\cite{Peskin:1995ev}, we now determine the multiplicative
renormalization of the susceptibility in the massless limit.
The running of the coupling and of $Z_\Gamma$ are governed by the beta
function and the anomalous dimension,
\begin{align}
\label{eq:beta_def}
&\beta(g) \equiv \mu\frac{dg}{d\mu} = -\beta_0\,\frac{g^3}{16\pi^2} + O(g^5), \\
&\gamma_\Gamma(g) \equiv \mu\frac{d\ln Z_\Gamma}{d\mu}
= \gamma_\Gamma^{(0)}\,\frac{g^2}{16\pi^2} + O(g^4),
\label{eq:gamma_def}
\end{align}
with $\beta_0 = 11 - 2N_f/3$ (for $N_c=3$) and $\gamma_\Gamma^{(0)}$ the
leading-order anomalous dimension, which depends on the Dirac structure
$\Gamma$.

Since $\chi_\Gamma^{\rm reg}$ is a bare-theory quantity, independent of
the renormalization scale $\mu$, and $Z_\Gamma$ is $T$-independent,
applying $\mu\,d/d\mu$ to eq.~(\ref{eq:chi_reg_renorm}) yields the
Callan--Symanzik equation for the difference
$\chi_\Gamma^{R}(\mu,T)-\chi_\Gamma^{R}(\mu,T_r)$. Since the reference
temperature $T_r$ is arbitrary, the renormalized susceptibility at each
temperature satisfies it separately:
\begin{equation}
\left[\mu\frac{\partial}{\partial\mu}
+ \beta(g)\frac{\partial}{\partial g}
+ 2\gamma_\Gamma(g)\right]\chi_\Gamma^{R}(\mu,T) = 0,
\label{eq:CS}
\end{equation}
where the term $2\gamma_\Gamma = \mu\,d\ln Z_\Gamma^{2}/d\mu$ arises from
the derivative of the prefactor $Z_\Gamma^{2}$, the factor of two
reflecting the two insertions of $O_\Gamma$ in the susceptibility.

Equation~(\ref{eq:CS}) is written for the massless susceptibility, where
the only scales are $\mu$ and $T$. At nonzero mass the full
Callan--Symanzik equation carries an additional term
$\gamma_m\, m\,\partial/\partial m$, with $\gamma_m$ the mass anomalous
dimension. This term does not affect the determination of $Z_\Gamma$. In a
mass-independent scheme $Z_\Gamma$ and $\gamma_\Gamma$ are
mass-independent, since the ultraviolet divergences come from the
short-distance region $|x|\sim a$, where $m|x|\sim ma \ll 1$ and the
propagator is effectively massless. The mass enters the divergent
coefficients only through corrections analytic in $ma$, which vanish as
$a\to 0$, and is absent from $Z_\Gamma$ and $\gamma_\Gamma$.
The invariance used below is invariance
under the renormalization scale $\mu$. The ratio $\kappa_{AB}$ is
evaluated at a fixed physical mass, common to both partners. For the
scalar density, the renormalization-group invariance of the mass term
$m\bar qq$ fixes $\gamma_S = -\gamma_m$, so $\chi_S$ renormalizes with
$Z_S^{2} = Z_m^{-2}$, reproducing the standard result.

The structure of eq.~(\ref{eq:CS}) is constrained by dimensional
analysis. In the massless theory at temperature $T$, the only scales
available to $\chi_\Gamma^{R}$ are $\mu$ and $T$, with $\Lambda_{\rm QCD}$
entering through $g(\mu)$, so
\begin{equation}
\chi_\Gamma^{R}(\mu,T) = T^2\,G\big(g(\mu),\ln(\mu/T)\big),
\label{eq:dim_analysis}
\end{equation}
exactly.
This is a statement about scales, not about the temperature dependence.
The dependence on $\mu$ is fixed by eq.~(\ref{eq:CS}).
Integrating it from $\mu=T$ gives
\begin{align}
\chi_\Gamma^{R}(\mu,T)
&= \exp\!\left[-2\int_{g(T)}^{g(\mu)}\!dg\,
\frac{\gamma_\Gamma(g)}{\beta(g)}\right] T^2\,F(T/\Lambda_{\rm QCD})
\notag\\
&= \left[\frac{\alpha_s(\mu)}{\alpha_s(T)}\right]^{\gamma_\Gamma^{(0)}/\beta_0}
T^2\,F(T/\Lambda_{\rm QCD})\,\big[1+O(\alpha_s)\big],
\label{eq:CS_solution}
\end{align}
with $F(T/\Lambda_{\rm QCD})\equiv G(g(T),0)$ the susceptibility
renormalized at $\mu=T$, in units of $T^2$.
The prefactor carries the entire $\mu$ dependence.
Hence $G$ depends on $\mu$ unless $\gamma_\Gamma=0$, and $F$ is a single
function of $T/\Lambda_{\rm QCD}$.
Expanded to first order in $g^2(\mu)$, the prefactor gives the logarithm
$1-\gamma_\Gamma^{(0)}(g^2/8\pi^2)\ln(\mu/T)$, which is
eq.~(\ref{eq:NLO_div}) below.
Perturbation theory expands $F$ in $\alpha_s(T)$.
To all orders, therefore, $F$ varies with $T$ only logarithmically.
Contributions such as powers of $\Lambda_{\rm QCD}/T$, being
$\propto\exp[-8\pi^2 n/(\beta_0 g^2(T))]$, are invisible at every order
and constitute the nonperturbative content of $F$.
The prefactor $T^2$ thus carries no implication about the sign or the
monotonicity of $\chi_\Gamma^R$.
Both are properties of $F$.

Perturbation theory supplies the asymptotic form of $G$ for
$T\gg\Lambda_{\rm QCD}$,
$G = \tilde c_0 + \tilde c_1\,g^2\,\ln(\mu/T) + O(g^4)$, i.e.\
$F=\tilde c_0+O(\alpha_s(T))$.
Already at this level the sign is instructive, and it is channel
dependent. The fermionic thermal parts of the master integrals are
$T\sum_n\!\int\!\frac{d^3p}{(2\pi)^3}\frac{1}{\omega_n^2+p^2}
-\int\!\frac{d^4p}{(2\pi)^4}\frac{1}{p^2}=-\frac{T^2}{24}$ and, for a
spatial index $k$, $\bigl[\int p_k^2/p^4\bigr]_{\rm th}=-T^2/48$ and
$\bigl[\int \omega_n^2/p^4\bigr]_{\rm th}=+T^2/48$. Combined with the
Dirac traces $\tr_D[\Gamma\slashed p\,\bar\Gamma\slashed p]$ they give, in the
manifestly positive normalization (Appendix~\ref{app:thermal_coeff}),
\begin{equation}
\tilde c_0^{S,P}=-\tilde c_0^{T,X}=-\frac{N_c\tr_F[t^at^a]}{6},\qquad
\tilde c_0^{V,A}=0,
\label{eq:c0_channels}
\end{equation}
i.e.\ $-1/4,0,+1/4$ for $N_c=3$. The corresponding results at nonzero and
nondegenerate quark masses are given in
Appendix~\ref{app:thermal_mass_nondeg}. At leading order the scalar and
pseudoscalar susceptibilities therefore \emph{decrease} with $T$, the
spatial vector and axial-vector ones are $T$-independent at $O(T^2)$, and
the tensor and axial-tensor ones \emph{increase}. Correspondingly, for
$T<T_r$, $\chi_\Gamma^{\rm reg}(T;T_r)=Z_\Gamma^2\tilde c_0(T^2-T_r^2)$ is
positive for $S,P$ and negative for $T,X$, and for $V,A$ it is of order
$\alpha_s T^2$ or nonperturbative. The same pattern is found in the free lattice theory with a chirally
symmetric action at the values of $N_t$ used in practice. It must be
stressed, however, that eq.~(\ref{eq:c0_channels}) is a statement about
the noninteracting theory, i.e.\ about the asymptotic regime
$T\gg\Lambda_{\rm QCD}$ in which the expansion of $G$ applies. In the
window $T_c\lesssim T\lesssim$ a few $T_c$ the interactions turn on the
nonperturbative part of $F$, dominated by the low-lying modes of the
Dirac operator, and this can override the $O(T^2)$ coefficient.
In particular, $F$ need not be monotonic in this window, and the vector
pair, whose $O(T^2)$ coefficient vanishes, acquires its slope from the
interactions alone.
Near $T_c$ the scalar and pseudoscalar susceptibilities fall
considerably faster than $T_r^2-T^2$.
Which behavior is realized in a given window is a dynamical question, to
be settled by the data.
In the $N_f=2+1+1$ data of Ref.~\cite{Chiu:2026sxy}, with
$T_r=658$~MeV and $164\le T\le 385$~MeV at three lattice spacings,
$\chi_\Gamma^{\rm reg}(T;T_r)$ is positive for $S,P$ and negative for
$V,A$ and $T,X$ at every temperature.
The scalar and pseudoscalar susceptibilities thus decrease with $T$,
while the vector- and tensor-type ones increase.
The expansion is
used below for a single purpose: to fix the coefficient of the one-loop
logarithm and verify that it is controlled by the anomalous dimension.
Nothing in the construction of $\kappa_{AB}$ depends on the values of
$\tilde c_0$ or $\tilde c_1$, nor on the range of validity of the
expansion. Inserting eq.~(\ref{eq:dim_analysis}), with this expansion of $G$, into
(\ref{eq:CS}) and equating the coefficients of $g^2$ [using eqs.~(\ref{eq:beta_def}) and (\ref{eq:gamma_def})]:
\begin{equation}
\tilde c_1 + 2\,\frac{\gamma_\Gamma^{(0)}}{16\pi^2}\,\tilde c_0 = 0
\quad\Longrightarrow\quad
\tilde c_1 = -\frac{\gamma_\Gamma^{(0)}\,\tilde c_0}{8\pi^2},
\label{eq:c1}
\end{equation}
so that
\begin{equation}
\chi_\Gamma^{R}(\mu,T) = \chi_\Gamma^{R(0)}(T)\left[
1 - \gamma_\Gamma^{(0)}\,\frac{g^2(\mu)}{8\pi^2}\,\ln\frac{\mu}{T}
+ O(g^4)\right],
\label{eq:NLO_div}
\end{equation}
with $\chi_\Gamma^{R(0)}(T) = \tilde c_0\,T^2$ the asymptotic
high-temperature form. The one-loop log coefficient is thus fixed to be
proportional to the leading-order anomalous dimension
$\gamma_\Gamma^{(0)}$, as anticipated.

Integrating the definition~(\ref{eq:gamma_def}) from the matching scale
$\mu = 1/a$, where $Z_\Gamma = 1$, down to $\mu$ at fixed coupling gives
$Z_\Gamma^{2} = 1 + \gamma_\Gamma^{(0)}\,(g^2/8\pi^2)\ln(a\mu) + O(g^4)$.
The all-orders solution with running coupling is derived in
Sec.~\ref{sec:Z}. Substituting this together with eq.~(\ref{eq:NLO_div})
into eq.~(\ref{eq:chi_reg_renorm}), the arbitrary scale $\mu$ cancels
between $Z_\Gamma^{2}$ and the renormalized susceptibilities, as it must,
leaving
\begin{align}
&\chi_\Gamma^{\rm reg}(T;T_r,a) =
\big[\chi_\Gamma^{R(0)}(T) - \chi_\Gamma^{R(0)}(T_r)\big] + \notag \\
&\gamma_\Gamma^{(0)}\frac{g^2}{8\pi^2}
\big[\chi_\Gamma^{R(0)}(T)\ln(aT) - \chi_\Gamma^{R(0)}(T_r)\ln(aT_r)\big]
+ O(g^4),
\label{eq:NLO_bare}
\end{align}
with $\chi_\Gamma^{R(0)}(T) = \tilde c_0\,T^2$ the asymptotic
high-temperature form and $\gamma_\Gamma^{(0)}$ the leading-order
anomalous dimension. Two features
are worth noting. First, the residual dependence on the cutoff is
\emph{logarithmic}, not power-like: the additive divergences have
been removed by the temperature subtraction, and what survives is the
multiplicative renormalization, appearing here as the coefficient
$\gamma_\Gamma^{(0)}(g^2/8\pi^2)\ln(aT)$ multiplying the finite
susceptibility. For an operator with vanishing anomalous dimension
($\gamma_\Gamma^{(0)}=0$) even this logarithm is absent and
$\chi_\Gamma^{\rm reg}$ is finite. Second, the ultraviolet logarithm pairs
the cutoff with the physical infrared scale, here the temperature,
just as the mass logarithm of eq.~(\ref{eq:mass_div}) pairs it
with the quark mass.

\subsubsection*{All-orders structure and the need for resummation}

At order $g^{2n}$, the renormalized susceptibility
$\chi_\Gamma^{R}(\mu,T)$ receives contributions proportional
to $g^{2n}\ln^n(\mu/T)$ (leading-logs), $g^{2n}\ln^{n-1}(\mu/T)$
(subleading-logs), etc.
Equivalently, in the regularized susceptibility
$\chi_\Gamma^{\rm reg}$ these appear as powers of $\ln(a\mu)$ carried by
$Z_\Gamma^{2}$.
On the lattice, the lattice scale is $1/a$, while the physical
renormalization scale $\mu$ is much smaller.
Hence $\ln(a\mu)$ is large
and negative, and the leading-log series
$\sum_n (g^2 \ln(a\mu))^n$ is not small. Fixed-order perturbation theory
in $g^2$ is therefore inadequate, and the large logs must be resummed.
The renormalization group provides this resummation through the
multiplicative renormalization constant $Z_\Gamma(\mu, a)$, which
we now derive.

\subsection{Multiplicative renormalization}
\label{sec:Z}

We now solve the renormalization group equation for $Z_\Gamma$ and verify
that it resums the leading-log series found at NLO. From the definition of
the anomalous dimension in eq.~(\ref{eq:gamma_def}),
\begin{equation}
\mu \frac{d}{d\mu}\ln Z_\Gamma = \gamma_\Gamma(g(\mu)),
\label{eq:RGE}
\end{equation}
which restates the definition~(\ref{eq:gamma_def}) along the running
coupling $g = g(\mu)$.
The physical input is the $\mu$-independence
of the regularized susceptibility, eq.~(\ref{eq:chi_reg_renorm}). Converting the
$\mu$-derivative to a $g$-derivative via the chain rule
$\mu\,d/d\mu = \beta(g)\,d/dg$, the RGE becomes
\begin{equation}
\beta(g)\,\frac{d\ln Z_\Gamma}{dg} = \gamma_\Gamma(g),
\end{equation}
which integrates to
\begin{align}
\ln Z_\Gamma(\mu, a)
&= \int_{g(1/a)}^{g(\mu)} \frac{\gamma_\Gamma(g')}{\beta(g')}\,dg' \notag \\
&= -\frac{\gamma_\Gamma^{(0)}}{2\beta_0}\,\ln\!\left[\frac{g^2(\mu)}{g^2(1/a)}\right]
+ O(g^2),
\end{align}
where the lower limit $g(1/a)$ corresponds to the coupling
at the UV cutoff scale $1/a$, and we used
$\gamma_\Gamma/\beta = -\gamma_\Gamma^{(0)}/(\beta_0 g)$ at leading
order. Exponentiating:
\begin{equation}
Z_\Gamma(\mu, a) = \left[
\frac{\alpha_s(\mu)}{\alpha_s(1/a)}
\right]^{-\gamma_\Gamma^{(0)}/(2\beta_0)},
\label{eq:Z_solution}
\end{equation}
where $\alpha_s = g^2/(4\pi)$ and $\beta_0 = 11 - 2N_f/3$ (for $N_c=3$)
is the leading coefficient of the beta function.
Equation~(\ref{eq:Z_solution}) is the leading-log resummation, controlled
by the one-loop coefficients $\gamma_\Gamma^{(0)}$ and $\beta_0$.
Subleading-logarithms, governed by the higher-order coefficients
$\gamma_\Gamma^{(1)}$, $\beta_1$, and beyond, modify $Z_\Gamma$ at higher
order. They do not affect $\kappa_{AB}$, in which $Z_\Gamma$ cancels
exactly between symmetry partners (Sec.~\ref{sec:kappa}), so the
leading-log form suffices for our purposes.

For operators with $\gamma_\Gamma^{(0)} \neq 0$ (scalar, pseudoscalar,
tensor vector, axial-tensor vector), $Z_\Gamma$ is logarithmically divergent
as $a \to 0$. Using the leading-log running
$\alpha_s(1/a) = 2\pi/[\beta_0\ln(1/(a\Lambda_{\rm QCD}))]$:
\begin{equation}
Z_\Gamma(\mu, a) \sim
\left[\frac{\alpha_s(\mu)\,\beta_0}{2\pi}\,\ln\frac{1}{a\Lambda_{\rm QCD}}\right]^{-\gamma_\Gamma^{(0)}/(2\beta_0)}.
\end{equation}

We now verify that $Z_\Gamma^{2}$, as defined by (\ref{eq:Z_solution}),
indeed resums the leading-log series identified in
Sec.~\ref{subsec:subtraction}, eq.~(\ref{eq:NLO_div}). Expanding
(\ref{eq:Z_solution}) to leading order in $g^2(\mu)$ using
$\alpha_s(1/a) = \alpha_s(\mu)/[1 - \alpha_s(\mu)\,\beta_0\ln(a\mu)/(2\pi) + \cdots]$:
\begin{align}
Z_\Gamma^{2}(\mu, a)
&= \left[\frac{\alpha_s(\mu)}{\alpha_s(1/a)}\right]^{-\gamma_\Gamma^{(0)}/\beta_0}
\notag\\
&= \left[1 - \alpha_s(\mu)\,\frac{\beta_0}{2\pi}\,\ln(a\mu)\right]^{-\gamma_\Gamma^{(0)}/\beta_0}
\notag\\
&= 1 + \frac{\gamma_\Gamma^{(0)}\,\alpha_s(\mu)}{2\pi}\,\ln(a\mu)
+ O(\alpha_s^2).
\label{eq:Z2_expansion}
\end{align}
Using $\alpha_s = g^2/(4\pi)$, the one-loop coefficient of $\ln(a\mu)$
in $Z_\Gamma^{2}$ is $+\gamma_\Gamma^{(0)}\,g^2(\mu)/(8\pi^2)$, which
matches exactly the $\ln(\mu/T)$ correction in eq.~(\ref{eq:NLO_div})
with opposite sign, so that the product
$Z_\Gamma^{2}\,\chi_\Gamma^{R}(\mu,T)$ is $\mu$-independent. This is the
expected agreement: since
$\chi_\Gamma^{\rm reg} = Z_\Gamma^{2}[\chi_\Gamma^R(T)-\chi_\Gamma^R(T_r)]$
[eq.~(\ref{eq:chi_reg_renorm})], the explicit one-loop
log is generated by $Z_\Gamma^{2}$ acting on the
renormalized susceptibility, and the cancellation of $\mu$ fixes the
coefficient. This
confirms that the one-loop log in the OPE coefficient is the leading term of
the geometric series summed by (\ref{eq:Z_solution}). The all-orders
leading-log series $\chi_\Gamma^R\sum_n [\,\gamma_\Gamma^{(0)}\alpha_s\ln(a\mu)]^n$
becomes
$[\alpha_s(\mu)/\alpha_s(1/a)]^{-\gamma_\Gamma^{(0)}/\beta_0}\,\chi_\Gamma^R =
Z_\Gamma^2\,\chi_\Gamma^R$ after resummation, valid for
$a\Lambda_{\rm QCD}\ll 1$. The large logarithms are thereby absorbed into
$Z_\Gamma^2$, which retains the residual dependence
$Z_\Gamma^2 \sim [\ln(1/(a\Lambda_{\rm QCD}))]^{-\gamma_\Gamma^{(0)}/\beta_0}$
on the lattice spacing. This factor is common to symmetry partners and
cancels in $\kappa_{AB}$.

\subsection{Quark-disconnected contributions}
\label{subsec:disconnected}

One further feature is specific to the flavor singlet.
Its susceptibility receives quark-disconnected contractions in addition to
the connected ones. Summed over the four-volume, the disconnected
contraction is a product of single-propagator loops,
\begin{align}
\chi_{\Gamma,\rm disc} = \frac{1}{V} \Big\{
&\vev{\left(\Tr[\Gamma\,(D_c+m)^{-1}]\right)^2}  \notag \\
&-\vev{\Tr[\Gamma\,(D_c+m)^{-1}]}^2 \Big\},
\label{eq:chi_disc}
\end{align}
where $(D_c+m)^{-1}$ is the valence quark propagator in lattice QCD with
exact chiral symmetry~\cite{Chiu:1998eu}, $\Tr$ denotes the trace over
color, Dirac, and site indices, and $V$ is the four-volume. With the
$O_\Gamma^\dagger$ of eq.~(\ref{eq:Ct}) the second loop carries
$\bar\Gamma=\pm\Gamma$ [eq.~(\ref{eq:Gammabar})], so the relative sign
between the connected and the disconnected parts is this $\pm$; it has been
suppressed in eq.~(\ref{eq:chi_disc}).

Equation~(\ref{eq:chi_disc}) is a variance, and this makes its tree-level
content immediate. In the free theory the gauge field is absent, the
propagator is a fixed c-number matrix, and the loop
$L_\Gamma \equiv \Tr[\Gamma\,(D_c+m)^{-1}]$ is a number rather than a random
variable. The two terms of eq.~(\ref{eq:chi_disc}) are then equal,
$\vev{L_\Gamma^2} = \vev{L_\Gamma}^2 = L_\Gamma^2$, and the disconnected
susceptibility vanishes identically.

The disconnected contribution is therefore generated entirely by gauge-field
fluctuations. It is absent from the free-field analysis of
Sec.~\ref{subsec:additive}, which is why the coefficients $\alpha_\Gamma$ and
$c_m^\Gamma$ computed there are quark-connected quantities, and why that
analysis is nonetheless complete at tree level. Beyond tree level the
disconnected contractions contribute to the Wilson coefficient of the
identity operator $\Unit$ in the same operator product expansion, and therefore
shift $\alpha_\Gamma$ and $c_m^\Gamma$ rather than introduce new divergence
structures. The multiplicative constant is unaffected: the disconnected
part involves the same bilinear $O_\Gamma$ and carries the same
$Z_\Gamma$. Neither the values of the coefficients nor that of $Z_\Gamma$
is needed, the former being removed by the temperature subtraction and the
latter canceling in $\kappa_{AB}$.

For the Lorentz-index-carrying singlets,
$\Gamma = \gamma_k,\ i\gamma_5\gamma_k,\ \gamma_4\gamma_k,\
i\gamma_5\gamma_4\gamma_k$, the loop $\Tr[\Gamma\,(D_c+m)^{-1}]$ does
\emph{not} vanish on a given gauge configuration: in a gauge background
$D_c^2$ is not a Dirac scalar (it contains $\sigma_{\mu\nu}$
field-strength terms), so no site-local Dirac-trace argument applies
beyond tree level. What vanishes is the ensemble average,
$\vev{\Tr[\Gamma\,(D_c+m)^{-1}]} = 0$ for all four channels. Each of these
operators carries a single free spatial index, while the thermal ensemble
singles out no spatial direction, and the only three-vector invariant under
the spatial rotation group is zero. The thermal state supports no vector,
axial-vector, or tensor condensate. This is the vacuum-expectation side of the mixing
statement of Sec.~\ref{subsec:subtraction}: taking the expectation value
of eq.~(\ref{eq:O_renorm_def}) gives
$\vev{O_\Gamma^{\rm bare}} = Z_\Gamma\vev{O_\Gamma^{R}} + b_\Gamma$, so
$b_\Gamma$ is the divergent part of the averaged loop, and a nonzero
$b_\Gamma$, like a nonzero $\vev{O_\Gamma}$, requires vacuum quantum
numbers, which the scalar singlet alone possesses. Consequently the VEV-subtraction term of
eq.~(\ref{eq:chi_disc}) is absent for these channels, while their
disconnected correlators, being even under all three symmetries, are
nonzero in general, in QCD with dynamical quarks and in quenched QCD
alike, generated by gauge-field fluctuations.

For the pseudoscalar singlet, $\Gamma=\gamma_5$, the loop is fixed by the
index theorem. The exact zero modes of the overlap operator saturate the
trace,
\begin{equation}
\Tr[\gamma_5\,(D_c+m)^{-1}] = \frac{Q_t}{m},
\label{eq:index_loop}
\end{equation}
where $Q_t = n_+ - n_-$ is the topological charge, an integer counting the
difference of the numbers of exact zero modes of positive and negative
chirality. Inserting eq.~(\ref{eq:index_loop}) into
eq.~(\ref{eq:chi_disc}) and averaging over gauge configurations of all topological
sectors, gives the disconnected susceptibility of the pseudoscalar singlet,
\begin{align}
\chi_{5,\rm disc}(T) &= \frac{\chi_t(T)}{m^2}, \notag \\
\chi_t(T) &= \frac{1}{V}\left(\vev{Q_t^2}_T - \vev{Q_t}_T^2\right),
\label{eq:chi_t}
\end{align}
where $\chi_t(T)$ is the topological susceptibility, defined here as the
variance of the topological charge per unit four-volume in the thermal
ensemble at temperature $T$. The subscript $T$ on the expectation values
is suppressed elsewhere. By parity the expectation value of $Q_t$ vanishes
upon averaging over all topological sectors,
\begin{align}
\vev{Q_t} =
\frac{\sum_{n=-\infty}^{+\infty} \vev{n|Q_t|n}}{\sum_{n=-\infty}^{+\infty} \vev{n|n}}=0,
\end{align}
so that $\chi_t=\vev{Q_t^2}/V$ in the infinite-statistics limit. At finite
statistics $\vev{Q_t}^2$ is a parity-violating fluctuation, and the variance
form is retained.

Both $Q_t$ and $\chi_t$ are RG-invariant, since $Q_t$ is an integer obtained
by counting exact zero modes and carries no anomalous dimension. The entire
scale dependence of $\chi_{5,\rm disc}$ therefore resides in $1/m^2$. Using
$m = Z_S^{-1} m_R$, which follows from the RG-invariance of $m\bar q q$, and
the relation $Z_S = Z_P$ for Ginsparg-Wilson fermions, the disconnected part
renormalizes exactly as the connected one,
\begin{align}
\chi_{5,\rm disc}^R(T,\mu) &= Z_P^{-2}\,\chi_{5,\rm disc}^{\rm reg} \notag \\
&= \frac{\chi_t(T)-\chi_t(T_r)}{Z_S^2\,m^2}
 = \frac{\chi_t(T)-\chi_t(T_r)}{m_R^2}.
\label{eq:chi5_disc_R}
\end{align}
The same $Z_P^{-2}$ that renormalizes the connected part also renormalizes
the disconnected part, so the singlet susceptibility renormalizes
multiplicatively as a whole, and the decomposition into connected and
disconnected pieces is preserved under renormalization. The renormalized
relation $\chi_{5,\rm disc}^R=\chi_t(T)/m_R^2$ is recovered only if
$\chi_t(T_r)\ll\chi_t(T)$, i.e., for $T_r\gg T_c$.

The same condition applies to the partner splitting
$\chi_\pi-\chi_\delta$, which upon effective restoration of
$SU(2)_L\times SU(2)_R$ equals $\chi_{5,\rm disc}$ for two light
flavors [Sec.~\ref{sec:sym_rel_conseq}, eq.~(\ref{eq:pidelta_chit})].
The temperature subtraction~(\ref{eq:chi_reg}) removes the additive
divergences of $\chi_\pi$ and $\chi_\delta$, but it also subtracts the
topological contribution $\chi_t(T_r)/m^2$ to the reference splitting
$\chi_\pi(T_r)-\chi_\delta(T_r)$.
With eq.~(\ref{eq:chi_reg_renorm}), $Z_P=Z_S$ and $m=Z_S^{-1}m_R$, one
obtains
\begin{equation}
\big[\chi_\pi^R(T)-\chi_\delta^R(T)\big]
-\big[\chi_\pi^R(T_r)-\chi_\delta^R(T_r)\big]
\simeq\frac{\chi_t(T)-\chi_t(T_r)}{m_R^2},
\label{eq:pidelta_reg_Tr}
\end{equation}
with the renormalization scale suppressed.
The regularized splitting thus measures $\chi_t(T)-\chi_t(T_r)$ rather
than $\chi_t(T)$, and the renormalized relation
$\chi_\pi^R-\chi_\delta^R\simeq\chi_t/m_R^2$ is recovered only for
$T_r\gg T_c$.

For the scalar singlet, $\Gamma=\Id$, the loop is the chiral condensate,
$\frac{1}{V}\Tr[(D_c+m)^{-1}] = \vev{\bar q q}$, which is nonzero at any
temperature. This is the one channel in which the VEV-subtraction term of
eq.~(\ref{eq:chi_disc}) does not drop out: it removes the disconnected
square $\vev{\bar q q}^2$ point by point in temperature, leaving the genuine
fluctuation of the condensate,
$\vev{(\Tr\,(D_c+m)^{-1})^2} - \vev{\Tr\,(D_c+m)^{-1}}^2$. Unlike the
pseudoscalar case, it is not fixed by an index theorem.

The ultraviolet structure of the disconnected parts is the same in every
channel. The loop $\Tr[\Gamma\,(D_c+m)^{-1}]$, built from the propagator at
coincident points, is power divergent, and its variance inherits
short-distance coincident-point divergences. These are of the same
short-distance origin as the connected ones, arising from the
$|x|\to 0$ region where the coefficient functions of the operator product
expansion take their vacuum form. This is a statement about the operator
product and its Wilson coefficients, and is insensitive to the quark-line
topology of the contractions. The temperature independence of the
divergences (Sec.~\ref{subsec:subtraction}) therefore applies to the
disconnected ones as well, and the temperature
subtraction~(\ref{eq:chi_reg}) removes them together with the connected
ones.
Multiplicatively, the disconnected contribution involves the same bilinear
$O_\Gamma$ and renormalizes with the same $Z_\Gamma$ as the connected
part of that channel, so the partner equalities of Sec.~\ref{sec:chains}
and the cancellation in $\kappa_{AB}$ are unaffected.
No additive divergence of any susceptibility, singlet or nonsinglet,
survives the subtraction.

\subsubsection*{The pseudoscalar connected susceptibility and the condensate}

Exact chiral symmetry relates the connected pseudoscalar susceptibility to the
chiral condensate. Since $D_c = \gamma_\mu D_c^\mu$ is a pure Dirac vector, it anticommutes
with $\gamma_5$, $\gamma_5 D_c \gamma_5 = -D_c$, so that $\gamma_5 (D_c+m)^{-1}\gamma_5 = (m - D_c)^{-1}$ and
\begin{align}
&\Tr[\gamma_5 (D_c+m)^{-1}\gamma_5 (D_c+m)^{-1}] \notag \\
=& \Tr[(m^2 - D_c^2)^{-1}]
= \frac{\Tr[(D_c+m)^{-1}]}{m},
\label{eq:g5_identity}
\end{align}
where we have used
$(D_c+m)^{-1} = (m - D_c)(m^2 - D_c^2)^{-1}$ and
$\tr(\gamma_\mu)=0$ at each site.
Hence
\begin{align}
\chi_{5,\rm conn}
&= \frac{1}{V}\,\vev{\Tr[\gamma_5 (D_c+m)^{-1}\gamma_5 (D_c+m)^{-1}]}
\notag\\
&= \frac{1}{V}\,\frac{\vev{\Tr[(D_c+m)^{-1}]}}{m}
\;\equiv\; \frac{\Sigma}{m},
\label{eq:5_conn}
\end{align}
with $\Sigma$ the chiral condensate.

Equation~(\ref{eq:5_conn}) is consistent with the renormalization established
above, and provides an independent check of it. The condensate renormalizes as
$\Sigma = Z_S\,\Sigma_R$ and the mass as $m = Z_S^{-1} m_R$, so that
\begin{align}
Z_P^2\,\chi_{5,\rm conn}^{R}(T;T_r,\mu)
&= \chi_{5,\rm conn}^{\rm reg} \notag \\
&= \chi_{5,\rm conn}(T) - \chi_{5,\rm conn}(T_r) \notag \\
&= \frac{\Sigma(T) - \Sigma(T_r)}{m} \notag \\
&= Z_S^2\,\frac{\Sigma_R(T;T_r,\mu)}{m_R(\mu)} .
\end{align}
Using $Z_S = Z_P$ (Sec.~\ref{subsec:ZAZB_lat}), the renormalized form of
eq.~(\ref{eq:5_conn}) follows,
\begin{equation}
\chi_{5,\rm conn}^{R}(T;T_r,\mu)
= \frac{\Sigma_R(T;T_r,\mu)}{m_R(\mu)} .
\label{eq:5_conn_R}
\end{equation}
The relation therefore holds in the same form before and after
renormalization, which is possible only because $Z_S = Z_P$. Conversely,
eq.~(\ref{eq:g5_identity}) constrains the two constants without assuming their
equality. Its two sides carry $Z_P^{2}$ and $Z_S^{2}$ respectively, so
$\chi_{5,\rm conn}^{R} = (Z_S/Z_P)^2\,\Sigma_R/m_R$, and since both
$\chi_{5,\rm conn}^{R}$ and $\Sigma_R/m_R$ are finite, the ratio $Z_S/Z_P$
must be cutoff-independent: the logarithms cancel between them, and
\begin{equation}
\gamma_S = \gamma_P .
\label{eq:gS_gP}
\end{equation}
The same conclusion follows from the index relation~(\ref{eq:index_loop}),
which gives $(Z_P/Z_S)\,m_R \Tr[\gamma_5 (D_c+m)^{-1}]_R = Q_t$ with $Q_t$ an
integer, hence cutoff-independent. Neither argument fixes the constant
$Z_S/Z_P$ to unity. That is supplied by the exact axial rotation of the DWF/GW
action (Sec.~\ref{subsec:ZAZB_lat}). What they do provide is an independent
check that the anomalous dimensions agree, as they must if
eq.~(\ref{eq:5_conn_R}) is to hold at every scale.

\subsubsection*{Independence of the contact-term prescription}
\label{subsec:prescription}

The renormalization established in this paper does not depend on whether the
$t=0$ slice is retained in the bare susceptibility. The two admissible
prescriptions, eq.~(\ref{eq:chi_bare_lat}) with the contact term and
eq.~(\ref{eq:chi_bare_lat_ne0}) without it, were introduced in
Sec.~\ref{sec:susceptibility}.

Under eq.~(\ref{eq:chi_bare_lat_ne0}) the connected and disconnected parts
of any channel are the same double lattice sums as before, restricted to
pairs of sites lying in different time slices,
\begin{align}
&\chi_{\Gamma,\rm conn}\big|_{t\neq 0}  \notag \\
=& -\frac{1}{V} \sum_{\substack{x,y \\ x_4 \neq y_4}}
 \vev{\tr_{DC}[\Gamma\,(D_c+m)^{-1}_{x,y}
 \bar\Gamma\,(D_c+m)^{-1}_{y,x}]}, \notag \\
&\chi_{\Gamma,\rm disc}\big|_{t\neq 0}  \notag \\
=& \frac{1}{V} \sum_{\substack{x,y \\ x_4 \neq y_4}}
\Big\{ \vev{\tr_{DC}[\Gamma\,(D_c+m)^{-1}_{x,x}]\;
     \tr_{DC}[\bar\Gamma\,(D_c+m)^{-1}_{y,y}]} \notag \\
&\quad- \vev{\tr_{DC}[\Gamma\,(D_c+m)^{-1}_{x,x}]}\,
   \vev{\tr_{DC}[\bar\Gamma\,(D_c+m)^{-1}_{y,y}]} \Big\} ,
\label{eq:chi_disc_ne0}
\end{align}
where $\tr_{DC}$ denotes the trace over Dirac and color indices at a single
site.
Here $\bar\Gamma$ and the minus sign of the Wick contraction are written out.

The disconnected part of eq.~(\ref{eq:chi_disc_ne0}) is again a covariance
of the same operator $O_\Gamma$. The operators being correlated are
unchanged, so the temperature
subtraction and the multiplicative renormalization proceed exactly as before,
\begin{align}
&\chi_{\Gamma,\rm disc}^{\rm reg}\big|_{t\neq 0}
= Z_\Gamma^2\,\big[\chi_{\Gamma,\rm disc}^{R}(T)
 - \chi_{\Gamma,\rm disc}^{R}(T_r)\big]\big|_{t\neq 0}\, , \notag \\
&\chi_{\Gamma,\rm conn}^{\rm reg}\big|_{t\neq 0}
= Z_\Gamma^2\,\big[\chi_{\Gamma,\rm conn}^{R}(T)
 - \chi_{\Gamma,\rm conn}^{R}(T_r)\big]\big|_{t\neq 0}\, .
\end{align}
The renormalized quantities differ from those of
eq.~(\ref{eq:chi_bare_lat}) by a finite amount, namely the contribution of
the $t=0$ slice, but they carry the same $Z_\Gamma$. Both parts therefore
renormalize multiplicatively with the same factor, and so does their sum,
\begin{equation}
\chi_{\Gamma}^{\rm reg}\big|_{t\neq 0}
= Z_\Gamma^2\,\big[\chi_{\Gamma}^{R}(T)
 - \chi_{\Gamma}^{R}(T_r)\big]\big|_{t\neq 0}\, .
\end{equation}

The reason is general and worth stating plainly. The factor $Z_\Gamma$
renormalizes the operator $O_\Gamma$, whereas the choice of prescription
selects which pairs of points are summed. Restricting the sum removes an
additive contribution. It does not alter the operators being correlated. The
two prescriptions therefore define bare susceptibilities that differ by a
finite, temperature-dependent amount, but they carry the same $Z_\Gamma$, obey
the same Callan-Symanzik equation, and yield the same $Z_A = Z_B$. Every
statement in this paper, including the renormalization-group invariance of
$\kappa_{AB}$, holds for both eq.~(\ref{eq:chi_bare_lat}) and
eq.~(\ref{eq:chi_bare_lat_ne0}). The two agree in the continuum limit, where
the weight $a\,C_\Gamma(0,T)$ of the omitted slice contributes to
$\chi_\Gamma^{\rm reg}$ only through its temperature-dependent part, which is
finite and therefore vanishes as $a\to 0$.

\subsection{The complete UV decomposition}
\label{subsec:UV_decomp}

The NLO result~(\ref{eq:NLO_bare}) and the all-orders structure of
Sec.~\ref{subsec:subtraction} showed that, once the additive divergences are removed by
the temperature subtraction, the residual UV dependence of
$\chi_\Gamma^{\rm reg}$ is the multiplicative factor $Z_\Gamma^{2}(\mu,a)$.
This is the content of eq.~(\ref{eq:chi_reg_renorm}), which was established
for every channel from the operator renormalization~(\ref{eq:O_renorm_def})
and the VEV subtraction~(\ref{eq:vev_sub_mult}).
It is instructive to display the decomposition at a single temperature,
before the difference is formed.
The VEV-subtracted bare susceptibility splits into its additive
divergences and the multiplicatively renormalized remainder,
\begin{align}
\chi_\Gamma^{\rm bare}(T, a) &= \frac{\alpha_\Gamma}{2a^2}
+ c_m^\Gamma\, m^2\ln\frac{1}{am} \notag \\
&+ Z_\Gamma^{2}(\mu, a)\,\chi_\Gamma^R(\mu, T, m) + O(a).
\label{eq:UV_decomp}
\end{align}
Both additive terms are temperature-independent, so forming the difference
$\chi_\Gamma^{\rm bare}(T,a)-\chi_\Gamma^{\rm bare}(T_r,a)$ cancels them and
reproduces eq.~(\ref{eq:chi_reg_renorm}), with
$Z_\Gamma^{2}\,[\chi_\Gamma^R(\mu,T,m) - \chi_\Gamma^R(\mu,T_r,m)]$.
In~(\ref{eq:UV_decomp}):
\begin{itemize}
\item $\alpha_\Gamma/(2a^2)$ is the leading additive power divergence
arising from the identity operator $\Unit$ in the OPE. The power $1/a^2$
is \textbf{universal} across all bilinears, while the coefficient
$\alpha_\Gamma$ is channel-dependent (determined by the Dirac trace
$\eta_\Gamma$). Both are \emph{temperature-independent}.
\item $c_m^\Gamma\, m^2\ln(1/(am))$ is the mass-dependent additive
divergence (Sec.~\ref{subsec:additive}, eq.~\eqref{eq:mass_div}). It is
chirally even, and therefore present even for Ginsparg--Wilson fermions and
already in the continuum, and it is likewise temperature-independent.
\item $Z_\Gamma(\mu, a)$, given by eq.~\eqref{eq:Z_solution}, encodes
the all-orders leading-log resummation governed by the anomalous
dimension $\gamma_\Gamma$.
\item $\chi_\Gamma^R(\mu, T, m)$ is the renormalized, UV-finite,
temperature-dependent susceptibility.
\end{itemize}

Because $\alpha_\Gamma$ and $c_m^\Gamma$ are additive coincident-point
divergences, they
do not enter the multiplicative constant $Z_\Gamma$, which renormalizes
the operator $O_\Gamma$ and depends on its Dirac structure alone.
Their channel dependence (e.g.\ $c_m^A \neq c_m^B$) therefore does not
threaten the equality $Z_A = Z_B$. The decomposition~(\ref{eq:UV_decomp})
is a statement about operator structure and holds nonperturbatively. What
perturbation theory supplies is only the \emph{values} of $\alpha_\Gamma$,
$c_m^\Gamma$, and $Z_\Gamma$, none of which enters the construction of
$\kappa_{AB}$.

\subsection{Choice of the reference temperature}
\label{subsec:T_sub}

Since all additive divergences are $T$-independent, they cancel exactly
in the temperature subtraction that defines
$\chi_\Gamma^{\rm reg}(T;T_r,a)$ [eq.~(\ref{eq:chi_reg})]. Combined with
the multiplicative renormalization~(\ref{eq:chi_reg_renorm}),
\begin{equation}
\chi_\Gamma^{\rm reg}(T;T_r,a) = Z_\Gamma^{2}(\mu,a)\left[
\chi^R_\Gamma(\mu,T) - \chi^R_\Gamma(\mu,T_r)\right],
\end{equation}
the regularized susceptibility is \textbf{free of all additive UV
divergences} and remains multiplicatively renormalizable, with the
factor $Z_\Gamma^{2}$ of eq.~\eqref{eq:Z_solution} carrying all
the logarithmic UV dependence. The temperature subtraction removes the
additive divergences without any knowledge of $\alpha_\Gamma$ or
$c_m^\Gamma$ being required.

The reference temperature is a genuine choice, and we now establish in
what sense $\kappa_{AB}(T;T_r)$ is independent of it.
From eq.~(\ref{eq:chi_reg_renorm}), define the partner difference and
sum of the renormalized susceptibilities,
\begin{align}
d(T)&\equiv \chi_A^R(\mu,T)-\chi_B^R(\mu,T), \notag\\
s(T)&\equiv \chi_A^R(\mu,T)+\chi_B^R(\mu,T).
\label{eq:ND_def}
\end{align}
In terms of these the ratio~(\ref{eq:kappa_def}) is, for partners of a
common sign,
\begin{equation}
\kappa_{AB}(T;T_r)
= \left|\frac{\chi_A^{\rm reg}-\chi_B^{\rm reg}}{\chi_A^{\rm reg}+\chi_B^{\rm reg}}\right|
= \frac{|d(T)-d(T_r)|}{|s(T)-s(T_r)|} ,
\label{eq:kappa_ND}
\end{equation}
where the common factor $Z_A^2=Z_B^2\equiv Z^2$ (Sec.~\ref{sec:chains}) has
cancelled between numerator and denominator. The ratio is
$\mu$-independent although $d$ and $s$ separately are not, since a change
of scale rescales $\chi_A^R$ and $\chi_B^R$ by the common factor
$[Z_\Gamma(\mu,a)/Z_\Gamma(\mu',a)]^2$ ($Z_A=Z_B$), which cancels. The
$T_r$-dependence of $\kappa_{AB}$ thus resides entirely in $d(T_r)$ and
$s(T_r)$.

At $m\neq 0$ the symmetry is explicitly broken, so the renormalized
partner splitting $d(T)$ does not vanish identically at any temperature,
even in the continuum limit. Above the transition it is a small
explicit-breaking remnant, $O(m^2)$ at tree level
[eq.~(\ref{eq:A_deg})], that would vanish only in the chiral limit.
``Effective restoration'' therefore means $d(T)$ small within the
precision of the data, not $d(T)=0$. A reference temperature is
\emph{admissible} if two conditions hold: it lies far enough above the
crossover $T_c$ of the sector that $d(T_r)$ is consistent with zero within
errors, and on the lattice it retains enough time slices, $N_t=1/(aT_r)$,
that the denominator $s(T)-s(T_r)$ keeps a definite sign over the whole
range.

The first condition is a requirement on the quark masses as well as on
the temperature.
Above the crossover the residual splitting $d(T_r)$ is of order
$m_1m_2$ for the quark flavors of the pair
[eq.~(\ref{eq:A_nondeg}) at tree level], against a partner sum of order
$T_r^2$.
$T_r$ must therefore be chosen such that these $O(m^2)$ effects are small
compared with the precision of the data.
For the light sector this holds at any $T_r$ above $T_c$.
For a heavy--light pair the splitting is proportional to $m_lm_Q$ and
remains small.
For a pair of two heavy quarks, such as $\bar cc$, the first condition
requires $m_c^2/T_r^2\ll1$, i.e., $T_r\gg m_c\simeq1.2$~GeV.
Whether this holds depends on the lattice.
With $T_r=1/(N_ra)$ and $N_r\ge4$, it requires $a\ll1/(4m_c)\simeq0.04$~fm,
which is not met for most lattices in use.
Admissibility is a property of the reference for each pair, since
$d(T_r)$ and $s(T_r)$ enter $\kappa_{AB}$ pair by pair
[eq.~(\ref{eq:kappa_ND})].
A reference that fails the first condition only for the $\bar cc$ pairs
therefore remains admissible for all other pairs.
Only the $\bar cc$ ratios then carry a bias of order $m_c^2/T_r^2$.

The second condition is a requirement on the data, not a consequence of
the theory.
Perturbation theory fixes the sign of $s(T)-s(T_r)$ only at
asymptotically high temperature, through eq.~(\ref{eq:c0_channels}):
positive for the scalar--pseudoscalar pair, negative for the tensor pair,
and undetermined at $O(T^2)$ for the vector pair, whose leading
coefficient vanishes.
In the window above $T_c$ the nonperturbative part of $F$ need not be
monotonic (Sec.~\ref{subsec:subtraction}).
If $s(T)$ is not monotonic, $s(T)-s(T_r)$ can vanish at some $T\neq T_r$.
The ratio then exceeds unity near that temperature, and the reference is
inadmissible for the range that contains it.
The constancy of the sign must therefore be verified channel by channel.
In the data of Ref.~\cite{Chiu:2026sxy} it holds in all three pairs,
$(P,S)$, $(V,A)$ and $(T,X)$, at every temperature and lattice spacing
(Sec.~\ref{subsec:subtraction}).
At $N_t=2$, where the single surviving time slice no longer resolves the
Dirac structures and all six channels approach a common value, the second
condition fails and the reference is inadmissible.

Numerator and denominator must use the \emph{same} reference: only then
are both free of additive divergences with the same $Z^2$, so that
channels are compared on a common scale, $0\le\kappa_{AB}\le1$. The
partners share a sign well above $T_c$, but need not near it: the
connected scalar susceptibility is not positive definite (the near-zero
modes with $|\lambda|<m$ contribute with the sign of $\lambda^2-m^2$,
through the spectral weight $(\lambda^2-m^2)/(\lambda^2+m^2)^2$), so
near $T_c$ $\chi_\delta^{\rm reg}$ can turn negative while
$\chi_\pi^{\rm reg}>0$.
The exact zero modes act in
the same direction, contributing $+1/(2Vm^2)$ to $\chi_\pi$ and
$-1/(2Vm^2)$ to $\chi_\delta$. The partners then straddle zero, and
$\kappa_{AB}=1$, since the splitting is then as large as the
susceptibilities themselves. The zero of the
splitting $\chi_A^{\rm reg}-\chi_B^{\rm reg}$ is unaffected. For the vector--axial-vector pair a further point
enters: since $\tilde c_0^{V,A}=0$, $s(T)-s(T_r)$ can be small for a
plasma reference, making $\kappa_{VA}$ a ratio of two small numbers.
The practical choice is therefore $T_r$ well above $T_c$, so that the
partners are degenerate, yet not so high that the denominator ceases to
exceed the numerator substantially. The window above $T_c$ leaves an
ample range. In practice $T_r$ is the highest admissible temperature of
the set, so that $\kappa_{AB}$ is evaluated for $T<T_r$.
The numerator of eq.~(\ref{eq:kappa_ND}) is $d(T)-d(T_r)$, and since
$d(T_r)$ is consistent with zero for every admissible reference, the
numerators obtained with two admissible references $T_r$ and $T_r'$
differ only by $d(T_r)-d(T_r')$, the difference of two quantities each
consistent with zero: within errors the numerator is $d(T)$ for every
admissible choice. Three consequences follow.

First, the zero of $\kappa_{AB}$ sits at the temperature $T_0$ where
$d(T_0)=d(T_r)$, i.e.\ where the splitting has fallen to its residual
value at the reference. For two admissible references the zeros differ by
\begin{equation}
\delta T_0 \simeq \frac{d(T_r)-d(T_r')}{d'(T_0)},
\label{eq:dT0}
\end{equation}
the numerator of which is the difference of two residual splittings, each
consistent with zero within errors. Thus $\delta T_0$ is of the order of
that residual, vanishes in the chiral limit, and lies within the precision
that defines effective restoration. The zero is common to all admissible
$T_r$.

Second, for all $T\ge T_0$, where $d(T)$ is itself consistent with zero,
$\kappa_{AB}(T;T_r)$ is consistent with zero for every admissible $T_r$.
The values obtained with different references are therefore compatible
throughout the effectively restored region.

Third, below $T_0$, where $d(T)\neq 0$, the sign of the partner splitting
$\chi_A^{\rm reg}-\chi_B^{\rm reg}$ is fixed by $d(T)$ and is common to
all admissible $T_r$, but the magnitude of $\kappa_{AB}$ is not:
the denominator $s(T)-s(T_r)$ is a normalization of definite sign whose
value depends on $T_r$ through $s(T_r)$. At fixed lattice spacing the regularized scalar and pseudoscalar
susceptibilities are monotonically decreasing, positive functions of
$T$, falling toward zero as $T\to T_r$ from below.
The vector- and tensor-type ones are negative and rise toward zero in
the same data of Ref.~\cite{Chiu:2026sxy} (Sec.~\ref{subsec:subtraction}).
In every channel the denominator $s(T)-s(T_r)$ has a definite,
$T_r$-dependent magnitude, and two references give
ratios that differ below $T_0$ by the factor
$|s(T)-s(T_r')|/|s(T)-s(T_r)|$. Equivalently, the relative difference
between the two references is $|s(T_r')-s(T_r)|/|s(T)-s(T_r)|$, the change
of the partner sum between the references relative to the regularized
sum. It is largest in the channels where the partner sum still varies
between the two references while the regularized sum is small: for the
vector--axial-vector pair, whose $\tilde c_0^{V,A}=0$ makes the regularized
sum the smallest, and to a lesser degree for the tensor pair, whose
partner sum rises with $T$ well above $T_c$. For the scalar--pseudoscalar
pair the sum falls steeply from $T$ to $T_r$ and is nearly flat between
admissible references, so the spread is the smallest. The magnitude of
$\kappa_{AB}$ below $T_0$ is thus meaningful only at a stated $T_r$, at
finite lattice spacing and in the continuum alike, and comparisons
between channels or ensembles are made at a common $T_r$. The sign of the splitting, the
zero of $\kappa_{AB}$, and its vanishing above $T_0$ do not depend on $T_r$.

These statements hold at each lattice spacing.
For the continuum limit $T_r$ must be held fixed in physical units, since
at fixed $aT_r$ the reference moves to infinite temperature.

The $T_r$-independence can then be tested by comparing $\kappa_{AB}$ for
two admissible references. Since the reference enters only as a constant,
this needs no new measurements at the temperatures of interest, only the
bare susceptibilities at the second reference. The present paper supplies
the analytic argument. The numerical comparison, with three admissible
references at three lattice spacings in $N_f=2+1+1$ lattice QCD, is given
in Sec.~IV\,A of Ref.~\cite{Chiu:2026sxy}. It confirms the common sign of
the partner splitting at every temperature and lattice spacing and
exhibits the $T_r$ dependence of the magnitude through the factor above,
in the order vector, tensor, scalar--pseudoscalar. The zero $T_0$ lies
above the simulated range there, so its reference independence remains to
be tested. The second admissibility condition is essential to such a comparison: a
reference with too few time slices, such as $N_t^{r}=2$, sets the
normalization by cutoff effects rather than physics.

The subtraction at $T_r$ is also what makes the numerator finite, not
merely a normalization. At the bare level the numerator is
$\chi_A^{\rm reg}-\chi_B^{\rm reg}
=[\chi_A^{\rm bare}(T)-\chi_B^{\rm bare}(T)]
-[\chi_A^{\rm bare}(T_r)-\chi_B^{\rm bare}(T_r)]$,
and the bracket at $T_r$ does \emph{not} vanish: since $c_m^A\neq c_m^B$,
\begin{equation}
\chi_A^{\rm bare}(T_r)-\chi_B^{\rm bare}(T_r)
=(c_m^A-c_m^B)\,m^2\ln\tfrac{1}{am} + Z^2\,d(T_r),
\end{equation}
even when the renormalized partners are degenerate there. Its
divergent part is exactly the temperature-independent mass logarithm
carried by the partner difference at $T$, so subtracting the $T_r$ bracket
removes that divergence from the numerator and leaves the finite
$Z^2[d(T)-d(T_r)]$, the numerator of eq.~(\ref{eq:kappa_ND}) before
the cancellation of $Z^2$. The subtraction at a thermal
reference thus does double duty: it removes the additive divergence and it
sets the baseline against which restoration is measured.

The choice $T_r = 0$ illustrates the point in reverse. In the vacuum the
partners are split, $\chi_A^R(0)\neq\chi_B^R(0)$, so even where the
partners are effectively degenerate at $T>T_c$ the ratio does not vanish
but approaches the temperature-dependent residual
\begin{equation}
\kappa_{AB}(T;0) \;\longrightarrow\;
\frac{-\,[\chi_A^R(0)-\chi_B^R(0)]}
{\chi_A^R(T)+\chi_B^R(T)-\chi_A^R(0)-\chi_B^R(0)} .
\label{eq:kappa_Tr0}
\end{equation}
Restoring the vanishing above $T_c$ would require adding this offset back,
which needs the renormalized vacuum splitting $\chi_A^R(0)-\chi_B^R(0)$ as
a separate input. At nonzero quark mass this splitting is not directly
measurable from the bare data: the bare partner difference at a single
temperature retains the divergence $(c_m^A-c_m^B)\,m^2\ln(1/(am))$ of
Sec.~\ref{subsec:additive}, and isolating the splitting requires a scheme
choice for its removal. The choice $T_r \gg T_c$ avoids all of this: the
offset vanishes automatically with the degeneracy of the partners at the
reference temperature.

Now we examine a conventional way to investigate symmetry restoration by
measuring the difference of renormalized susceptibilities between symmetry
partners,
\begin{align}
\Delta_{AB}
=&\,Z_{A}^{-2}\,\chi_A^{\rm reg}(T;0) - Z_{B}^{-2}\,\chi_B^{\rm reg}(T;0)
\notag \\
=&\,Z_A^{-2}\,[\chi_A^{\rm bare}(T)-\chi_A^{\rm bare}(0)] \notag \\
 &-\,Z_B^{-2}\,[\chi_B^{\rm bare}(T)-\chi_B^{\rm bare}(0)] \notag \\
=&\,\chi_A^R(T)-\chi_A^R(0)-\chi_B^R(T)+\chi_B^R(0) ,
\end{align}
where each channel carries its own renormalization constant, so that $Z_A$ and
$Z_B$ must both be determined nonperturbatively. They are independent in
general, and coincide only for a regularization with exact chiral symmetry
(Sec.~\ref{subsec:ZAZB_lat}). Accordingly $\Delta_{AB}$ inherits the scheme
and scale dependence of both.
Obviously, $\Delta_{AB}$ approaches the nonzero constant
$\chi_B^R(0)-\chi_A^R(0)$ for $T>T_c$ once the symmetry has been
effectively restored, $\chi_A^R(T)\approx\chi_B^R(T)$.
Even worse, this nonzero constant is channel dependent, which makes it
difficult to compare the symmetry restoration patterns among different
channels, let alone to determine which symmetry is restored at a lower or
higher temperature, leading to tensions between the $\Delta_{AB}$ of
different channels, even in the continuum limit. To remove these
ambiguities, a common normalization of all channels is required, and the
RG-invariant symmetry ratio $\kappa_{AB}$ constructed in
Ref.~\cite{Chiu:2026sxy} fulfills the requirement.

\subsection{Characteristic temperatures from a single channel}
\label{subsec:T_char}

The two-part decomposition yields practical corollaries for the
determination of characteristic temperatures from a single channel.

\subsubsection*{Scalar channel: the pseudocritical temperature}
\label{subsec:T_char_S}

Consider the RG-invariant combination
$m^2\,[\chi_\sigma(T)-\chi_\sigma(T_r)]$, where $\chi_\sigma$ denotes
the flavor-singlet scalar (chiral) susceptibility and the factor $m^2$
cancels the multiplicative renormalization, since $m\,\bar{q}q$ is
RG-invariant. Its peak position in $T$ is independent of the choice of
$T_r$: for two reference temperatures $T_r$ and $T_r'$, the two curves
differ by the $T$-independent constant
$m^2\,[\chi_\sigma(T_r)-\chi_\sigma(T_r')]$, a uniform vertical
displacement that leaves the shape of the curve, and hence the location
of its maximum, unchanged. Any thermal ensemble may therefore serve as
the reference, and no zero-temperature subtraction is required. The
thermal reference is also advantageous in practice. Zero-temperature
ensembles are computationally far more expensive than finite-temperature
ensembles of the same spatial volume and lattice spacing. At
$T_r \gg T_c$ the finite-volume effects of $\chi_\sigma(T_r)$ are
exponentially suppressed by screening masses of order $2\pi T_r$.
Moreover, since the topological susceptibility is strongly suppressed at
$T_r \gg T_c$, the reference measurement is insensitive to the sampling
of topological sectors, whereas a zero-temperature reference on fine
lattices is exposed to the well-known freezing of the topological
charge. The discretization effects at
the reference temperature remain controlled provided the number of time
slices $1/(a T_r)$ stays moderate, which bounds $T_r$ from above at a
given lattice spacing. The
invariance holds for the subtracted susceptibility itself, and survives
normalization by any temperature-independent physical scale. A convenient
dimensionless choice is
$m^2\,[\chi_\sigma(T)-\chi_\sigma(T_r)]/m_\pi^4$ with $m_\pi$
the zero-temperature pion mass, available from the scale setting. By the
GMOR relation $m^2/m_\pi^4$ approaches a finite constant in the chiral
limit, so this normalization does not suppress the observable as
$m \to 0$. If instead the subtracted susceptibility is
divided by a temperature-dependent normalization such as $T^4$, the
constant $-\,m^2\chi_\sigma(T_r)$ becomes the $T$-dependent term
$-\,m^2\chi_\sigma(T_r)/T^4$, and the peak position then depends on
$T_r$. Such normalized definitions remain consistent, as in the
$m^2\,[\chi_\sigma(T)-\chi_\sigma(0)]/T^4$ of
Ref.~\cite{Aoki:2006br}, but the reference temperature is then part of
the definition and must be specified together with it. At a crossover the pseudocritical temperature is in any case
observable dependent. The peak of
$m^2\,[\chi_\sigma(T)-\chi_\sigma(T_r)]$ defines a reference-free and
normalization-free choice.

\subsubsection*{Pseudoscalar channel: the anomalous sector}
\label{subsec:T_char_P}

The same construction applies to the anomalous sector. Since
$Z_P = Z_S$, the identical dressing renormalizes the disconnected
pseudoscalar susceptibility, and by the index relation
eq.~(\ref{eq:chi_t}) the observable is the subtracted topological
susceptibility,
$m^2\,[\chi_{5,\rm disc}(T)-\chi_{5,\rm disc}(T_r)]
=\chi_t(T)-\chi_t(T_r)$. This quantity is monotonic in $T$, so the
characteristic temperature of the anomalous sector is defined by its
inflection point rather than by a maximum. The derivative annihilates
the $T_r$ constant identically, so the inflection point is independent
of $T_r$. This temperature marks the steepest suppression of
the anomalous fluctuations, a feature of the curve $\chi_t(T)$ itself.
It is not identical to an effective restoration temperature defined by
the vanishing of a $U(1)_A$-breaking quantity: such definitions are
threshold statements on the decaying tail of $\chi_t(T)$, they depend on
the resolution with which the tail can be distinguished from zero, and
they generally lie above the inflection point. The quantity
$\chi_{5,\rm disc} = \chi_t/m^2$ here pertains to the singlet (anomalous)
sector. The degeneracy of the nonsinglet pair $(\pi,\delta)$ is a
distinct question, governed by the quark-connected correlators and taken
up next.

\section{Equality of renormalization constants}
\label{sec:chains}

\subsection{General principle}
\label{subsec:general}

If two operators $O_A$ and $O_B$ are related by a symmetry
transformation that is respected by the regularized action,
their renormalization constants are equal: $Z_A = Z_B$.
This is a purely UV property.
It depends only on the
short-distance structure of the regularized theory, not on the
state of the system. (The same transformations, applied to correlation
functions in a symmetric state, give the equalities between partner
correlators collected in Sec.~\ref{sec:sym_rel}.) In particular:

\begin{itemize}
\item \emph{Spontaneous symmetry breaking does not affect $Z_A = Z_B$.}
SSB is an infrared property of the vacuum state.
The UV counterterms are determined by the action alone.

\item \emph{The $U(1)_A$ anomaly does not affect the equalities used in
$\kappa_{AB}$.}
Anomaly contributions arise from closed quark loops and are flavor
traced. They therefore vanish for nonsinglet operators, and the
nonsinglet partner equalities that enter $\kappa_{AB}$ are unaffected.
The singlet operators tied to nonsinglets by the non-anomalous
$SU(N_f)_L\times SU(N_f)_R$ multiplets of Sec.~\ref{subsec:ZAZB_lat}
are equally protected.

\item \emph{Non-degenerate quark masses do not affect $Z_A = Z_B$}
in any mass-independent renormalization scheme
(e.g., RI-MOM~\cite{Martinelli:1995ty}).
\end{itemize}

The crucial requirement is that the \emph{regularization} preserves
the symmetry. For Ginsparg-Wilson fermions on the lattice, the
relation $\{D, \gamma_5\} = 2ra\, D \gamma_5 D$~\cite{Ginsparg:1981bj}
ensures that chiral symmetry is an exact symmetry of the lattice
action~\cite{Luscher:1998pqa}, guaranteeing
$Z_A = Z_B$ for all chiral partners.

Note that the singlet vector and axial-vector currents,
$V_\mu^0 = \bar q t^0\gamma_\mu q$ and
$A_\mu^0 = \bar q t^0\gamma_\mu\gamma_5 q$, are each invariant under both
$SU(N_f)_L\times SU(N_f)_R$ and $U(1)_A$.
Therefore they are not symmetry partners of any meson operators
and do not enter $\kappa_{AB}$ at all,
thus we omit $Z_A^s$ and $Z_V^s$ in the following.

\subsection{Nonperturbative proof for domain-wall and overlap fermions}
\label{subsec:ZAZB_lat}

Now we prove that $Z_A=Z_B$ for domain-wall fermions as well as for
Ginsparg--Wilson fermions.

\subsubsection{Domain-wall fermions}

For domain-wall fermions the quark fields $q$ and
$\bar q$ are defined through the boundary modes of the five-dimensional theory,
as given in Refs.~\cite{Furman:1994ky,Chiu:2003ir,Chen:2012jya}, and these
fields obey the ordinary continuum chiral projection independent of the gauge
field. Consequently the bilinear $\bar q\, t^a \Gamma q$ transforms under
chiral rotations exactly as its continuum counterpart, and its correlators are
expressed through the valence propagator $(D_c+m)^{-1}$, which carries the
continuum chiral properties.

Consider the scalar and pseudoscalar densities
$O_S = \bar q\, t^a q$ and $O_P = \bar q\, i\gamma_5 t^a q$, $a=0,\cdots,N_f^2-1$.
The factor $i$ renders $O_P$ Hermitian. The same factor accompanies the axial-vector and
axial-tensor densities $O_A$ and $O_X$ below, while $O_S$ and $O_T$ are
Hermitian without it. The vector density $O_V$ is kept in its conventional
form without a phase. This phase is immaterial to the additive divergences of
Secs.~\ref{subsec:additive} and~\ref{sec:Z}, where each correlator
$C_\Gamma$ contains two factors of $\Gamma$ and the phase squares to a channel
sign already absorbed in $\eta_\Gamma$. It matters only for the reality
properties of the operators and the form of the rotation formulas below. Under the
infinitesimal $U(1)_A$ rotation
\begin{equation}
q \to (1+ i\theta\gamma_5/2)\, q, \qquad
\bar q \to \bar q\,(1+i\theta\gamma_5/2),
\label{eq:U1A_rotation}
\end{equation}
$O_S$ and $O_P$ map into one another,
\begin{equation}
O_S \longrightarrow O_S + \theta\, O_P, \qquad
O_P \longrightarrow O_P - \theta\, O_S.
\label{eq:OSOP_rotation}
\end{equation}
In a mass-independent scheme the mass term enters only multiplicatively and
introduces no counterterm that could split the pair (Sec.~\ref{subsec:general}),
so the ultraviolet renormalization respects this rotation whether or not the
quark mass vanishes and whether or not the symmetry is realized in the vacuum.
The renormalized operators must therefore transform in the same way as the
bare ones, which is possible only if $O_S$ and $O_P$ carry a single
renormalization constant,
\begin{equation}
Z_S^s = Z_P^s, \qquad Z_S^{ns} = Z_P^{ns} \qquad \text{(exact, at finite $a$)}.
\label{eq:ZS_ZP_exact}
\end{equation}
The argument is nonperturbative, relying on no expansion in the coupling, and
holds for both the flavor singlet and the nonsinglet.

The equality $Z_S = Z_P$ deserves a comment. The pair $(O_S,O_P)$ is
related by the $U(1)_A$ rotation, the very symmetry whose breaking the
ratio $\kappa_{PS}$ is designed to measure. Yet $Z_S = Z_P$ does not
require that symmetry to be unbroken. As stressed in
Sec.~\ref{subsec:general}, the equality follows from the symmetry of the
regularized \emph{action}, not of the vacuum: spontaneous breaking
shifts $\vev{\bar qq}$ and the spectrum, the anomaly shifts the
topological sector and $\chi_t$, and neither introduces an ultraviolet
counterterm that could split $Z_S$ from $Z_P$. The equality therefore
persists into the $U(1)_A$-broken phase, which is what allows
$\kappa_{PS}$ to isolate the breaking cleanly.
Similar comments apply to the renormalization constants of the other
symmetry partners of $U(1)_A$ or $SU(N_f)_L \times SU(N_f)_R$.

The tensor and axial-tensor densities
$(O_T)_k = \bar q\,\gamma_4\gamma_k\, t^a q$ and
$(O_X)_k = \bar q\,i\gamma_5\gamma_4\gamma_k\, t^a q$ are exchanged by the same
$U(1)_A$ rotation. Because $\gamma_4\gamma_k$ contains an even number of
$\gamma$-matrices it commutes with $\gamma_5$, so
$\gamma_5\gamma_4\gamma_k = \gamma_4\gamma_k\gamma_5$ and
\begin{align}
(O_T)_k &\longrightarrow (O_T)_k + \theta\,(O_X)_k, \notag\\
(O_X)_k &\longrightarrow (O_X)_k - \theta\,(O_T)_k,
\end{align}
so that $O_T$ and $O_X$ share a single renormalization constant,
for singlet and nonsinglet alike.
\begin{equation}
Z_T^s = Z_X^s, \qquad Z_T^{ns} = Z_X^{ns} \qquad \text{(exact, at finite $a$)}.
\label{eq:ZT_ZX_exact}
\end{equation}

Under the nonsinglet axial rotations the same commuting property connects
the singlet and nonsinglet sectors, and it is here that the general-$N_f$
structure must be made explicit. Consider the infinitesimal nonsinglet
axial rotation of $SU(N_f)_L\times SU(N_f)_R$
\begin{equation}
q \to (1+i\gamma_5\theta^c t^c/2)\, q, \qquad
\bar q \to \bar q\,(1+i\gamma_5\theta^c t^c/2),
\label{eq:SUNfA_rotation}
\end{equation}
with $c=1,\ldots,N_f^2-1$, acting on $O_\Gamma^b = \bar q\,\Gamma\,t^b q$.
For any Dirac structure that commutes with $\gamma_5$, the two terms of the
variation add and the flavor anticommutator survives,
\begin{equation}
\delta O_\Gamma^b = \frac{i\theta^c}{2}\,\bar q\,(\gamma_5\Gamma)\,\{t^c,t^b\}\,q .
\label{eq:anticomm_var}
\end{equation}
For $SU(N_f)$ generators in the normalization $\tr_F(t^at^b)=\delta^{ab}/2$,
with $t^0=\Id/\sqrt{2N_f}$,
\begin{equation}
\{t^c,t^b\} = \sqrt{\tfrac{2}{N_f}}\;\delta^{cb}\,t^0 + d^{cbd}\,t^d,
\qquad
\{t^c,t^0\} = \sqrt{\tfrac{2}{N_f}}\;t^c ,
\label{eq:anticomm}
\end{equation}
where $d^{cbd}$ are the totally symmetric structure constants of
$SU(N_f)$, which vanish identically for $N_f=2$. For the tensor pair,
$\gamma_5\,\gamma_4\gamma_k = -i\,(i\gamma_5\gamma_4\gamma_k)$ and
$\gamma_5\,(i\gamma_5\gamma_4\gamma_k) = i\,\gamma_4\gamma_k$, so
eqs.~(\ref{eq:anticomm_var}) and (\ref{eq:anticomm}) give
\begin{align}
&\delta (O_T^{ns})_k^b = \tfrac12\theta^c\Big[\sqrt{\tfrac{2}{N_f}}\,\delta^{cb}\,(O_X^{s})_k
 + d^{cbd}\,(O_X^{ns})_k^d\Big], \notag\\
&\delta (O_X^{ns})_k^b = -\tfrac12\theta^c\Big[\sqrt{\tfrac{2}{N_f}}\,\delta^{cb}\,(O_T^{s})_k
 + d^{cbd}\,(O_T^{ns})_k^d\Big], \notag\\
&\delta (O_X^{s})_k = -\tfrac12\sqrt{\tfrac{2}{N_f}}\;\theta^c\,(O_T^{ns})_k^c, \notag\\
&\delta (O_T^{s})_k = +\tfrac12\sqrt{\tfrac{2}{N_f}}\;\theta^c\,(O_X^{ns})_k^c .
\label{eq:TX_cross}
\end{align}
In a mass-independent scheme with chirally symmetric regularization the
renormalization commutes with the non-anomalous
$SU(N_f)_L\times SU(N_f)_R$, so two operators connected by a nonvanishing
variation carry the same renormalization constant. The $\delta^{cb}$
terms of eq.~(\ref{eq:TX_cross}) connect each nonsinglet operator with
the singlet of the partner channel, giving $Z_T^{ns}=Z_X^{s}$ and
$Z_X^{ns}=Z_T^{s}$ for every $N_f$. Combined with the $U(1)_A$
relations~(\ref{eq:ZT_ZX_exact}), $Z_T^{ns}=Z_X^{ns}$ and
$Z_T^{s}=Z_X^{s}$, which likewise hold for every $N_f$, this closes the
four constants into a single chain,
\begin{equation}
\boxed{Z_T^{ns} = Z_X^{ns} = Z_T^s = Z_X^s \equiv Z_{TX},}
\label{eq:TX_chain}
\end{equation}
for any $N_f\ge 2$ and, in a mass-independent scheme, for any quark
masses. For $N_f\ge 3$ the $d^{cbd}$ terms of eq.~(\ref{eq:TX_cross})
give the nonsinglet equality $Z_T^{ns}=Z_X^{ns}$ from the chiral group
alone as well, consistently with eq.~(\ref{eq:ZT_ZX_exact}).

The scalar and pseudoscalar densities, for which $\Gamma=\Id$ likewise
commutes with $\gamma_5$, follow the same pattern. Equation~(\ref{eq:anticomm_var}) with
$\gamma_5\Id = -i\,(i\gamma_5)$ gives the analogue of eq.~(\ref{eq:TX_cross})
with $(O_T,O_X)\to(O_S,O_P)$: the cross multiplets $(O_S^{ns},O_P^{s})$
and $(O_S^{s},O_P^{ns})$, for $N_f=2$ the familiar $(\delta,\eta)$ and
$(\sigma,\pi)$ chiral partners, give $Z_S^{ns}=Z_P^{s}$ and
$Z_S^{s}=Z_P^{ns}$, and with the $U(1)_A$
relations~(\ref{eq:ZS_ZP_exact}),
\begin{equation}
\boxed{Z_S^{ns} = Z_P^{ns} = Z_S^s = Z_P^s \equiv Z_{SP},}
\label{eq:SP_chain}
\end{equation}
for any $N_f\ge 2$ and any quark masses in a mass-independent scheme.

Notably, the singlet--nonsinglet equalities in both sectors, $Z^{ns}=Z^{s}$
across the cross multiplets, follow from the non-anomalous
$SU(N_f)_L\times SU(N_f)_R$ symmetry alone, untouched by the $U(1)_A$
anomaly, while the equalities within a fixed flavor representation
follow from the $U(1)_A$ rotation, whose anomaly affects neither
(Sec.~\ref{subsec:ZAZB_lat}). The two symmetries together give the
chains for every $N_f\ge 2$. The singlet scalar $O_S^0$ does mix additively with the
identity operator (Sec.~\ref{subsec:subtraction}), but this does not affect
its multiplicative constant, which is the statement made here.

Consider nonsinglet vector and axial-vector currents
$(O_V^{ns})_k^a= \bar q\,\gamma_k t^a q$ and
$(O_A^{ns})_k^a= \bar q\,i\gamma_5\gamma_k t^a q$.
Under the transformation~(\ref{eq:SUNfA_rotation}) the two terms of the
variation now subtract, since $\gamma_k\gamma_5 = -\gamma_5\gamma_k$, and
the flavor commutator $[t^c,t^b]=i f^{cbd}t^d$ survives instead of the
anticommutator, so that the nonsinglet vector and axial-vector operators
rotate into one another,
\begin{align}
&\delta (O_V^{ns})_k^a= +\tfrac{i}{2}\theta^b f^{bac}\,(O_A^{ns})_k^c, \notag \\
&\delta (O_A^{ns})_k^a= -\tfrac{i}{2}\theta^b f^{bac}\,(O_V^{ns})_k^c,
\label{eq:VA_rotation}
\end{align}
with $f^{bac}$ the structure constants of $SU(N_f)$
($f^{bac}=\epsilon^{bac}$ for $N_f=2$). Thus $O_V^{ns}$ and $O_A^{ns}$
belong to a single multiplet of the chiral group and carry a common
renormalization constant,
\begin{equation}
\boxed{\;Z_V^{ns} = Z_A^{ns} \;\equiv\; Z_{VA},\;}
\label{eq:VA_chain}
\end{equation}
for any $N_f\ge 2$ and any quark masses in a mass-independent scheme.
The singlet vector and axial-vector currents $V^0_k$ and $A^0_k$ are
each invariant under $U(1)_A$ and under
$SU(N_f)_L\times SU(N_f)_R$. They therefore cannot serve as probes of
the effective restoration of either symmetry, and their renormalization
constants are irrelevant to the present study.

The common value of $Z_V^{ns} = Z_A^{ns}$ is
discussed below (\emph{The value of $Z_{VA}$}). Only the equality is used for $\kappa_{VA}$.

\subsubsection{Overlap fermions}

The same result holds for the overlap formulation on the four-dimensional
lattice, where the quark action is $\bar q\,(D_c + m)(1+raD_c)^{-1} q$. Here
$r$ is fixed by the negative mass in the Wilson kernel of the overlap operator,
$D_c = \gamma_\mu D_c^\mu$ is chirally symmetric, $D_c \gamma_5 + \gamma_5 D_c=0$,
and $D=D_c(1+raD_c)^{-1}$ satisfies the Ginsparg--Wilson relation
$D\gamma_5+\gamma_5 D=2ra\, D\gamma_5 D$. The meson operator
$O_\Gamma = \bar q\,\Gamma\, t^a (1+raD_c)^{-1} q$ is then nonlocal, unlike its
continuum and domain-wall counterparts. The factor
$(1+raD_c)^{-1} = 1-raD$ inserted between $\bar q$ and $q$ ensures that, after
the Wick contraction in the correlator $\vev{O_\Gamma(x)\,O_\Gamma^\dagger(y)}$,
the result is expressed through the valence propagator $(D_c+m)^{-1}$ rather
than the inverse of the sea-quark operator $D(m)\equiv(D_c+m)(1+raD_c)^{-1}$.

The rotation argument goes through cleanly if the chiral rotation is taken in
the asymmetric L\"uscher form, in which only the $q$ field carries
the modification. For the singlet $U(1)_A$ rotation,
\begin{equation}
\delta q = \tfrac{i}{2}\theta\,\gamma_5\,(1-raD_c)(1+raD_c)^{-1} q, \qquad
\delta\bar q = \tfrac{i}{2}\theta\,\bar q\,\gamma_5,
\label{eq:GW_rotation}
\end{equation}
where $(1-raD_c)(1+raD_c)^{-1} = (1-2raD)$.
Since $D_c=\gamma_\mu D_c^\mu$ anticommutes with $\gamma_5$,
\begin{equation}
(1+raD_c)^{-1}\gamma_5 = \gamma_5\,(1-raD_c)^{-1} .
\label{eq:K_g5}
\end{equation}
Writing $K\equiv(1+raD_c)^{-1}$ for the nonlocal factor in
$O_\Gamma = \bar q\,\Gamma\,t^a K q$, the scalar density varies as
\begin{equation}
\delta O_S = \frac{i\theta}{2}\,
\bar q\,t^a\big[\gamma_5 K + K\,\gamma_5(1-raD_c)K\big] q .
\end{equation}
By eq.~(\ref{eq:K_g5}), $K\gamma_5(1-raD_c)K = \gamma_5 K$, so the two terms are
equal and
\begin{equation}
\delta O_S = i\theta\,\bar q\,t^a\,\gamma_5 K\, q = \theta\, O_P ,
\end{equation}
with no residual factor. The pseudoscalar density varies into the scalar in the
same way, $\delta O_P = -\theta\, O_S$, so the rotation closes on the pair
$(O_S,O_P)$ and they share a single renormalization constant, $Z_S=Z_P$ and (\ref{eq:ZS_ZP_exact})
exactly as in the continuum and domain-wall cases.
The tensor pair works identically. Since $\gamma_4\gamma_k$ commutes with
$\gamma_5$, the same rotation~(\ref{eq:GW_rotation}) gives
$\delta O_T = \theta\, O_X$ and $\delta O_X = -\theta\, O_T$, hence
$(O_T,O_X)$ share a single renormalization constant, $Z_T=Z_X$ and (\ref{eq:ZT_ZX_exact})
exactly as in the continuum and domain-wall cases.

The nonsinglet axial rotation of $SU(N_f)_L\times SU(N_f)_R$ is obtained
by inserting a flavor generator,
\begin{equation}
\delta q = \tfrac{i}{2}\theta^c\,\gamma_5\, t^c\,(1-raD_c)K\, q, \qquad
\delta\bar q = \tfrac{i}{2}\theta^c\,\bar q\,\gamma_5\, t^c .
\label{eq:GW_rotation_ns}
\end{equation}
Since $K$ and $D_c$ are flavor independent, $t^c$ commutes with them, so the
Dirac algebra is untouched and the cancellation
$K\gamma_5(1-raD_c)K=\gamma_5 K$ applies exactly as before. Acting on
$O_\Gamma^b = \bar q\,\Gamma\, t^b K q$,
\begin{equation}
\delta O_\Gamma^b = \frac{i\theta^c}{2}\,
\bar q\,\big[\gamma_5\Gamma\, t^ct^b + \Gamma\gamma_5\, t^bt^c\big] K q ,
\label{eq:GW_ns_variation}
\end{equation}
again with no residual factor. This is eq.~(\ref{eq:anticomm_var}) of the
domain-wall case with the nonlocal factor $K$ inserted, and the flavor
algebra is identical: the channel dependence enters only through whether
$\Gamma$ commutes with $\gamma_5$.

For $\Gamma = \Id,\ i\gamma_5,\ \gamma_4\gamma_k,\
i\gamma_5\gamma_4\gamma_k$, which commute with $\gamma_5$, the two terms of
eq.~(\ref{eq:GW_ns_variation}) add and the flavor anticommutator
$\{t^c,t^b\}$ of eq.~(\ref{eq:anticomm}) survives,
\begin{equation}
\delta O_\Gamma^b = \frac{i\theta^c}{2}\,
\bar q\,(\gamma_5\Gamma)\,\{t^c,t^b\}\,K q .
\end{equation}
The rotation exchanges the Dirac structure, $S\leftrightarrow P$ and
$T\leftrightarrow X$, and connects the singlet and nonsinglet flavor
representations through the $\delta^{cb}t^0$ term of eq.~(\ref{eq:anticomm})
and, for $N_f\ge 3$, the nonsinglets among themselves through $d^{cbd}$,
exactly as in eq.~(\ref{eq:TX_cross}). Since the nonlocal factor $K$ is
flavor blind, the operators $\bar q\,\Gamma\,t^a K q$, $a=0,\ldots,N_f^2-1$,
transform under $SU(N_f)_L\times SU(N_f)_R$ exactly as their local
counterparts, and the argument of the domain-wall case applies verbatim:
the closure on the cross multiplets $O_S^{ns}\leftrightarrow O_P^{s}$,
$O_S^{s}\leftrightarrow O_P^{ns}$, $O_T^{ns}\leftrightarrow O_X^{s}$,
$O_T^{s}\leftrightarrow O_X^{ns}$, combined with the $U(1)_A$
rotation~(\ref{eq:GW_rotation}), reproduces eqs.~(\ref{eq:SP_chain})
and~(\ref{eq:TX_chain}) for the nonlocal overlap operators for any
$N_f\ge 2$.

For $\Gamma=\gamma_k$ and $i\gamma_5\gamma_k$, which anticommute with
$\gamma_5$, the two terms of eq.~(\ref{eq:GW_ns_variation}) subtract and the
flavor commutator survives instead,
\begin{equation}
\delta O_\Gamma^b = \frac{i\theta^c}{2}\,
\bar q\,(\gamma_5\Gamma)\,[t^c,t^b]\,K q .
\end{equation}
Since $[t^c,t^b]=if^{cbd}t^d$ is nonsinglet, the rotation closes within
the nonsinglet sector, $O_V^{ns}\leftrightarrow O_A^{ns}$, giving
$Z_V^{ns}=Z_A^{ns}$ and eq.~(\ref{eq:VA_chain}) for any $N_f$. Because $[t^c,t^0]=0$ the
singlet vector and axial-vector do not rotate at all, and the same subtraction
makes them invariant under eq.~(\ref{eq:GW_rotation}) as well. Being invariant
under both rotations, they are not symmetry partners and do not enter
$\kappa_{AB}$.

Thus the full set of symmetry partner equalities holds for the nonlocal overlap operators,
by the same mechanism as for the local operators of the continuum and domain-wall
formulations.

\subsubsection{The value of $Z_{VA}$}
The value of $Z_V$ depends on which current is used. The conserved
vector current of the lattice action satisfies its Ward identity
exactly and has $Z_V = 1$ at finite $a$. It is point-split for
DWF~\cite{Furman:1994ky,Chen:2012jya}, the splitting being required by
the gauge links of the one-link hopping, and nonlocal for the
four-dimensional Ginsparg--Wilson fermion~\cite{Ginsparg:1981bj}, in
keeping with the extended kernel of $D$. The local vector current used
here is not the conserved current: its renormalization constant is
finite, since $\gamma_V = 0$, but differs from unity at finite coupling,
$Z_V = 1 + O(g^2(1/a))$, and approaches unity only in the continuum
limit. The residual chiral breaking at finite $N_s$ contributes further
$O(m_{\rm res})$ corrections. Together with eq.~(\ref{eq:VA_chain}),
\begin{equation}
Z_V^{ns} = Z_A^{ns} = Z_{VA} \;\to\; 1
\qquad (a\to 0,\; N_s\to\infty).
\label{eq:VA_unity}
\end{equation}
More generally, what $\kappa_{AB}$ requires, however, is only the equality $Z_A = Z_B$
for symmetry partners,
which holds exactly at finite $a$ in the $N_s\to\infty$ limit by the
chiral rotations of Sec.~\ref{subsec:ZAZB_lat}, for every symmetry
partner pair in eqs.~(\ref{eq:SP_chain}), (\ref{eq:TX_chain}), and
(\ref{eq:VA_chain}).
At finite $N_s$ the residual chiral breaking induces an $O(m_{\rm res})$ splitting between
the partners, which vanishes exponentially as $N_s\to\infty$.
The value itself never enters the ratio $\kappa_{AB}$.

\section{RG invariance of $\kappa_{AB}$}
\label{sec:kappa}

We now prove that the symmetry ratio $\kappa_{AB}$
(eq.~\eqref{eq:kappa_def}) is exactly RG-invariant and scheme-independent
within the class of mass-independent schemes with chirally symmetric
regularization, in which the partner equalities $Z_A = Z_B$ hold.

The proof has two steps. First, by the temperature subtraction
(eq.~\eqref{eq:chi_reg}),
the regularized susceptibilities $\chi_A^{\rm reg}$ and
$\chi_B^{\rm reg}$ are free of all additive UV divergences.
This holds for every channel, singlet and nonsinglet alike, including the
scalar singlet, whose bare operator is not multiplicatively renormalizable
but whose VEV-subtracted susceptibility is (Sec.~\ref{subsec:subtraction}).
Both the numerator $\chi_A^{\rm reg} - \chi_B^{\rm reg}$
and the denominator $\chi_A^{\rm reg} + \chi_B^{\rm reg}$
are therefore UV-finite in the additive sense.
Second, for symmetry partners with $Z_A = Z_B \equiv Z$,
\begin{align}
\kappa_{AB} &= \frac{Z^{2}\,\bigl|\,\chi^R_A(T) - \chi^R_A(T_r)
- \chi^R_B(T) + \chi^R_B(T_r)\,\bigr|}
{Z^{2}\,|\chi^R_A(T) - \chi^R_A(T_r)|
+ Z^{2}\,|\chi^R_B(T) - \chi^R_B(T_r)|} \notag \\
&= \frac{\bigl|\,\chi^R_A(T) - \chi^R_A(T_r)
- \chi^R_B(T) + \chi^R_B(T_r)\,\bigr|}
{|\chi^R_A(T) - \chi^R_A(T_r)|
+ |\chi^R_B(T) - \chi^R_B(T_r)|},
\label{eq:kappa_cancel}
\end{align}
and the common factor $Z^{2}>0$ cancels between numerator and denominator.
The cancellation requires only the equality $Z_A = Z_B$, which is guaranteed
by the exact chiral symmetry for every pair of symmetry partners. The
individual value of $Z$ is never needed and need not be computed. This is
what makes $\kappa_{AB}$ well-defined without any non-perturbative
renormalization, and it applies to all channels alike.
Since $Z^{2}$ cancels in the ratio, $\kappa_{AB}$ depends on
neither the renormalization scheme ($\msbar$, RI-MOM, etc.)
nor the renormalization scale $\mu$.
It is a well-defined physical observable in the continuum limit.
Table~\ref{tab:uv_cancel} summarizes the UV structure at each stage
of the construction.

\begin{table}[t]
\caption{\label{tab:uv_cancel}%
UV structure at each stage of the construction.
The additive power divergence cancels between symmetry partners in the
numerator (when $\alpha_A = \alpha_B$), while at $m\neq 0$ the
mass-dependent logarithm survives in the difference.
The denominator
retains the full $1/a^2$ divergence in the bare (non-subtracted)
construction. The temperature subtraction removes all additive terms
entirely.}
\begin{ruledtabular}
\begin{tabular}{@{}lcc@{}}
Quantity & Additive div. & Multiplicative \\
\colrule
$\chi_\Gamma^{\rm bare}(T)$
  & $\alpha_\Gamma/(2a^2)$,\ $c_m^\Gamma m^2\ln\!\frac{1}{am}$ & $Z_\Gamma^{2}$ \\
$\chi_A^{\rm bare} - \chi_B^{\rm bare}$
  & $\propto m^2\ln\!\frac{1}{am}$\,$^{\dagger}$ & $Z^{2}$ if $Z_A=Z_B$ \\
$\kappa_{AB}^{\rm bare}$ (no subtraction)
  & $\to 0$ as $a \to 0$ & --- \\
\colrule
$\chi_\Gamma^{\rm reg}(T;T_r)$
  & None & $Z_\Gamma^{2}$ \\
$\chi_A^{\rm reg} - \chi_B^{\rm reg}$
  & None & $Z^{2}$ if $Z_A=Z_B$ \\
$\kappa_{AB}$
  & None & None \\
\end{tabular}
\end{ruledtabular}
\begin{flushleft}
\footnotesize $^{\dagger}$ The power divergences cancel because
$\alpha_A = \alpha_B$, which follows from the same symmetry relating the
partners that gives $Z_A = Z_B$, and holds in a
symmetry-preserving regularization. At $m\neq 0$ the mass-dependent
logarithm survives in the difference, since the $m^2$ term breaks the
symmetry explicitly and $c_m^A \neq c_m^B$ in general.
It is removed by
the temperature subtraction.
\end{flushleft}
\end{table}

\section{Symmetry relations and the $U(1)_A$ probes}
\label{sec:sym_rel}

The chiral rotations of Sec.~\ref{sec:chains} act on the operators
$O_\Gamma^a$ and were used there to establish the equality of
renormalization constants of symmetry partners, an operator statement
that holds for any quark masses in a mass-independent scheme. The same
rotations relate the correlators of partner operators. Many of the
resulting relations for two and three massless flavors are long
known. Cohen and Ji~\cite{Cohen:1996sb} organized them comprehensively,
for mesons and baryons, by enumerating the chiral multiplets in which
the interpolating currents fall and reading off the equalities implied
by a common multiplet. We recall the meson relations here in the
notation of this paper, derived instead by applying the explicit chiral
rotations of Sec.~\ref{sec:chains} to the correlators. The two routes
give the same equalities, but the explicit transformation is what makes
the accompanying statement about renormalization constants available:
the rotation that relates $C_A$ to $C_B$ relates $O_A$ to $O_B$ and hence
$Z_A$ to $Z_B$. This is what underlies the use of $\kappa_{AB}$ across the
hierarchical flavor windows of a theory with nondegenerate masses,
discussed below. The corresponding relations between
susceptibilities follow by summation.

\subsection{Setup}

Let $C_\Gamma^{a}(x)\equiv\vev{O_\Gamma^a(x)\,O_\Gamma^{a\dagger}(0)}$ with
$O_\Gamma^a=\bar q\,\Gamma\,t^a q$ (no sum over $a$), and write
$C_\Gamma^{ns}$ for a nonsinglet component $a\neq 0$ and $C_\Gamma^{s}$
for the singlet $a=0$. For a flavor-off-diagonal $t^a$ the correlator
has only the quark-connected contraction. The singlet has in addition
the disconnected contraction, whose four-volume sum is
eq.~(\ref{eq:chi_disc}), with the relative sign fixed by
$\bar\Gamma=\gamma_4\Gamma^\dagger\gamma_4=\pm\Gamma$ as noted there. In the
normalization $\tr_F(t^at^b)=\delta^{ab}/2$ the connected contraction is
the same for all $a$, so that
$C_\Gamma^{s}(x)=C_\Gamma^{ns}(x)\pm C_{\Gamma,\rm disc}(x)$.

The essential idea is simple. If two operators $O_A$ and $O_B$ are
related by a symmetry transformation, then they belong to the same
multiplet of that symmetry, and effective restoration of that symmetry
gives $C_A(x)\simeq C_B(x)$ for all $x$, and hence, term by term
in the sums (\ref{eq:chi_bare_lat}) and (\ref{eq:chi_bare_lat_ne0}),
$\chi_A\simeq\chi_B$, $\chi_A^{\rm reg}\simeq\chi_B^{\rm reg}$, and
$\kappa_{AB}\simeq 0$. The relations below are the statements of this
kind that follow from the rotations of Sec.~\ref{sec:chains}. At nonzero
quark mass they hold up to explicit-breaking corrections, which is the
sense of ``effective restoration'' used in Sec.~\ref{subsec:T_sub}. The
equalities are written with an equal sign for brevity.

\subsection{$SU(N_f)_L\times SU(N_f)_R$ relations}

The nonsinglet axial rotations, eq.~(\ref{eq:SUNfA_rotation}), give the
variations~(\ref{eq:TX_cross}) for tensor--axial-tensor and their
scalar--pseudoscalar analogues, and~(\ref{eq:VA_rotation}) for the vector
pair. For $N_f=2$
the $d$-terms vanish and the rotations close on the cross multiplets,
so invariance of the state under $SU(2)_L\times SU(2)_R$ gives
\begin{align}
C_S^{ns}(x)&=C_P^{s}(x),\qquad C_S^{s}(x)=C_P^{ns}(x),
\label{eq:C_SU2_SP}\\
C_T^{ns}(x)&=C_X^{s}(x),\qquad C_T^{s}(x)=C_X^{ns}(x),
\label{eq:C_SU2_TX}\\
C_V^{ns}(x)&=C_A^{ns}(x),
\label{eq:C_SU2_VA}
\end{align}
the scalar/pseudoscalar lines being the familiar $(\delta,\eta')$ and
$(\sigma,\pi)$ relations, the tensor lines the relations of the
$(\rho_T,h_{1T})$ and $(b_1,\omega_T)$ currents, and the last the
$(\rho,a_1)$ relation. For $N_f\ge 3$
the $d^{cbd}$ terms of eq.~(\ref{eq:TX_cross}) rotate the nonsinglet
operators among themselves as well, so that in addition
\begin{equation}
C_S^{ns}(x)=C_P^{ns}(x),\qquad C_T^{ns}(x)=C_X^{ns}(x)\qquad (N_f\ge 3).
\label{eq:C_SU3_ns}
\end{equation}
Combining eqs.~(\ref{eq:C_SU2_SP}), (\ref{eq:C_SU2_TX}), and
(\ref{eq:C_SU3_ns}), the four correlators of each quartet coincide under
the chiral group alone,
\begin{align}
C_S^{ns}=C_P^{ns}=C_S^{s}=C_P^{s},\qquad
C_T^{ns}=C_X^{ns}=C_T^{s}=C_X^{s}\notag\\
(N_f\ge 3).
\label{eq:C_SU3_chain}
\end{align}

Equations~(\ref{eq:C_SU2_SP})--(\ref{eq:C_SU3_chain}) reproduce the
meson relations of Ref.~\cite{Cohen:1996sb}: for $N_f=2$ the
$(\tfrac12,\tfrac12)\oplus(\tfrac12,\tfrac12)$ multiplet gives the
scalar/pseudoscalar and tensor relations and the $(0,1)\oplus(1,0)$
multiplet the vector relation, while for $N_f=3$ the
$(3,\bar3)\oplus(\bar3,3)$ multiplet gives the quartet chain
(\ref{eq:C_SU3_chain}). Here they
follow from the explicit rotations of Sec.~\ref{sec:chains} acting on the
operators $O_\Gamma^a$, so that each correlator relation is accompanied
by the equality of renormalization constants derived there. The two
together make $\kappa_{AB}$ a finite, cutoff-independent measure of the
corresponding partner splitting at any quark mass.

\subsection{$U(1)_A$ relations}
\label{subsec:U1A_rel}

The singlet axial rotation, eq.~(\ref{eq:U1A_rotation}) for DWF and
eq.~(\ref{eq:GW_rotation}) for the overlap, acts identically on all
flavor components and maps $O_S^a\to O_P^a$, $O_T^a\to O_X^a$ for every
$a$, while $O_V^a$ and $O_A^a$ are invariant. Effective restoration of
$U(1)_A$ therefore gives, for every $N_f$,
\begin{align}
C_S^{ns}(x)&=C_P^{ns}(x),\qquad C_S^{s}(x)=C_P^{s}(x),
\label{eq:C_U1A_SP}\\
C_T^{ns}(x)&=C_X^{ns}(x),\qquad C_T^{s}(x)=C_X^{s}(x).
\label{eq:C_U1A_TX}
\end{align}
For the
connected nonsinglet correlators eq.~(\ref{eq:C_U1A_SP}) is
$C_\pi(x)=C_\delta(x)$, the pointwise form of $\chi_\pi=\chi_\delta$, and
eq.~(\ref{eq:C_U1A_TX}) the corresponding $(\rho_T,b_1)$ statement for
the tensor pair.

Whether these $U(1)_A$ relations provide information that is not
already implied by the restoration of $SU(N_f)_L\times SU(N_f)_R$ depends
on $N_f$. For $N_f=2$ the chiral group relates
$O_S^{ns}$ only to $O_P^{s}$ and $O_P^{ns}$ only to $O_S^{s}$, and
likewise for the tensor pair [eqs.~(\ref{eq:C_SU2_SP}) and
(\ref{eq:C_SU2_TX})]. It does not relate the two members of a
$U(1)_A$ pair to each other, so the equalities (\ref{eq:C_U1A_SP}) and
(\ref{eq:C_U1A_TX}) are genuine $U(1)_A$ statements. For $N_f\ge 3$ the quartet chain
(\ref{eq:C_SU3_chain}) already equates all four correlators of each
quartet under the chiral group, so every relation in
eqs.~(\ref{eq:C_U1A_SP}) and (\ref{eq:C_U1A_TX}) is then implied by the
restoration of $SU(N_f)_L\times SU(N_f)_R$ alone.

The same conclusion follows directly from the flavor structure. Collect
the scalar and pseudoscalar densities into $\Phi=\bar q_R q_L$, an
$N_f\times N_f$ matrix transforming as $(N_f,\bar N_f)$ under
$SU(N_f)_L\times SU(N_f)_R$, with $\Phi=S+iP$. Then
$\Phi\Phi^\dagger=S^2+P^2+i(PS-SP)$ and $\Phi\Phi=S^2-P^2+i(SP+PS)$, so
the sum of the scalar and pseudoscalar correlators is
$\langle\Phi\Phi^\dagger\rangle$ while their $U(1)_A$-breaking difference
is ${\rm Re}\langle\Phi\Phi\rangle$, the correlator of $\Phi$ with itself.
When the chiral group is restored a correlator can be nonzero only if the
operator product includes a singlet of $SU(N_f)_L\times SU(N_f)_R$. Now
$\Phi\Phi^\dagger\in(N_f\otimes\bar N_f,\bar N_f\otimes N_f)$ always
contains a singlet, whereas $\Phi\Phi\in(N_f\otimes N_f,\bar N_f\otimes
\bar N_f)$ contains one only when $N_f\otimes N_f$ does, i.e.\ for $N_f=2$
where $\mathbf 2\simeq\bar{\mathbf 2}$. Hence the $U(1)_A$-breaking
difference can survive chiral restoration only for two flavors,
reproducing the result above. Consequently, for three or more massless
flavors there are no two simple currents in the same chiral multiplet
that transform differently under $U(1)_A$, and the restoration of
$U(1)_A$ cannot be probed by any two-point correlator of quark bilinears,
as noted by Cohen and Ji~\cite{Cohen:1996sb}. The $U(1)_A$ information
then resides in the disconnected, topological sector
(Sec.~\ref{subsec:disconnected}) or in higher-point functions, an example of
the latter being given in Ref.~\cite{Cohen:1996sb}.
Cohen and Ji's conclusion holds for massless quarks.
In physical QCD the quark masses are nonzero and hierarchical,
$0<m_u<m_d<m_s<m_c<m_b$, and the chain (\ref{eq:C_SU3_chain}) holds only
up to terms proportional to the quark masses.

For physical QCD the flavor sectors restore hierarchically
(Sec.~\ref{sec:intro}), with
$T_c^{ud}<T_c^{uds}<T_c^{udsc}<T_c^{udscb}$ and
$T_1^{ud}<T_1^{uds}<T_1^{udsc}<T_1^{udscb}$.
Here $T_c^{N_f}$ and $T_1^{N_f}$ stand for $T_c^{ud},\ldots,T_c^{udscb}$
and $T_1^{ud},\ldots,T_1^{udscb}$, for $N_f=2,\ldots,5$.
The probes of $SU(N_f)_L\times SU(N_f)_R$ are $\kappa_{VA}^a$ for
$a=1,\cdots,N_f^2-1$, and those of $U(1)_A$ are $\kappa_{PS}^a$ and
$\kappa_{TX}^a$ for $a=0,1,\cdots,N_f^2-1$, where $a=0$ denotes the
singlet.
For a given threshold $\epsilon$, if $\kappa_{VA}^a\le\epsilon$ for
$T\ge T_c^{N_f}$ and all $a=1,\cdots,N_f^2-1$, then
$SU(N_f)_L\times SU(N_f)_R$ is effectively restored at the level
$\epsilon$.
Similarly, if $\kappa_{PS}^a\le\epsilon$ and $\kappa_{TX}^a\le\epsilon$
for $T\ge T_1^{N_f}$ and all $a=0,\cdots,N_f^2-1$, then the $U(1)_A$ of the
$N_f$ flavors is effectively restored at the level $\epsilon$.
This prescription has been applied in Ref.~\cite{Chiu:2026tot} for
$N_f=2+1+1$ lattice QCD at the physical point, where the disconnected
parts of the singlet and the diagonal nonsinglets have been neglected.
For the light block this is justified above $T_c^{ud}$.
There the connected parts give the $U(1)_A$ splitting of the isovector pair in
place of that of the light singlets, which has the same magnitude up to the
$SU(2)_L\times SU(2)_R$ splittings of $(\sigma,\pi)$ and
$(\delta,\eta')$~\cite{Chiu:2026spec}.
The neglected heavy-flavor terms are the topological terms, suppressed by
$m_l/m_h$, and the covariances of the light and heavy scalar
loops~\cite{Chiu:2026spec}.
In the window $(T_c^{ud},T_c^{uds})$, $m_s$ breaks
$SU(3)_L\times SU(3)_R$, so the degeneracy of a $(u,d)$ pair is not
enforced by the enlarged chiral group, and by the $N_f=2$ argument it
reports on $U(1)_A$.
Above $T_c^{uds}$, the chain (\ref{eq:C_SU3_chain}) is broken at order
$m_s$ for the pairs involving the $s$ quark.
Their splittings then carry the $U(1)_A$ breaking, suppressed by the quark
masses and together with explicit chiral breaking of comparable order.
These $\kappa_{PS}^a$ and $\kappa_{TX}^a$ are therefore $U(1)_A$ probes
of the $(u,d,s)$ sector, with a sensitivity that vanishes in the massless
limit, in agreement with Ref.~\cite{Cohen:1996sb}.
The same holds for the $c$ quark in $(u,d,s,c)$ and the $b$ quark in
$(u,d,s,c,b)$.


The singlet pairs $(C_P^{s},C_S^{s})$ carry no independent information.
Once the chiral group of the sector is effectively restored, the singlet
relations of eqs.~(\ref{eq:C_SU2_SP}) and (\ref{eq:C_SU2_TX}) give
$C_S^{s}-C_P^{s}=C_P^{ns}-C_S^{ns}$, i.e.\ the singlet $U(1)_A$ splitting
equals the nonsinglet one, both being the disconnected correlator
$C_{P,\rm disc}$ (Sec.~\ref{sec:sym_rel_conseq}).
For $N_f\ge 3$ massless flavors both vanish with the chiral group.
For the $(u,d)$ sector the singlet pair therefore probes $U(1)_A$ in the
same window as the nonsinglet pair, at the cost of the disconnected
contractions.

\subsection{Consequences}
\label{sec:sym_rel_conseq}

Summing over the lattice, each pointwise relation becomes the
corresponding relation between susceptibilities. The equalities $\chi_\pi=\chi_\delta$
and $\chi_T^{ns}=\chi_X^{ns}$ follow from $U(1)_A$ for every $N_f$, and from
$SU(N_f)_L\times SU(N_f)_R$ for $N_f\ge 3$. The equality $\chi_V^{ns}=\chi_A^{ns}$
follows from $SU(N_f)_L\times SU(N_f)_R$ for every $N_f$. These are the relations
that enter the ratios of Ref.~\cite{Chiu:2026sxy}, for the $(u,d)$ sector
in $N_f=2+1+1$ lattice QCD at the physical point. The two nonsinglet ratios that probe
$U(1)_A$ are $\kappa_{PS}$, from $\chi_P$ and $\chi_S$ (for $N_f=2$,
$\kappa_{\pi\delta}$ from $\chi_\pi$ and $\chi_\delta$), and $\kappa_{TX}$,
from $\chi_T$ and $\chi_X$. In each the two operators are related by the
$U(1)_A$ rotation but sit in the same chiral multiplet, and their equal
renormalization, $Z_P=Z_S$ and $Z_T=Z_X$ (Sec.~\ref{sec:chains}), makes
the ratio finite and cutoff-independent for every $N_f$
[eq.~(\ref{eq:kappa_cancel})].
As shown above, they are genuine $U(1)_A$ probes for $N_f=2$, and in the
hierarchical flavor windows of physical QCD at nonzero quark masses
(Sec.~\ref{subsec:U1A_rel}), while for $N_f\ge 3$ massless flavors their
vanishing is already implied by the chiral group.
Note that in $N_f=2+1+1$ QCD the
strange and charm sea quarks also dress the $u,d$ valence propagators and
enter the configuration weight through their determinants, so $T_c^{ud}$
of $N_f=2+1+1$ QCD differs from that of $N_f=2$ QCD.

Two further consequences connect the correlator relations with the
topological sector (Sec.~\ref{subsec:disconnected}).
First, for $N_f=2$ the difference of the
two relations in eq.~(\ref{eq:C_SU2_SP}), with
$C_P^{s}=C_P^{ns}-C_{P,\rm disc}$, gives the pointwise identity
$C_\pi(x)-C_\delta(x)=C_{P,\rm disc}(x)$ in the
$SU(2)_L\times SU(2)_R$-symmetric state.
Its four-volume sum, with eq.~(\ref{eq:chi_t}), is
\begin{equation}
\chi_\pi-\chi_\delta=\chi_{5,\rm disc}=\frac{\chi_t}{m^2}.
\label{eq:pidelta_chit}
\end{equation}
Second, for $N_f\ge 3$ eqs.~(\ref{eq:C_SU2_SP}) and (\ref{eq:C_SU3_ns})
together give $C_P^{ns}=C_P^{s}$, i.e.\ $C_{P,\rm disc}(x)=0$ and hence
$\chi_t=0$ in the $SU(N_f)_L\times SU(N_f)_R$-symmetric state with three
or more massless flavors, in accordance with the vanishing of the
topological susceptibility whenever a quark is massless.

\section{Contrast with Wilson fermions}
\label{sec:wilson}

For Wilson-type fermions, which break chiral symmetry at $O(a)$,
the UV divergence structure is qualitatively different. The explicit
breaking induces an additive mass renormalization $m_c \sim O(1/a)$, whose
dimension-three counterpart is the well-known $1/a^3$ divergence of the
condensate $\vev{\bar qq}$, and, in the susceptibility, a chirally odd
linear-mass divergence $\propto m/a$ that is forbidden for GW fermions by
the exact lattice chiral symmetry (the Dirac-trace protection of
Sec.~\ref{subsec:additive}). These divergences are summarized in
Table~\ref{tab:wilson}.

\begin{table}[t]
\caption{\label{tab:wilson}%
Additive divergences in the bare susceptibility $\chi^{\rm bare}$ for
Wilson versus Ginsparg--Wilson (GW) fermions. The chirally odd
linear-mass divergence is allowed for Wilson but forbidden for GW by the
exact lattice chiral symmetry (the Dirac-trace argument of
Sec.~\ref{subsec:additive}). The chirally even $m^2$ term (marked
$^{\ast}$) is a spurion-allowed \emph{logarithmic} divergence, present
for either discretization and already in the continuum. Being
short-distance, all entries are temperature-independent and are removed
by the $T$-subtraction.}
\begin{ruledtabular}
\begin{tabular}{lcc}
Additive divergence & Wilson & GW \\
\colrule
$\alpha_\Gamma/(2a^2)$ (identity operator $\Unit$, OPE) & Present & Present \\
$\propto m/a$ (chirally odd) & Present & Absent \\
$c_m^\Gamma\, m^2\ln(1/(am))$ (chirally even) & Present & Present$^{\ast}$ \\
\end{tabular}
\end{ruledtabular}
\end{table}

For Wilson fermions, the hard breaking of chiral symmetry by the Wilson
term generates a chirally even piece of the quark propagator that is not
proportional to $m$. This both produces the additive mass renormalization
(the analogue, at the level of the condensate, of the $1/a^3$ mixing of
$\bar{q}q$ with the identity operator) and lifts the Dirac-trace protection,
allowing the linear-mass divergence $\propto m/a$ in the susceptibility.
These additive divergences are still removed by the temperature
subtraction, as before. The essential difference is that the explicit
chiral-symmetry breaking of the Wilson term also invalidates the equality
$Z_A = Z_B$ between symmetry partners (for instance $Z_S\neq Z_P$), which is
the property on which the multiplicative cancellation
in $\kappa_{AB}$ relies. Restoring a
controlled construction then requires non-perturbative subtraction and
improvement procedures that are computationally demanding and introduce
additional systematic uncertainties.
The tensor pair is an exception. Since
$i\gamma_5\gamma_4\gamma_k\propto\epsilon_{kij}\gamma_i\gamma_j$, the densities
$(O_T)_k$ and $(O_X)_k$ of Sec.~\ref{subsec:ZAZB_lat} are the electric and
magnetic components of one antisymmetric tensor $\bar q\sigma_{\mu\nu}t^aq$,
which is irreducible under the hypercubic group. Hence, on an isotropic
lattice, $Z_T=Z_X$ at finite lattice spacing~\cite{Chiu:2026tot}, and after the temperature subtraction
$\kappa_{TX}$ is free of renormalization constants. Only lattice artifacts,
of $O(a)$ without improvement and of $O(a^2)$ with it, split the two channels.

For GW fermions, the exact lattice chiral symmetry forbids the
$m$-independent hard breaking: there is no additive mass renormalization,
and the chirally odd linear-mass divergence $\propto m/a$ is absent
(the Dirac-trace argument of Sec.~\ref{subsec:additive}).
The chirally even term $c_m^\Gamma\, m^2\ln(1/(am))$ is, by
contrast, \emph{not} forbidden by chiral symmetry: the quark mass is
itself a spurion of chiral-symmetry breaking, and $m^2$ is chirally even,
so this logarithmic divergence is present in general, as it already is in
the continuum.
It does not spoil $\kappa_{AB}$ because it is short-distance and
therefore temperature-independent: together with the
term $\alpha_\Gamma/(2a^2)$ it cancels exactly in $\chi(T)-\chi(T_r)$.
What survives the subtraction is purely multiplicative
($Z_\Gamma^{2}$, controlled by $\gamma_\Gamma$), and for symmetry
partners the equality $Z_A = Z_B$ makes $Z^{2}$ cancel in
$\kappa_{AB}$.
The construction of $\kappa_{AB}$ is therefore protected by the
temperature subtraction together with the equality $Z_A = Z_B$, not by
the absence of the mass-dependent logarithmic divergence.

These obstructions are visible in practice. A recent large-scale study of
$U(1)_A$ restoration with Wilson-clover fermions on highly anisotropic
lattices~\cite{Aarts:2026kpq} finds that the bare susceptibility
difference $\chi_\pi - \chi_\delta$ never crosses zero at any
temperature, as expected from the unremoved additive divergences and the
multiplicative mismatch $Z_P \neq Z_S$. The restoration signal is instead
extracted from correlators normalized at the thermal midpoint, with smeared
sources: mostly as a $\tau$-dependent ratio, and, for the estimate
$T_{U(1)_A}=319(22)$~MeV, summed from a cutoff $\tau_{\rm min}$. The summed ratio is
finite and free of the multiplicative constants, but it equals the
difference over the sum of the bare truncated susceptibilities, with one
channel reweighted by the midpoint ratio
$G_\pi(N_\tau/2)/G_\delta(N_\tau/2)$, which is unity only at degeneracy.

Since each channel enters only through the correlator normalized by its
own midpoint value, the observable is invariant under independent
rescalings of $G_\pi$ and $G_\delta$. It therefore tests only whether the
two correlators are proportional, $G_\pi(\tau)=c\,G_\delta(\tau)$ for all
$\tau$, i.e., whether their spectral functions are proportional. The
constant $c$, which contains $Z_P^2/Z_S^2$ together with the ratio of the
physical amplitudes, drops out, so a vanishing ratio is necessary for
degeneracy but not sufficient.
For a single pole, $G_H(\tau) = A_H\cosh[m_H(\tau-N_\tau/2)]$, the argument
of the hyperbolic cosine vanishes at the midpoint, so $G_H(N_\tau/2) = A_H$
and dividing by it leaves $\cosh[m_H(\tau-N_\tau/2)]$. The normalized
correlator then depends only on the mass $m_H$, and the comparison reduces
to the equality of the masses. With several states it depends on the masses
and on the relative weights of the states, but not on the overall
amplitude. The summed ratio is weaker still, since only a weighted sum over
$\tau\ge\tau_{\rm min}$ has to vanish. It can vanish with unequal masses,
when the mismatch in the masses is compensated by a mismatch in the weights. The converse failure is equally
concrete: in the free theory the spatial correlators of the tensor-type
channels carrying the Euclidean time index fall as $e^{-2\pi Tz}/z^2$, while
those of the other channels fall as $e^{-2\pi Tz}/z$~\cite{Chiu:2023hnm}, so
the two sets share the screening mass $2\pi T$ and are never degenerate; a
comparison of screening masses alone would miss the splitting that the
amplitudes carry. None of this depends on the
midpoint being the normalization point. Dividing by the correlator at any
fixed $\tau_0$ removes only the overall amplitude, so the conclusion applies
to any self-normalized ratio.
The susceptibility difference $\chi_\pi-\chi_\delta$ is a different
quantity. It integrates the unnormalized correlators, and so retains
precisely the amplitude information that the shape comparison discards.
Shape degeneracy and susceptibility degeneracy are therefore distinct
necessary conditions for restoration, neither implying the other. Moreover, the temporal cutoff
removes the additive divergences only if it is held fixed in physical
units and wide compared with both the lattice spacing and the range of
the chirally odd short-distance artifacts. A cutoff of a few time slices
retains a substantial residue.

With exact chiral symmetry none of these
prescriptions is needed: the temperature subtraction removes the
divergences with every time slice retained and nothing discarded, and
$\kappa_{AB}$ is exactly RG-invariant.

Any comparison of the resulting degeneracy temperatures between the two
formulations must take two further differences into account. In
Ref.~\cite{Aarts:2026kpq} the nonsinglet $\pi$--$\delta$ degeneracy sets
in at $T_{U(1)_A}=319(22)$~MeV, at a single lattice spacing and
$m_\pi\simeq 380$~MeV, whereas $\kappa_{AB}$ is intended for
Ginsparg--Wilson fermions at the physical point. First, the
shape ratio of Ref.~\cite{Aarts:2026kpq} and $\kappa_{AB}$ are, as shown
above, different observables: one compares the normalized shapes of the
correlators, the other their integrated amplitudes. Both vanish at exact
degeneracy, but away from it they measure different things, and there is
no reason for the temperatures at which they fall below a given threshold
to coincide, even in the continuum limit. Second, both observables are
dominated by the infrared: the partner splitting they probe is an
explicit-breaking effect controlled by the quark mass, or equivalently the
pion mass, and the two studies are at very different pion masses. The
results of Ref.~\cite{Aarts:2026kpq}, the $SU(2)_L\times SU(2)_R$
restoration at $T_c\simeq 180$~MeV and the $\pi$--$\delta$ degeneracy at
$319(22)$~MeV, are obtained at $m_\pi\simeq 380$~MeV. The heavy pion
raises the entire temperature scale, as reflected in
$T_c\simeq 180$~MeV compared with $T_c\simeq 156$~MeV at the physical
point, and enlarges the explicit $U(1)_A$ breaking, both pushing the
quoted degeneracy temperature upward. The pion-mass difference is
therefore expected to be the primary source of any discrepancy between
determinations at these two masses, with the choice of observable and the
continuum extrapolation as further contributions. It should be stressed that the obstructions discussed in
this section concern the construction of a renormalization-free partner
ratio, not the location of the crossover itself: on the same anisotropic
Wilson-clover ensembles the FASTSUM collaboration has determined the
pseudocritical temperature from the renormalized chiral condensate,
finding $T_{\rm pc}^{\bar\psi\psi} = 181(2)$~MeV at $m_\pi = 384(4)$~MeV
and $164(2)$~MeV at $m_\pi = 236(2)$~MeV, and, extrapolating along the
pion mass with an $O(4)$-scaling fit, $T_{\rm pc}^{\bar\psi\psi}=158(3)$~MeV
at the physical point~\cite{Aarts:2020vyb}, consistent with the
chiral-condensate determinations of the Wuppertal--Budapest and HotQCD
collaborations obtained with staggered fermions (albeit without a
continuum extrapolation). A direct
comparison of $U(1)_A$ degeneracy temperatures between the two
formulations would require the Wilson observable at or near the physical
pion mass and in the continuum limit, which, in view of the obstructions
above, is a formidable task.
Note also that the two stages of Ref.~\cite{Aarts:2026kpq} both lie within
the nonsinglet sector. The relative ordering of the nonsinglet and singlet
(topological) restorations is a separate question, constrained by the
Ward identity $\chi_\pi-\chi_\delta\simeq\chi_t/m^2$ of
eq.~(\ref{eq:pidelta_chit}).

\section{Extension to continuum QCD}
\label{sec:continuum}

The results above are established in the continuum limit ($a\to0$) of
lattice QCD with exact chiral symmetry. Because the lattice sums become
integrals and the spacing enters only through the UV cutoff $1/a$, the
short-distance analysis is identical in form to continuum QCD with $1/a$
replaced by the continuum regulator $1/\eps$ of Sec.~\ref{sec:susceptibility}. Multiplicative renormalizability,
the renormalization-group equations, the temperature independence of
the additive UV divergences (a general property of thermal field
theory~\cite{Kislinger:1975ab,Landsman:1986uw,Dolan:1973qd,Weinberg:1974hy,Bernard:1974bq}),
and the partner equalities $Z_A = Z_B$ derived in
Sec.~\ref{sec:chains} therefore hold equally in the continuum theory,
provided the continuum scheme is mass independent and its
regularization preserves chiral symmetry. On the
lattice with Ginsparg--Wilson fermions the chiral Ward identities hold
nonperturbatively at finite spacing, and the equalities are exact
already at $a\neq 0$. Since $\kappa_{AB}$ is RG invariant and
scheme independent, its continuum value is the same regardless of the
regularization used to compute it.

\section{Summary}
\label{sec:summary}

We have presented a complete analysis of the UV divergence structure
of meson susceptibilities in QCD, proving that the symmetry ratio
$\kappa_{AB}$ is exactly RG-invariant and scheme-independent.
The proof rests on three pillars:

\begin{enumerate}
\item \textbf{Temperature subtraction removes the additive divergences.}
The power divergence $\alpha_\Gamma/(2a^2)$ and the mass-dependent
logarithm $c_m^\Gamma\, m^2\ln(1/(am))$ in the bare susceptibility
are temperature-independent and cancel exactly in
$\chi^{\rm reg}(T;T_r) = \chi(T) - \chi(T_r)$.
The remaining logarithmic UV dependence is multiplicative, residing
entirely in $Z_\Gamma^{2}$. The multiplicative structure is exact to all
orders, a standard consequence of the multiplicative renormalizability of
gauge-invariant quark bilinears~\cite{Collins:1984xc,Peskin:1995ev}. We
resum $Z_\Gamma$ explicitly at leading-log, but its precise form does not
affect $\kappa_{AB}$, which requires only $Z_A = Z_B$.

\item \textbf{Multiplicative renormalization cancels in the ratio.}
The factor $Z_A^2$ is common to both $\chi_A^{\rm reg}$
and $\chi_B^{\rm reg}$ when $Z_A = Z_B$, and cancels in $\kappa_{AB}$.

\item \textbf{$Z_A = Z_B$ for symmetry partners.}
This equality holds for any $N_f$ and any quark masses in a
mass-independent scheme with chirally symmetric regularization.
It is unaffected by spontaneous symmetry breaking or the $U(1)_A$
anomaly. The $Z$-factor equalities needed for $\kappa_{AB}$ are:
\begin{align}
Z_S^{ns} = Z_P^{ns} = Z_S^s = Z_P^s &\equiv Z_{SP}, \\
Z_T^{ns} = Z_X^{ns} = Z_T^s = Z_X^s &\equiv Z_{TX}, \\
Z_V^{ns} = Z_A^{ns} &\equiv Z_{VA}.
\end{align}
The vector and axial-vector constants are finite: unity for the
conserved current, and approaching unity in the continuum limit for the
local bilinears (Sec.~\ref{subsec:ZAZB_lat}). Only these equalities among
symmetry partners, not their values, are needed for $\kappa_{AB}$.
\end{enumerate}

The reference temperature entering $\kappa_{AB}$ is a genuine choice,
and Sec.~\ref{subsec:T_sub} makes precise in what sense the ratio is
independent of it. A reference is admissible if the partners are
degenerate there within errors, which requires the explicit breaking by
the quark masses to be small at $T_r$, and if the denominator of
$\kappa_{AB}$ keeps a definite sign over the range.
For all admissible choices the zeros of
$\kappa_{AB}$ and the sign of the partner splitting are the same, while the
magnitude of $\kappa_{AB}$ below the restoration point carries a
$T_r$-dependent normalization. The common sign and the $T_r$ dependence of
the magnitude are demonstrated numerically in Ref.~\cite{Chiu:2026sxy},
whose simulated range lies below the zero.
The equalities $Z_A=Z_B$ are established for general $N_f$ from the
variations of the operators under the non-anomalous
$SU(N_f)_L\times SU(N_f)_R$ combined with the $U(1)_A$ rotation.
The nonsinglet $\kappa_{\pi\delta}$ is identified as the two-point
probe of $U(1)_A$ for two light flavors.
In physical QCD, with nonzero and hierarchical quark masses, the
$\kappa_{PS}^a$ and $\kappa_{TX}^a$ of the pairs involving the heaviest
flavor of each sector are $U(1)_A$ probes of that sector, with a
sensitivity that vanishes in the massless limit (Sec.~\ref{sec:sym_rel}).
The Ward identity $\chi_\pi-\chi_\delta\simeq\chi_t/m^2$ of
eq.~(\ref{eq:pidelta_chit}) provides a direct test of
$SU(2)_L\times SU(2)_R$ restoration.

Beyond the ratio, the analysis yields reference-free determinations of
characteristic temperatures (Sec.~\ref{subsec:T_char}). Since both the
power divergence and the mass-dependent logarithm are temperature
independent, any thermal ensemble may serve as the reference: the peak
position of the RG-invariant combination
$m^2\,[\chi_\sigma(T)-\chi_\sigma(T_r)]$, optionally normalized by
$m_\pi^4$, is exactly independent of $T_r$ and defines the
pseudocritical temperature without any zero-temperature subtraction.
By the index relation, the analogous combination in the disconnected
pseudoscalar channel is the subtracted topological susceptibility,
$m^2\,[\chi_{5,\rm disc}(T)-\chi_{5,\rm disc}(T_r)]
=\chi_t(T)-\chi_t(T_r)$, whose inflection point defines a
$T_r$-independent characteristic temperature of the anomalous sector.

These results hold for any regularization preserving chiral symmetry
(GW fermions on the lattice) and extend to continuum QCD in any
chirally symmetric scheme.

\bigskip
\begin{acknowledgments}
This work was supported by the National Science and Technology Council
(Grants No.~108-2112-M-003-005, No.~109-2112-M-003-006, No.~110-2112-M-003-009),
and Academia Sinica Grid Computing Centre (Grant No.~AS-CF11-114-A11).

\end{acknowledgments}


\appendix

\section{Free thermal temporal correlator}
\label{app:thermal}

This appendix derives the free temporal meson correlator at finite
temperature as the periodic image sum of its zero-temperature form,
for both the massless and massive cases. We show that all channels share the
same functional dependence on $t$ and $T$, and that all additive divergences
are temperature-independent. The calculation is carried out in position
space, and the coefficients of the additive divergences for all channels are
obtained. All coefficients have been checked against direct numerical
integration of the image sums.

\subsection{Massless power divergence}
\label{app:thermal_massless}

At finite temperature, the free massless quark propagator is antiperiodic
in Euclidean time $x_4 = t$ with period $\beta \equiv 1/T$. It is obtained
from the zero-temperature propagator $S_0$ by summing over thermal images,
\begin{equation}
S_T(\vec x, t) = \sum_{n=-\infty}^{\infty} (-1)^n\,
S_0(\vec x,\, t + n\beta),
\label{eq:thermal_prop}
\end{equation}
where the sign $(-1)^n$ enforces antiperiodicity.
The temporal integral picks up the coincident-point singularity, which gives
the temperature-independent additive power divergence, as discussed in
Sec.~\ref{subsec:additive}. The singularity enters with multiplicity two in
either prescription, from the $n=0$ and $n=-1$ images in the
definition~(\ref{eq:chi_bare}) and twice from the $n=0$ image in the
two-interval form~(\ref{eq:chi_bare_two}) used below.

\subsubsection{Scalar and pseudoscalar channels}

By eq.~(\ref{eq:tree_corr}) the correlator involves
$\tr[\Gamma\,S\,\bar\Gamma\,S]$ with $\bar\Gamma=\pm\Gamma$. Below we quote
the traces $\tr[\Gamma\,\slashed X\,\Gamma\,\slashed X]$; the sign of
$\bar\Gamma$ relative to $\Gamma$ is the channel sign of the manifestly
positive normalization, and only the magnitudes of the traces enter the
constants $\alpha_\Gamma$.

For $\Gamma = \Id$ and $\Gamma = \gamma_5$, the Dirac trace is isotropic.
With the $n$-th image four-vector
$X_n = (\vec x,\, t+n\beta) = (x_1,x_2,x_3,\, x_4+n\beta)$,
the off-diagonal ($n\neq n'$) contributions to the integrand of the correlator
carry the Dirac trace
\begin{align}
&\tr[\Gamma\,\slashed{X}_n\,\Gamma\,\slashed{X}_{n'}]
= 4\,\eta_\Gamma\,(X_n\!\cdot\!X_{n'}), \notag \\
& X_n\!\cdot\!X_{n'} = |\vec x|^2 + (t+n\beta)(t+n'\beta), \notag \\
& \eta_{\Id} = +1,\ \ \eta_{\gamma_5} = -1,
\end{align}
in the numerator, and $X_n\!\cdot\!X_{n'}$ vanishes at $\vec{x}=0$ as either image
approaches its coincident point. The spatial integral is finite in closed
form. With $a_n \equiv t+n\beta$,
\begin{equation}
\int d^3x\,\frac{X_n\!\cdot\!X_{n'}}{(X_n^2)^2\,(X_{n'}^2)^2}
= \frac{\pi^2\left[1+\mathrm{sign}(a_n a_{n'})\right]}
{\big(|a_n|+|a_{n'}|\big)^3}.
\label{eq:cross_integral}
\end{equation}
Image pairs of opposite temporal orientation, $a_n a_{n'}<0$, vanish
identically, and pairs of the same orientation give
$2\pi^2/(|a_n|+|a_{n'}|)^3$. As image $n$ approaches its coincident point,
$a_n\to 0$, the same-orientation pairs tend to the finite value
$2\pi^2/|(n'-n)\beta|^3$, set by the separation of the two images, so the
cross-image terms belong to the finite part of the correlator.
On the other hand, the diagonal ($n = n'$) contributions to the integrand of the correlator
carry the Dirac trace
\begin{equation}
\tr[\Gamma\,\slashed{X}_n\,\Gamma\,\slashed{X}_n]
	= 4\,\eta_\Gamma\,X_n^2, \qquad X_n^2 = |\vec x|^2 + (t+n\beta)^2,
\end{equation}
in the numerator, so the integrand is proportional to $X_n^2/(X_n^2)^4 = 1/(X_n^2)^3$.
The spatial integral of each diagonal image is
\begin{equation}
\int d^3x\,\frac{1}{(|\vec x|^2 + (t+n\beta)^2)^3}
= \frac{\pi^2}{4\,|t+n\beta|^3},
\end{equation}
Summing the diagonal images gives the singular part of the correlator, the
two antiperiodic propagator signs combining to $(-1)^{2n}=+1$ on the diagonal
$n=n'$,
\begin{equation}
C_\Gamma(t,T)\Big|_{\rm diag} = \frac{\alpha_\Gamma}{2}
\sum_{n=-\infty}^{\infty}\frac{1}{|t+n\beta|^3}.
\label{eq:K_sum}
\end{equation}
Only the $n=0$ image is singular at the coincident point, reducing there to
the zero-temperature $\alpha_\Gamma/(2|t|^3)$. The $n\neq 0$ images have
their poles at $t=-n\beta$ outside the fundamental interval and are regular.
The cross-image pairs $n\neq n'$ retain the alternating signs
$(-1)^{n+n'}$ and, by eq.~(\ref{eq:cross_integral}), sum to the regular
remainder
\begin{equation}
\Delta C_\Gamma(t,T) = \frac{\alpha_\Gamma}{2}
\sum_{n\neq n'}(-1)^{n+n'}\,
\frac{4\left[1+\mathrm{sign}(a_n a_{n'})\right]}
{\big(|a_n|+|a_{n'}|\big)^3}
\label{eq:cross_sum}
\end{equation}
for the scalar and pseudoscalar channels, with $a_n = t+n\beta$. The sum
converges absolutely and is finite for all $t$. The full free correlator is
\begin{equation}
C_\Gamma(t,T) = C_\Gamma(t,T)\Big|_{\rm diag} + \Delta C_\Gamma(t,T),
\label{eq:C_full}
\end{equation}
which is eq.~(\ref{eq:Ct_thermal}). It is even in $t$, periodic, and
symmetric about $t=\beta/2$, as the bosonic meson correlator must be, and
its entire coincident-point singularity resides in the $n=n'=0$ term.

\medskip
The power divergence of the bare susceptibility follows by integrating
the correlator~(\ref{eq:C_full}) over the two intervals of
eq.~(\ref{eq:chi_bare_two}). Only the $n=0$ diagonal image contributes the
singularity. On the positive interval,
\begin{equation}
\int_\eps^{1/(2T)}\frac{dt}{t^3}
= \frac{1}{2\eps^2} - 2T^2 = \frac{1}{2\eps^2} + (\text{finite}),
\label{eq:image_integral}
\end{equation}
while every $n\neq 0$ image and the cross-image
remainder~(\ref{eq:cross_sum}) are finite and contribute only to the finite
part. On
the negative interval the correlator is even, so its $n=0$ image is $1/|t|^3$
and
\begin{equation}
\int_{-1/(2T)}^{-\eps}\frac{dt}{|t|^3} = \frac{1}{2\eps^2}+(\text{finite}),
\label{eq:neg_interval}
\end{equation}
equal to the positive interval. The two sum to $1/\eps^2$, giving
$\chi_\Gamma^{\rm bare}\supset\alpha_\Gamma/(2\eps^2)$, as in eq.~(\ref{eq:tree_div}),
which on the lattice ($\eps\to a$) is the power divergence $\alpha_\Gamma/(2a^2)$.

\medskip
The definition~(\ref{eq:chi_bare}) gives the identical result by a different
counting. There the interval $[\eps,\,1/T-\eps]$ straddles two coincident points,
and the divergence comes from two images: the $n=0$ image at $t\to \eps$ and the
$n=-1$ image at $t\to 1/T-\eps$, each contributing $1/(2\eps^2)$. In the
two-interval form~(\ref{eq:chi_bare_two}) the single coincident point at $t=0$
is cut off at both near ends $t\to\pm \eps$, so its $n=0$ image contributes
twice. Either way the coincident-point singularity enters with total
multiplicity two, and the divergence is $1/\eps^2$.

\subsubsection{Vector, axial-vector, tensor vector, and axial-tensor vector channels}

For the remaining channels the Dirac trace is anisotropic. We use the
Euclidean convention
\begin{align}
&\gamma_k = \begin{pmatrix} 0 & \sigma_k \\ \sigma_k & 0\end{pmatrix}
\ (k=1,2,3),\quad
\gamma_4 = \begin{pmatrix} 0 & i\,\Id_2 \\ -i\,\Id_2 & 0\end{pmatrix}, \notag \\
&\gamma_5 = \gamma_1\gamma_2\gamma_3\gamma_4
= \begin{pmatrix} \Id_2 & 0 \\ 0 & -\Id_2\end{pmatrix},
\end{align}
with $\sigma_{1,2,3}$ the Pauli matrices, satisfying
$\{\gamma_\mu,\gamma_\nu\}=2\delta_{\mu\nu}$. The temporal-correlator
operators are the vector $V_k=\gamma_k$, axial-vector $A_k=\gamma_5\gamma_k$,
tensor vector $T_k=\gamma_4\gamma_k$, and axial-tensor vector
$X_k=\gamma_5\gamma_4\gamma_k$ with $k=1,2,3$~\cite{Chiu:2026sxy}. With
$X_n=(\vec x,\,t+n\beta)$ and $\slashed X_n=\gamma_\mu X_n^\mu$, the traces
$\tr[\Gamma\,\slashed X_n\,\Gamma\,\slashed X_n]$ are linear combinations of
the squared components $X_{n,\rho}^2$. For example,
\begin{align}
V_3=\gamma_3:&\quad
4\big(X_{n,3}^2 - X_{n,1}^2 - X_{n,2}^2 - X_{n,4}^2\big), \notag \\
A_3=\gamma_5\gamma_3:&\quad
4\big(X_{n,3}^2 - X_{n,1}^2 - X_{n,2}^2 - X_{n,4}^2\big), \notag \\
T_3=\gamma_4\gamma_3:&\quad
4\big(X_{n,3}^2 + X_{n,4}^2 - X_{n,1}^2 - X_{n,2}^2\big), \notag \\
X_3=\gamma_5\gamma_4\gamma_3:&\quad
4\big(X_{n,1}^2 + X_{n,2}^2 - X_{n,3}^2 - X_{n,4}^2\big),
\end{align}
and analogously for $k=1,2$. Unlike the scalar and pseudoscalar traces,
these are not proportional to $X_n^2$. Under the spatial integral the three
spatial components average isotropically,
\begin{equation}
\int d^3x\,\frac{x_k^2}{(X_n^2)^4}
= \frac13\int d^3x\,\frac{|\vec x|^2}{(X_n^2)^4}
= \frac{\pi^2}{24\,|t+n\beta|^3},
\end{equation}
while the temporal component keeps $X_{n,4}^2=(t+n\beta)^2$,
\begin{align}
\int d^3x\,\frac{X_{n,4}^2}{(X_n^2)^4} &= \frac{\pi^2}{8\,|t+n\beta|^3}, \notag \\
\int d^3x\,\frac{1}{(X_n^2)^3} &= \frac{\pi^2}{4\,|t+n\beta|^3},
\end{align}
consistent with $3\cdot\tfrac{\pi^2}{24}+\tfrac{\pi^2}{8}=\tfrac{\pi^2}{4}$.
Each anisotropic trace therefore collapses, after the spatial integral, to
a channel-dependent constant times $1/|t+n\beta|^3$. Summing the diagonal
images gives the singular part~(\ref{eq:K_sum}) for every channel, with the
anisotropic traces and the temporal projection modifying only the constant
$\alpha_\Gamma$. The cross-image traces of these channels are linear
combinations of $|\vec x|^2$ and $a_n a_{n'}$, so each pair integrates to
$\pi^2\left[c_1 + c_2\,\mathrm{sign}(a_n a_{n'})\right]/(|a_n|+|a_{n'}|)^3$
with channel-dependent constants $c_{1,2}$, regular for all $t$ as in the
scalar and pseudoscalar channels. The cross-image remainders contribute only
to the finite part. Taking the scalar value as the unit,
the relative constants are
\begin{align}
&\alpha_S : \alpha_P : \alpha_V : \alpha_A : \alpha_T : \alpha_X \notag \\
=& 1 : 1 : \tfrac23 : \tfrac23 : \tfrac13 : \tfrac13,
\end{align}
where $\alpha_V$, $\alpha_A$, $\alpha_T$, $\alpha_X$ are per spatial
polarization and are equal for $k=1,2,3$. The relative constants are magnitudes
of ratios of Dirac traces and are
independent of the representation of the $\gamma$-matrices. In particular
the coincident-point singularity is the
$T$-independent $1/|t|^3$ in every channel, so the additive power divergence
$\alpha_\Gamma/(2a^2)$ is temperature-independent throughout, as used in
Sec.~\ref{subsec:additive}.

\subsection{Mass-dependent divergence}
\label{app:thermal_mass}

The same image decomposition shows that the mass-dependent divergence is
also temperature-independent. The finite-temperature massive propagator is
the antiperiodic image sum of the closed-form zero-temperature massive
propagator~(\ref{eq:massive_prop}),
\begin{equation}
S_T^{(m)}(\vec x, t) = \sum_{n=-\infty}^{\infty}(-1)^n\,
S^{(m)}(\vec x,\, t+n/T).
\end{equation}
At $\vec{x}=0$, each image is singular only at its own coincident point
$t=-n/T$. The correlator is built from two such propagators and is therefore a
double image sum, and exactly as in the massless case only the diagonal
$n=n'=0$ term is singular at $t\to 0$, every other image and every cross-image
pair being regular there. That term is the zero-temperature massive
correlator, so the mass-dependent divergences are
temperature-independent.

Explicitly, the leading mass-dependent singularity is the $m^2$ term of
eq.~(\ref{eq:massive_expansion}), $\sim m^2/(x^2)^2$, whose spatial integral
is $\pi^2 m^2/|t|$
(Sec.~\ref{subsec:additive}). Including the prefactor of
eq.~(\ref{eq:tree_corr}) and the channel sign of the manifestly-positive
normalization, the $O(m^2)$ part of the zero-temperature correlator is
$C_\Gamma^{(m^2)}(t)\big|_{T=0} = c_m^\Gamma\, m^2/(2|t|)$, with the
channel-dependent coefficient $c_m^\Gamma$ computed in
Appendix~\ref{app:cm_coeff}. Summing the diagonal images gives the
periodic $O(m^2)$ kernel, even in $t$,
\begin{equation}
C_\Gamma^{(m^2)}(t,T)\Big|_{\rm diag}
= \frac{c_m^\Gamma\, m^2}{2}\sum_{n=-\infty}^{\infty}\frac{1}{|t+n\beta|},
\label{eq:K2_sum}
\end{equation}
the massive analogue of eq.~(\ref{eq:K_sum}), with $1/|t+n\beta|$ in place
of $1/|t+n\beta|^3$ and $c_m^\Gamma m^2/2$ in place of $\alpha_\Gamma/2$,
whose $n=0$ image is the coincident-point singularity $1/|t|$. The
cross-image terms are regular, as in the massless case, and belong to the
finite part. Over the two intervals of eq.~(\ref{eq:chi_bare_two}) this
even singularity is picked up at both ends, each interval giving
\begin{equation}
\int_\eps^{1/(2T)}\frac{dt}{|t|} = \int_{-1/(2T)}^{-\eps}\frac{dt}{|t|}
= \ln\frac{1}{2\eps T}.
\label{eq:log_interval}
\end{equation}
Multiplying by $c_m^\Gamma m^2/2$ and summing the two intervals, the $n=0$
image of eq.~(\ref{eq:K2_sum}) contributes
\begin{equation}
c_m^\Gamma\, m^2\ln\frac{1}{2\eps T}
= c_m^\Gamma\, m^2\ln\frac{1}{\eps m} + c_m^\Gamma\, m^2\ln\frac{m}{2T}
\label{eq:log_split}
\end{equation}
to $\chi_\Gamma^{\rm bare}$. The first term is the temperature-independent
logarithmic divergence of eq.~(\ref{eq:mass_div}), with $\eps\to a$ on the
lattice. The second is finite and belongs to the temperature-dependent
part. The $n\neq 0$ images are regular within each interval and integrate
to further finite, cutoff-independent contributions. The competition
between the scales $1/m$ and $1/T$ at the infrared end of the logarithm
thus resides entirely in the finite part, as the $m^2\ln(m/T)$ piece, and
never enters the divergent coefficient. The mass-dependent divergence of
eq.~(\ref{eq:mass_div}) is therefore temperature-independent, and is
removed by the temperature subtraction together with the power divergence.

\subsubsection{Coefficient of the mass logarithm}
\label{app:cm_coeff}

The coefficient $c_m^\Gamma$ in eq.~(\ref{eq:mass_div}) follows from the
$O(m^2)$ part of the one-loop susceptibility. As shown above, this additive
divergence is temperature-independent, so its coefficient may be computed at
zero temperature, where the finite-temperature Matsubara sum reduces to the
four-dimensional integral. With the operators normalized so that each
$\chi_\Gamma$ is manifestly positive (the power divergence $\alpha_\Gamma/2a^2$
is positive in every channel), the zero-temperature susceptibility is
\begin{align}
&\chi_\Gamma^{\rm bare} = \pm\, N_c\,\tr_F[t^a t^a]\int\!\frac{d^4p}{(2\pi)^4}\,
\tr_D\!\big[\Gamma\, S(p)\,\Gamma\, S(p)\big], \notag \\
&S(p) = \frac{m - i\slashed p}{p^2 + m^2},
\label{eq:chi_mom}
\end{align}
with $N_c$ the number of colors, $\tr_F[t^a t^a]$ the flavor factor, and the
channel sign fixed by positivity. This is the momentum-space counterpart of the position-space
propagator~(\ref{eq:massive_kernel}). The mass part of $S(p)$ is a Dirac
scalar and gives $\tr_D[\Gamma^2]$, the \emph{same} sign for scalar and
pseudoscalar, mirroring the $K_1^2$ term. The $\slashed p$ part is a Dirac
vector and gives $\tr_D[\Gamma\slashed p\,\Gamma\slashed p]$,
\emph{opposite} signs for the two channels, mirroring the
$\eta_\Gamma K_2^2$ term. Cross terms vanish by the same
odd-number-of-$\gamma$ argument. The two combine in the numerator Dirac
trace,
\begin{align}
&\tr_D\!\big[\Gamma(m-i\slashed p)\Gamma(m-i\slashed p)\big] \notag \\
=& \begin{cases}
4(m^2 - p^2), & \Gamma = \Id \ (\text{scalar}),\\[2pt]
4(m^2 + p^2), & \Gamma = \gamma_5\ (\text{pseudoscalar}).
\end{cases}
\end{align}
Using $4p^2 = 4(p^2+m^2) - 4m^2$, the two integrands reduce to
\begin{align}
\frac{N_S}{(p^2+m^2)^2} &= \frac{8m^2}{(p^2+m^2)^2} - \frac{4}{p^2+m^2},
\notag \\
\frac{N_P}{(p^2+m^2)^2} &= \frac{4}{p^2+m^2}.
\label{eq:SP_integrands}
\end{align}
With the Euclidean momentum cutoff $\Lambda = 1/\eps$, the master integrals are
\begin{align}
\int\!\frac{d^4p}{(2\pi)^4}\,\frac{1}{(p^2+m^2)^2}
&= \frac{1}{8\pi^2}\ln\frac{1}{\eps m} + \text{finite},
\notag \\
\int\!\frac{d^4p}{(2\pi)^4}\,\frac{1}{p^2+m^2}
&= \frac{1}{16\pi^2 \eps^2} - \frac{m^2}{8\pi^2}\ln\frac{1}{\eps m} + \text{finite}.
\end{align}
The quadratic divergence of the second integral is the power divergence
$\alpha_\Gamma/(2\eps^2)$, i.e.\ $\alpha_\Gamma/(2a^2)$ on the lattice. It is the same for both channels,
$\alpha_S = \alpha_P = N_c\,\tr_F[t^a t^a]/(2\pi^2) > 0$, and cancels in the
difference. Collecting the $m^2\ln(1/(\eps m))$ terms, which become
$m^2\ln(1/(am))$ on the lattice,
\begin{align}
c_m^S &= -N_c\,\tr_F[t^a t^a]\left(\frac{8}{8\pi^2} + \frac{4}{8\pi^2}\right)
= -\,\frac{3\,N_c\,\tr_F[t^a t^a]}{2\pi^2},
\notag \\
c_m^P &= -N_c\,\tr_F[t^a t^a]\,\frac{4}{8\pi^2}
= -\,\frac{N_c\,\tr_F[t^a t^a]}{2\pi^2},
\label{eq:cm_values}
\end{align}
where for $S$ the two terms come from $8m^2/(p^2+m^2)^2$ and
$-4/(p^2+m^2)$ respectively, and the overall sign is fixed by positivity.
The magnitudes are in the ratio $|c_m^S| : |c_m^P| = 3 : 1$. For $N_c = 3$ and
$\tr_F[t^a t^a] = \tfrac12$ these are $c_m^S = -9/(4\pi^2)$ and
$c_m^P = -3/(4\pi^2)$. The scalar coefficient agrees with the standard
one-loop chiral susceptibility obtained from
$\chi_S = -\,\partial\vev{\bar qq}/\partial m$, an independent check.


The same analysis extends to the remaining channels. For the vector,
axial-vector, tensor vector, and axial-tensor vector operators the coefficient of the
$m^2/(x^2)^2$ term is obtained from the spatial integrals that fix the
power-divergence constants of Appendix~\ref{app:thermal}. The indexed
channels single out the temporal direction, so the $m^2$ term of
eq.~(\ref{eq:massive_expansion}) is
projected with the same $\int d^3x$ measure, not by a four-dimensional
average. In the manifestly-positive normalization, the coefficients are
\begin{align}
&c_m^\Gamma = \zeta_\Gamma\,\frac{N_c\,\tr_F[t^a t^a]}{2\pi^2}, \notag \\
&(\zeta_S,\zeta_P,\zeta_V,\zeta_A,\zeta_T,\zeta_X) =
(-3,\,-1,\,0,\,-2,\,+1,\,-1).
\label{eq:cm_all}
\end{align}
For the vector-type channels $V$, $A$, $T$, $X$, the quoted value is that of a
single spatial component $k$, and is the same for $k=1,2,3$.
The scalar and pseudoscalar carry no such index.
The vector coefficient vanishes, $c_m^V = 0$,
consistent with the conservation of the vector current. Current
conservation is a Ward-identity statement about the correlator at separated
points. It protects the mass logarithm and the anomalous dimension
($c_m^V = 0$, $\gamma_V = 0$) but not the power divergence, which is a
short-distance coincident-point divergence. The vector channel therefore
retains a nonzero $\alpha_V$ (below).

The same calculation reproduces the power-divergence constants
$\alpha_S:\alpha_P:\alpha_V:\alpha_A:\alpha_T:\alpha_X = 1:1:\tfrac23:\tfrac23:
\tfrac13:\tfrac13$ (all positive, and equal within each partner pair), which
fixes the normalization.

For each pair of symmetry partners entering $\kappa_{AB}$, namely $(V,A)$,
$(S,P)$, and $(T,X)$, the power divergences are equal and cancel in the
difference, whereas the mass-logarithm coefficients differ,
$c_m^A \neq c_m^B$. The mass logarithm therefore does not cancel in
$\chi_A - \chi_B$, and is removed only by the temperature subtraction, which
acts on each channel separately.

\subsection{Thermal coefficients of the channels}
\label{app:thermal_coeff}

The leading thermal coefficients of eq.~(\ref{eq:c0_channels}) follow
from the momentum-space form (\ref{eq:chi_mom}) at $m=0$, i.e.\ at tree
level in the gauge coupling (the single quark loop of the Wick
contraction with free propagators), with the
finite-temperature measure $T\sum_n\int d^3p/(2\pi)^3$,
$p_4=\omega_n=(2n+1)\pi T$, and $S(p)=-i\slashed p/p^2$. With two
factors of $-i$, the Dirac trace is
$\tr_D[\Gamma S\Gamma S]=-\tr_D[\Gamma\slashed p\,\Gamma\slashed p]/p^4$,
and using $\gamma_\mu\gamma_\nu\gamma_\mu=-\gamma_\nu$ for $\nu\neq\mu$,
\begin{align}
\Gamma=\Id:&\quad \tr_D[\slashed p\slashed p]=4p^2,\notag\\
\Gamma=\gamma_5:&\quad \tr_D[\gamma_5\slashed p\gamma_5\slashed p]=-4p^2,\notag\\
\Gamma=\gamma_k:&\quad \gamma_k\slashed p\gamma_k=2p_k\gamma_k-\slashed p,\notag\\
&\; \tr_D[\ldots]=4(2p_k^2-p^2),\notag\\
\Gamma=\gamma_4\gamma_k:&\quad \gamma_4\gamma_k\slashed p\gamma_4\gamma_k
 =p_4\gamma_4+p_k\gamma_k\notag\\
&\qquad\qquad -\sum_{j\neq k}p_j\gamma_j,\notag\\
&\; \tr_D[\ldots]=4\Big(p_4^2+p_k^2-\sum_{j\neq k}p_j^2\Big),
\label{eq:A_traces}
\end{align}
the axial partners $i\gamma_5\gamma_k$ and $i\gamma_5\gamma_4\gamma_k$
giving the same structures. On the thermal circle the quark fields are antiperiodic, so $p_4$ is
quantized to the fermionic Matsubara frequencies and
$\int dp_4/(2\pi)\to T\sum_n$. The sum $T\sum_n\int d^3p/(2\pi)^3 f$ is
the bare susceptibility at temperature $T$. As $T\to 0$ it reverts to
$\int d^4p/(2\pi)^4 f$, the bare susceptibility at zero temperature.
Their difference,
\begin{equation}
[f]_{\rm th}\equiv T\sum_n\!\int\!\frac{d^3p}{(2\pi)^3}f
-\int\!\frac{d^4p}{(2\pi)^4}f ,
\label{eq:A_thdef}
\end{equation}
is $\chi^{\rm bare}(T)-\chi^{\rm bare}(0)$. It is ultraviolet finite,
since the divergence is the same at all temperatures
(Sec.~\ref{subsec:subtraction}) and cancels, and it carries the entire
temperature dependence. It is the quantity whose leading behavior
$\tilde c_0T^2$ is sought, and it is what enters the regularized
susceptibility, $\chi^{\rm reg}(T;T_r)=[f]_{\rm th}(T)-[f]_{\rm th}(T_r)$.
Three such master integrals are needed. The first is the standard
fermionic result, obtained by performing the Matsubara sum by contour
integration, which yields the vacuum integral plus a Fermi-function
term,
\begin{align}
\Big[\frac{1}{p^2}\Big]_{\rm th}
&=-\int\!\frac{d^3p}{(2\pi)^3}\frac{n_F(p)}{p} \notag\\
&=-\frac{1}{2\pi^2}\int_0^\infty\!\frac{p\,dp}{e^{p/T}+1}=-\frac{T^2}{24}.
\label{eq:A_I1}
\end{align}
The remaining two, $[\omega_n^2/p^4]_{\rm th}$ and
$[p_k^2/p^4]_{\rm th}$, are evaluated by a different route. The spatial
integrals at fixed $\omega$ are, in dimensional regularization,
$\int\!\frac{d^3p}{(2\pi)^3}\frac{1}{\omega^2+\vec p^{\,2}}=-\frac{|\omega|}{4\pi}$
and
$\int\!\frac{d^3p}{(2\pi)^3}\frac{\omega^2}{(\omega^2+\vec p^{\,2})^2}
=\frac{|\omega|}{8\pi}$, and hence
$\int\!\frac{d^3p}{(2\pi)^3}\frac{\vec p^{\,2}}{(\omega^2+\vec p^{\,2})^2}
=-\frac{3|\omega|}{8\pi}$. The fermionic Matsubara sum is then
$T\sum_n|\omega_n|=2\pi T^2\cdot 2\zeta(-1,\tfrac12)=\pi T^2/6$ in
zeta regularization. Unlike the evaluation of eq.~(\ref{eq:A_I1}), in
which the vacuum integral is subtracted explicitly, this route
(dimensionally regularized spatial integral followed by the
zeta-regularized frequency sum) discards the vacuum part through the
regularization itself: the linear divergence of the spatial integral
and the temperature-independent divergent part of $\sum_n|\omega_n|$ are
both set to zero by analytic continuation. This is legitimate because the
vacuum part is scaleless in dimensional regularization while the thermal
part is ultraviolet finite and hence scheme independent, but it must be
verified that what is discarded is only the vacuum part and not also a
finite temperature-dependent constant. The verification is to apply the
route to $1/p^2$, where the explicit subtraction of eq.~(\ref{eq:A_I1})
is available for comparison: the dimensionally regularized spatial
integral $-|\omega|/(4\pi)$ summed with $T\sum_n|\omega_n|=\pi T^2/6$
gives $-T^2/24$, in agreement with eq.~(\ref{eq:A_I1}). The same route is
then used for the two remaining integrals. Therefore
\begin{equation}
\Big[\frac{\omega_n^2}{p^4}\Big]_{\rm th}=+\frac{T^2}{48},\quad
\Big[\frac{\vec p^{\,2}}{p^4}\Big]_{\rm th}=-\frac{T^2}{16},\quad
\Big[\frac{p_k^2}{p^4}\Big]_{\rm th}=-\frac{T^2}{48}
\label{eq:A_I2}
\end{equation}
for each spatial component $k$, consistent with
$[1/p^2]_{\rm th}=[\omega_n^2/p^4]_{\rm th}+[\vec p^{\,2}/p^4]_{\rm th}$.
Inserting eqs.~(\ref{eq:A_traces}) and (\ref{eq:A_I2}), with the
overall sign of each channel fixed by the manifestly positive
normalization and $N_c\tr_F[t^at^a]$ factored out,
\begin{align}
S,P:&\quad 4\Big[\frac{1}{p^2}\Big]_{\rm th}=-\frac{T^2}{6},\notag\\
V,A:&\quad 4\Big(\Big[\frac{1}{p^2}\Big]_{\rm th}-2\Big[\frac{p_k^2}{p^4}\Big]_{\rm th}\Big)
 =4\Big(-\frac{1}{24}+\frac{2}{48}\Big)T^2=0,\notag\\
T,X:&\quad 4\Big(\Big[\frac{\omega_n^2}{p^4}\Big]_{\rm th}+\Big[\frac{p_k^2}{p^4}\Big]_{\rm th}
 -2\Big[\frac{p_j^2}{p^4}\Big]_{\rm th}\Big)=+\frac{T^2}{6},
\label{eq:A_c0}
\end{align}
which is eq.~(\ref{eq:c0_channels}). The vanishing of the vector
coefficient is exact at this order: the $-T^2/24$ of the $p^2$ term is
compensated by the two spatial components. These coefficients refer to
the full four-volume sum, which the momentum integral represents. The
equal-time slice omitted in eq.~(\ref{eq:chi_bare_lat_ne0}) is a
three-dimensional sum of a dimension-six correlator, so its thermal part
is $O(T^3)$. Weighted by the slice width $a$ it contributes
$aT^3=T^2/N_t$, suppressed by $1/N_t$ relative to $\tilde c_0T^2$. The
$t\neq 0$ sums therefore share the coefficients up to $O(1/N_t)$
corrections, which vanish in the continuum limit at fixed temperature.
The vacuum parts of the same momentum integrals are the power
divergences of the full four-volume sum. They differ from the
coefficients $\alpha_\Gamma$ of Appendix~\ref{app:thermal_massless}, which are
the power divergences of the $t\neq 0$ sum alone: the latter integrates
over three spatial dimensions at fixed $t$ and has no four-dimensional
symmetry, whereas the former averages each $p_\mu^2$ to $p^2/4$. For the
tensor channels this makes a qualitative difference: the numerator
$p_4^2+p_k^2-p_j^2-p_l^2$ integrates to $(1+1-1-1)p^2/4=0$ over the
four-dimensionally symmetric measure, so the full four-volume sum of the
tensor correlator has no power divergence at all, while
$\alpha_T=\alpha_S/3\neq 0$ for the $t\neq 0$ sum. The omitted equal-time
slice therefore carries a power divergence equal and opposite to
$\alpha_T$ and cancels it, which is why the equal-time slice of the
tensor correlator is large and of opposite sign to the rest.

\subsection{Massive and nondegenerate quarks}
\label{app:thermal_mass_nondeg}

The computation of Appendix~\ref{app:thermal_coeff} extends directly to
propagators of masses $m_1$ and $m_2$, as for a bilinear $\bar q_1\Gamma q_2$
with nondegenerate flavors. With $S_i(p)=(m_i-i\slashed p)/(p^2+m_i^2)$ the
cross terms are odd in $\slashed p$ and vanish, so
\begin{equation}
\tr_D\big[\Gamma S_1\Gamma S_2\big]
=\frac{m_1m_2\tr_D[\Gamma^2]-\tr_D[\Gamma\slashed p\Gamma\slashed p]}
{(p^2+m_1^2)(p^2+m_2^2)} .
\label{eq:A_nondeg_num}
\end{equation}
After the isotropic average $p_k^2\to\vec p^{\,2}/3$ under the spatial
integral, the numerators of eq.~(\ref{eq:A_traces}) and their axial
partners take, in the manifestly positive normalization, the common form
\begin{equation}
N_\Gamma=4\big[\epsilon_\Gamma\,m_1m_2+a_\Gamma\,\omega_n^2+b_\Gamma\,\vec p^{\,2}\big],
\label{eq:A_NGamma}
\end{equation}
with
\begin{equation}
(\epsilon_\Gamma,a_\Gamma,b_\Gamma)=
\begin{cases}
(+1,\,1,\,1) & P,\\ (-1,\,1,\,1) & S,\\
(+1,\,1,\,\tfrac13) & V,\\ (-1,\,1,\,\tfrac13) & A,\\
(+1,\,1,\,-\tfrac13) & T,\\ (-1,\,1,\,-\tfrac13) & X.
\end{cases}
\label{eq:A_eab}
\end{equation}
The chirally odd term $\epsilon_\Gamma m_1m_2$ changes sign between the
partners of each pair and is the source of their splitting at tree
level. The chirally even terms are common to the partners.

All Matsubara sums reduce, by partial fractions in $\omega_n^2$, to
\begin{align}
\mathcal A(E)&\equiv T\sum_n\frac{1}{\omega_n^2+E^2}=\frac{1-2n_F(E)}{2E},
\notag\\
\mathcal A_{\rm th}(E)&=-\frac{n_F(E)}{E},
\label{eq:A_calA}
\end{align}
with $E_i=\sqrt{\vec p^{\,2}+m_i^2}$. Using partial fractions,
\begin{align}
&[(\omega^2+E_1^2)(\omega^2+E_2^2)]^{-1} \notag \\
=&(E_2^2-E_1^2)^{-1} [(\omega^2+E_1^2)^{-1}-(\omega^2+E_2^2)^{-1}], \notag \\
&\omega^2[(\omega^2+E_1^2)(\omega^2+E_2^2)]^{-1}  \notag \\
=&(E_2^2-E_1^2)^{-1}[E_2^2(\omega^2+E_2^2)^{-1}-E_1^2(\omega^2+E_1^2)^{-1}], \notag
\end{align}
the thermal part of any channel is
\begin{align}
\chi_\Gamma^{\rm th}&=\frac{\mathcal N}{m_2^2-m_1^2}\!\int\!\frac{d^3p}{(2\pi)^3}
\Big\{a_\Gamma\big[E_2^2\mathcal A_{\rm th}(E_2)-E_1^2\mathcal A_{\rm th}(E_1)\big]
\notag\\
&\qquad+\big(\epsilon_\Gamma m_1m_2+b_\Gamma\vec p^{\,2}\big)
\big[\mathcal A_{\rm th}(E_1)-\mathcal A_{\rm th}(E_2)\big]\Big\},
\label{eq:A_nondeg}
\end{align}
with $\mathcal N\equiv 4N_c\tr_F[t^at^a]$ and $E_2^2-E_1^2=m_2^2-m_1^2$. For
degenerate masses the differences
become derivatives: with
$\mathcal B_{\rm th}(E)\equiv-\partial\mathcal A_{\rm th}/\partial E^2
=-\frac{n_F(E)}{2E^3}+\frac{n_F'(E)}{2E^2}$,
\begin{equation}
\chi_\Gamma^{\rm th}=\mathcal N\!\int\!\frac{d^3p}{(2\pi)^3}
\Big\{a_\Gamma\big[\mathcal A_{\rm th}-E^2\mathcal B_{\rm th}\big]
+\big(\epsilon_\Gamma m^2+b_\Gamma\vec p^{\,2}\big)\mathcal B_{\rm th}\Big\},
\label{eq:A_deg}
\end{equation}
so that, e.g., $\chi_P^{\rm th}=\mathcal N\int\mathcal A_{\rm th}$,
$\chi_S^{\rm th}=\mathcal N\int(\mathcal A_{\rm th}-2m^2\mathcal B_{\rm th})$,
and $\chi_T^{\rm th}=\mathcal N\int(\mathcal A_{\rm th}-\tfrac43\vec p^{\,2}\mathcal B_{\rm th})$.
In the massless limit $\int\frac{d^3p}{(2\pi)^3}[\mathcal A_{\rm th}-E^2\mathcal B_{\rm th}]
=[\omega_n^2/p^4]_{\rm th}=T^2/48$ and
$\int\frac{d^3p}{(2\pi)^3}\vec p^{\,2}\mathcal B_{\rm th}=[\vec p^{\,2}/p^4]_{\rm th}=-T^2/16$,
and eq.~(\ref{eq:A_deg}) collapses to
\begin{equation}
\tilde c_0^{\,\Gamma}=\frac{a_\Gamma-3b_\Gamma}{12}\,N_c\tr_F[t^at^a],
\label{eq:A_c0_general}
\end{equation}
which reproduces eq.~(\ref{eq:c0_channels}) for all six channels at
once. At the next order in $m^2/T^2$, expanding eq.~(\ref{eq:A_deg})
at $T\gg m$ gives
\begin{align}
\chi_\Gamma^{\rm th}=N_c\tr_F[t^at^a]\Big[\tilde c_0^{\,\Gamma}T^2
&-\frac{a_\Gamma+3b_\Gamma-2\epsilon_\Gamma}{4\pi^2}\,m^2\ln\frac{m}{T}
\notag\\
&+O(m^2)\Big],
\label{eq:A_m2log}
\end{align}
the coefficient of the logarithm being $(-1,-3,0,-2,+1,-1)/(2\pi^2)$ for
$(P,S,V,A,T,X)$. These are exactly the coefficients $c_m^\Gamma$ of the
mass logarithm in Appendix~\ref{app:cm_coeff}, as required by
eq.~(\ref{eq:log_split}): the vacuum logarithm $m^2\ln[1/(\eps m)]$ and
the thermal logarithm $m^2\ln(m/T)$ combine into $m^2\ln[1/(\eps T)]$,
which provides an independent check of the two appendices against each
other. At nonzero mass the partner splitting of each pair is
$\chi_A^{\rm th}-\chi_B^{\rm th}=2\mathcal N\,m_1m_2
\int\frac{d^3p}{(2\pi)^3}[\mathcal A_{\rm th}(E_1)-\mathcal A_{\rm th}(E_2)]/(m_2^2-m_1^2)$,
i.e.\ $2\mathcal N m^2\int\mathcal B_{\rm th}$ for degenerate masses. This
is the $O(m^2)$ explicit-breaking remnant at high temperature. Its vacuum
counterpart is the mass logarithm of Sec.~\ref{subsec:additive}, which for
nondegenerate quarks reads $m_1m_2\ln(1/(a\sqrt{m_1m_2}))$ at leading
order.

\end{document}